\documentclass[final,5p,times,twocolumn]{elsarticle}

\usepackage{graphicx}
\usepackage{amsmath}

\usepackage{amssymb}
\usepackage{lineno}
\usepackage{siunitx}
\usepackage{array}
\usepackage{subfigure}
\usepackage{booktabs}

\usepackage[colorlinks]{hyperref}
\biboptions{compress}
\usepackage{xspace}
\journal{Nucl. Instrum. Methods Phys. Res., Sectiont. A}

\begin{document}

\begin{frontmatter}



%\title{A Generalized Empirical Description of Gain and Gain Resolution in GEM-based Detectors Over Broad Reduced Field Ranges}
%\title{A Generalized Empirical Description of Gain and Gain Resolution in GEM-based Detectors Over Broad Parameter Space}
%\title{Application of a General Form of the First Townsend Coefficient to Avalanche Gain and Its Resolution in GEM-based Detectors Over Broad Parameter Space}

%\title{Parameterization of Measured Avalanche Gain and Its Resolution in GEM-based Detectors}

%\title{Avalanche Gain and Its Effect on Energy Resolution in GEM-based Detectors}

\title{Avalanche gain and its effect on energy resolution in GEM-based detectors}

%\title{High avalanche gain operation in GEM-based detectors and its effect on energy resolution}

%\title{Parameterizing Measurements of Avalanche Gain and Its Resolution in GEM-based Detectors}

%\title{Describing Gain and Gain Resolution Data Via the Application of the Generalized First Townsend Coefficient in GEM-based Detectors Over Broad Parameter Space}

%\author[]{T.~N.~Thorpe\corref{cor1} and \author{S.~E.~Vahsen}
%\author{T.~N.~Thorpe \corref{cor1} \coref{a}} %and S.~E.~Vahsen}
%\cortext[cor1]{Corresponding author. Tel.: +39 328 638 5202}
%\cortext[a]{test}

\author[]{T.~N.~Thorpe\corref{cor1}\fnref{label1}}
\ead{tnthorpe@g.ucla.edu}
\cortext[cor1]{Corresponding author. Tel.: +41227679281}
\fntext[label1]{Now at Department of Physics and Astronomy, University of California - Los Angeles, 475 Portola Plaza, Los Angeles, CA 90095}
%\fntext[label2]{Now at Laboratori Nazionali del Gran Sasso, Via G. Acitelli, 22, Assergi, AQ 67100, Italy}

%\address{Department of Physics \& Astronomy, University of Hawaii, 2505 Correa Road, Honolulu, HI 96822, USA}
%\address{Department of Astroparticle Physics, Gran Sasso Science Institute, Via F. Crispi, 7, L'Aquila, AQ 67100, Italy}
%\address{Laboratori Nazionali del Gran Sasso, Via G. Acitelli, 22, Assergi, AQ 67100, Italy}

\author{S.~E.~Vahsen}
\address{Department of Physics and Astronomy, University of Hawaii, Honolulu, HI 96822, USA}

\begin{abstract}

%Gas Electron Multipliers (GEMs) are gas avalanche devices that have been enabled by modern photo-lithography.  Compared to traditional multi wire proportional chambers, GEMs offer smaller feature size and higher rate capabilities.  When designing GEM-based detectors, often a large parameter space of gas mixtures, GEM specifications, and operational parameters needs to be considered.  In this context, a description which parameterizes the gain and the energy resolution to include these quantities is valuable for efficient detector design optimization and interpretation of results.  

We present avalanche gain and associated resolution measurements recorded with a $^4$He:CO$_2$ (70:30) gas mixture and pure SF$_6$, a Negative Ion (NI) gas.  SF$_6$ is of particular interest to the directional dark matter detection community, as its low thermal diffusion helps to retain recoil ionization track features over long drift lengths.  With the aid of a general form of the reduced first Townsend coefficient (RFTC), multiple GEM-based detector data sets are used to study the high-gain behavior of the $^4$He:CO$_2$ gas mixture.  The high-gain data is well described purely in terms of the reduced electric field strength and the number of GEMs, and the robust relationship between the RFTC and the average, reduced, electric field strength across the GEMs is emphasized.  The associated (pulse-height) resolution measurements are used to discuss the variance of the avalanche distribution and to describe and estimate the lower limits of energy resolution one should expect to measure using a simple relationship with the RFTC.  In the end, a description of avalanche gain, its effect on energy resolution, and the contributing experimental parameters in GEM-based detectors is developed over a broad parameter space for further use.  

\end{abstract}


\end{frontmatter}

%\tableofcontents

%\linenumbers

%%%%
\section{Introduction}\label{sec:intro}

\begin{figure*}[!h]
  \centering
 	
	%%%%%\subfigure {\includegraphics[width=7cm]{gain/images/dcube_milli_2_crop2.png}}
	%\subfigure {\includegraphics[width=7.04cm]{images/dcube_milli_2_crop.jpeg}}
	%\vspace{0pt}	
	%%%%%%%%\subfigure {\includegraphics[width=10.5cm]{gain/images/milli2_sources_crop2.png}}
	%\subfigure {\includegraphics[width=11.2cm]{images/milli2_sources_crop.jpg}}
	
	%\subfigure [3-tGEM]]{\includegraphics[width=6.37cm]{images/dcube_milli_2_crop.jpeg}}	
	%\subfigure [1-THGEM]{\includegraphics[width=11.65cm]{images/dcube_milli_4.jpg}}
	%\subfigure [2-THGEM]{\includegraphics[width=11.7cm]{images/milli2_sources_crop.jpg}}
	%\subfigure [2-tGEM]{\includegraphics[width=6.31cm]{images/milli2_thin.jpg}}
	
	\subfigure [3-tGEM: Setup with three \SI{50}{\mu m} GEMs.] {\includegraphics[width=6.345cm]{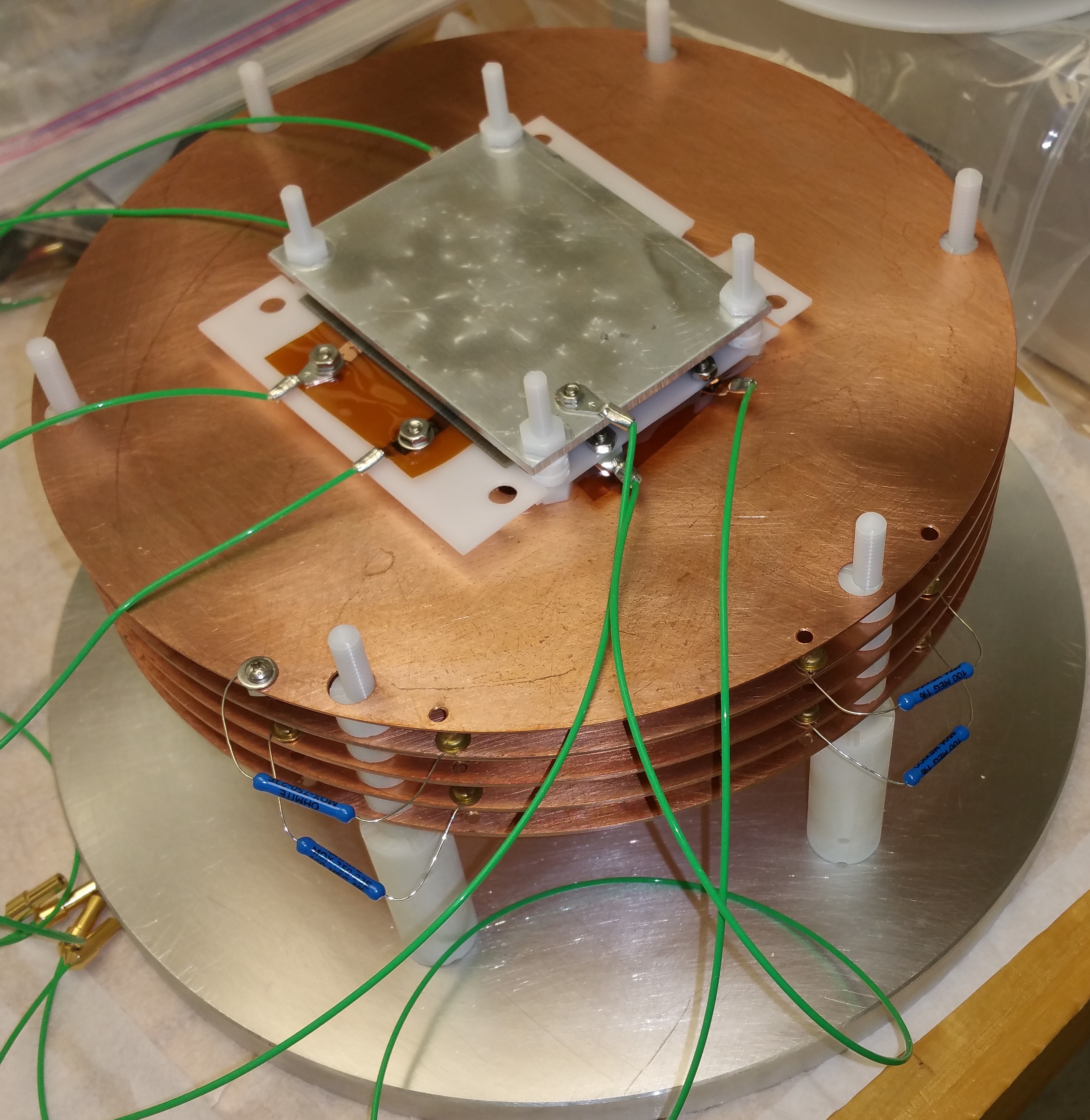}}	
	\subfigure [1-THGEM: Setup with one \SI{400}{\mu m} GEM.] {\includegraphics[width=11.595cm]{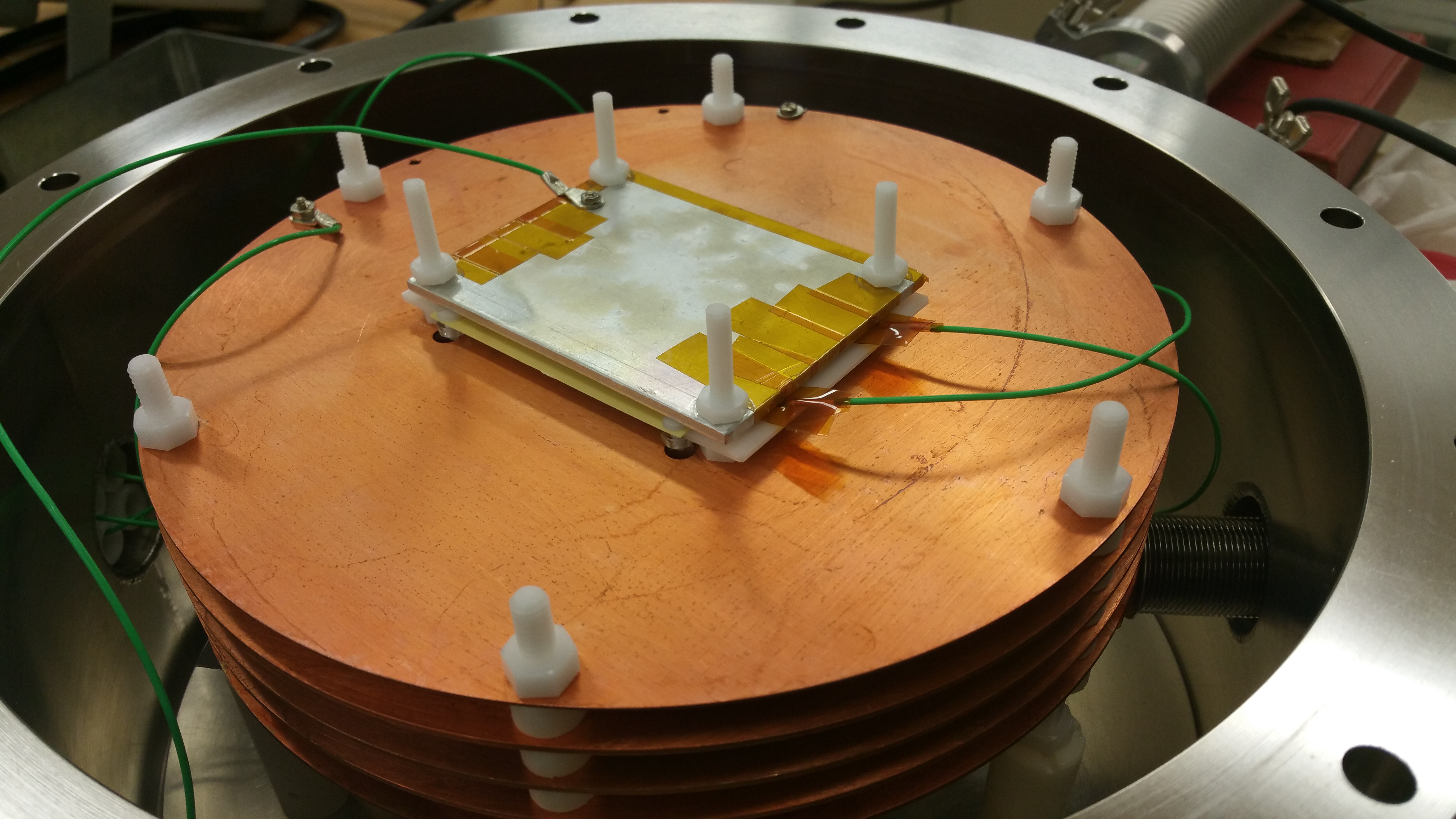}}
	\subfigure [2-THGEM: Setup with two \SI{400}{\mu m} GEMs.] {\includegraphics[width=11.66cm]{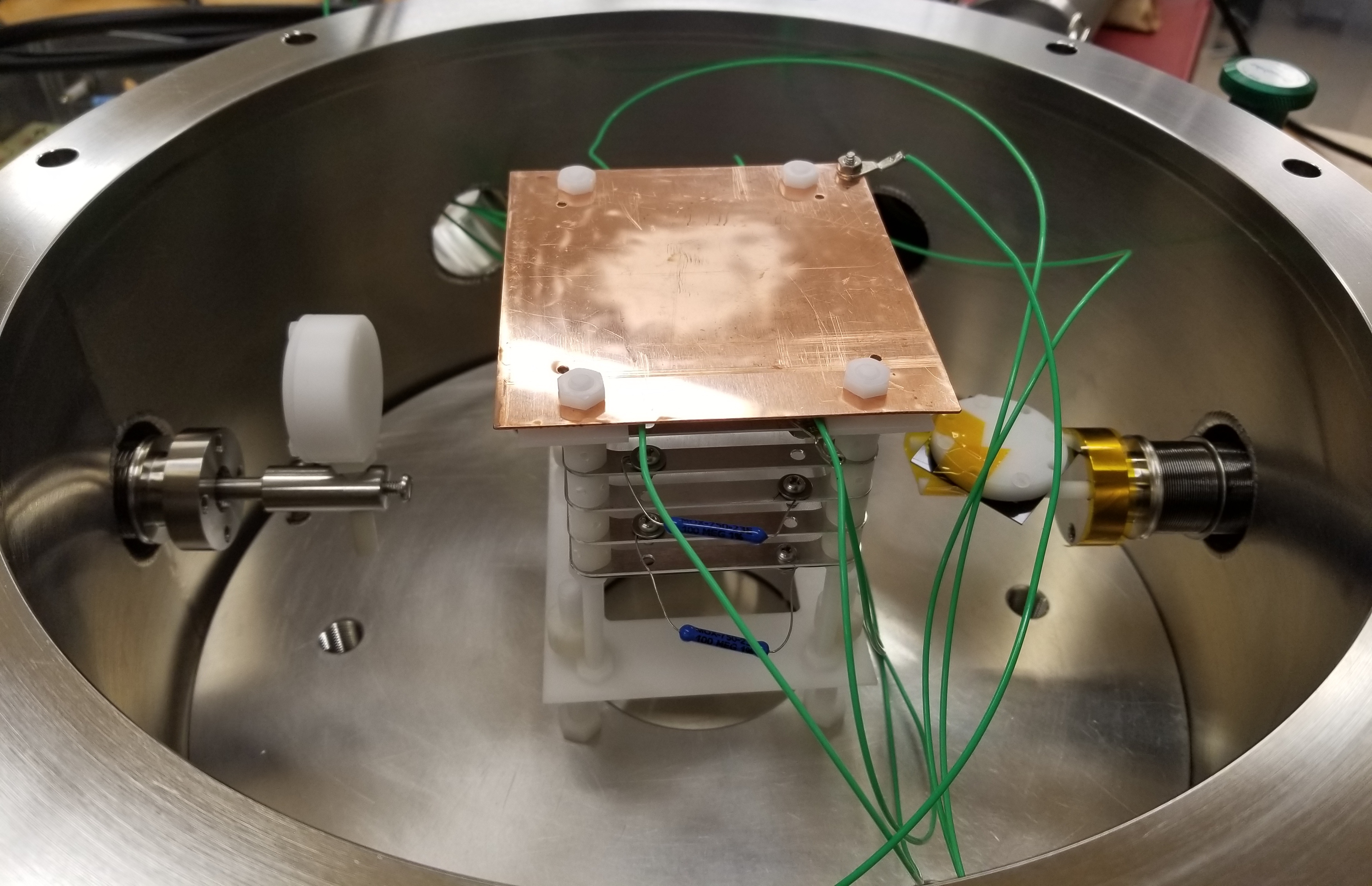}}
	\subfigure [2-tGEM: Setup with two \SI{50}{\mu m} GEMs.] {\includegraphics[width=6.28cm]{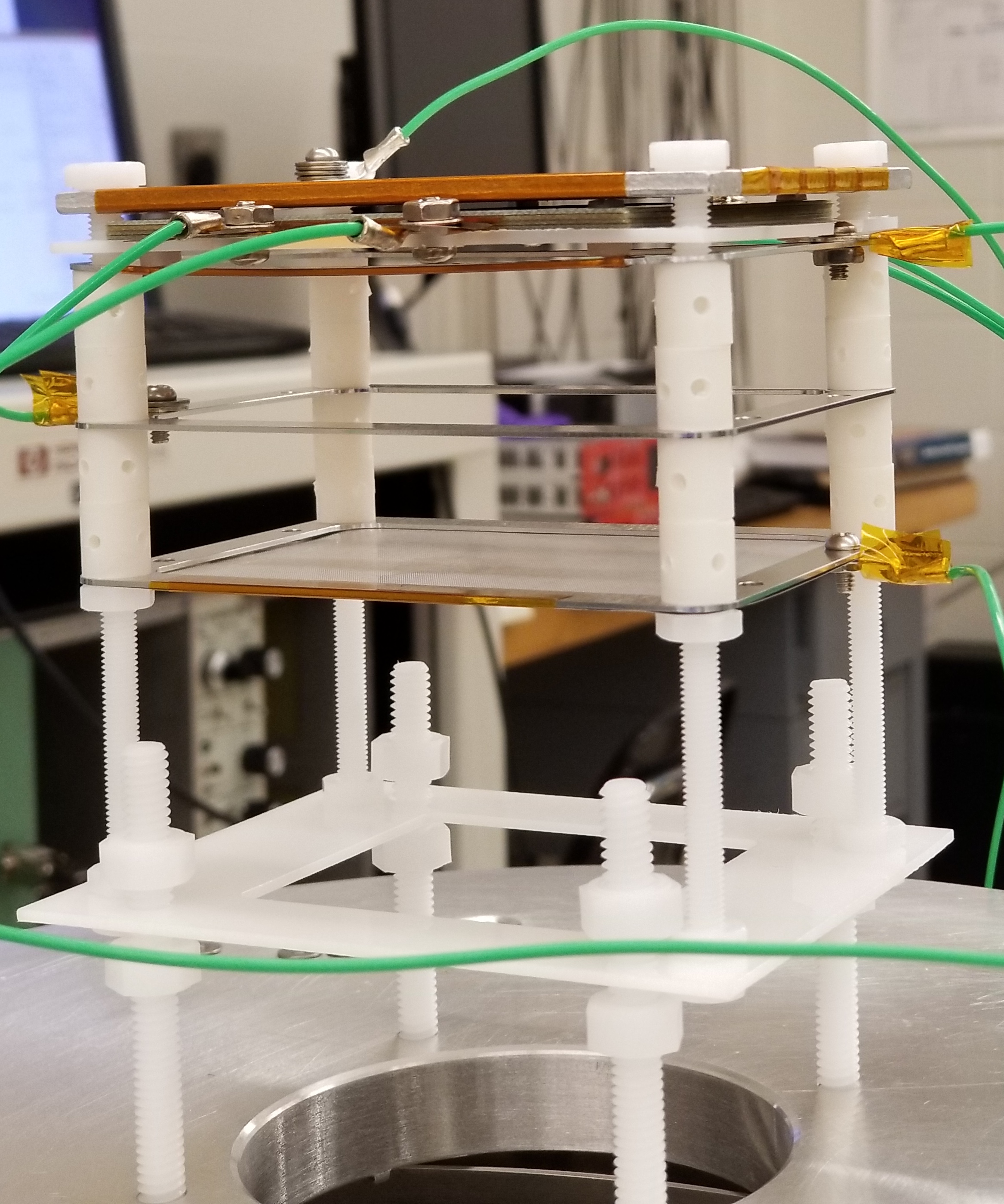}}
	
%	\subfigure {\includegraphics[width=7cm]{images/dcube_milli_2_crop.jpeg}}
%	\subfigure {\includegraphics[width=11.2cm]{images/milli2_sources_crop.jpg}}
%	\subfigure {\includegraphics[width=12.3cm]{images/dcube_milli_4.jpg}}
%	\subfigure {\includegraphics[width=5.8cm]{images/milli2_thin.jpg}}

  \setlength{\abovecaptionskip}{2pt}
  \caption{Experimental setups.  Top: (a) 3-tGEM and (b) 1-THGEM used larger copper field cage rings and suffered from higher material outgassing.  Bottom: (c) 2-THGEM and (d) 2-tGEM used thinner aluminum field cage rings and provided a very stable environment.  The white disk (Delrin) pictured facing upward in (c) is the collimated holder of a $^{55}$Fe source that can be moved laterally to perform either source or background measurements.  Negative charge drifts upwards in all setups, and the aluminum (top and lower right) and copper (lower left) charge collection plates can been seen mounted on top of the structures.}
 \setlength{\belowcaptionskip}{0pt}
 \label{fig:dcube_milli}
 \end{figure*}
\begin{table*}[!h]
\centering
  %\begin{tabular}{l | l | l | c | c | c | cr}
  \begin{tabular}{lll cccc}
  \toprule	
  Setup name & GEM assembly & Gas used (Sections discussed) & Drift  & Transfer 1 & Transfer 2 & Collection  \\
  \hline %\\
  \noalign{\vskip 2mm}
   %\rule{0pt}{5ex}
   %\vspace{0.2cm}
   %D$^{3} \mu$ & 2 thin & $^4$He:CO$_2$ (\ref{sec:heco2}) and Ar:CO$_2$ (\ref{sec:arco2}) & 0.92 & 0.28 & n/a & 0.22 \rule{0pt}{3ex} \\
   %2-tGEM* & 2 thin GEMs &  $^4$He:CO$_2$ (\ref{sec:heco2_arco2}) and Ar:CO$_2$ (\ref{sec:heco2_arco2}) & 0.92 & 0.28 & n/a & 0.22 \rule{0pt}{3ex} \\
   
   %2-tGEM & 2 thin GEMs & $^4$He:CO$_2$ (\ref{sec:heco2}, \ref{sec:heco2_analysis}) & \SI{4.53}{cm} & \SI{0.13}{cm} & n/a  & \SI{0.16}{cm} \rule{0pt}{5ex} \\
   2-tGEM & 2 thin GEMs & $^4$He:CO$_2$ (\ref{sec:heco2}, \ref{sec:heco2_analysis}) & \SI{4.53}{cm} & \SI{0.13}{cm} & n/a  & \SI{0.16}{cm} \\
   %D$^{3}$M1 & 3 thin & $^4$He:CO$_2$ (\ref{sec:heco2}) & 4.92 & 0.28 & 0.28 & 0.23 \\
   3-tGEM & 3 thin GEMs & $^4$He:CO$_2$ (\ref{sec:heco2}, \ref{sec:heco2_analysis}) & \SI{4.92}{cm} & \SI{0.28}{cm} & \SI{0.28}{cm} & \SI{0.23}{cm} \\
   %D$^{3}$M1 & 1 THGEM & $^4$He:CO$_2$ (\ref{sec:heco2}) and SF$_6$ (\ref{sec:sf6}) & 5.14  & n/a  & n/a & 0.18 \\
   %1-THGEM & 1 THGEM & $^4$He:CO$_2$ (\ref{sec:heco2}, \ref{sec:heco2_analysis}) and SF$_6$ (\ref{sec:sf6}, \ref{sec:sf6_analysis}) & 5.14  & n/a  & n/a & 0.18 \\
   1-THGEM & 1 THGEM & $^4$He:CO$_2$ (\ref{sec:heco2}, \ref{sec:heco2_analysis}) and SF$_6$ (\ref{sec:sf6}) & \SI{5.14}{cm}  & n/a  & n/a & \SI{0.18}{cm} \\
   %D$^{3}$M2 & 2 thin & $^4$He:CO$_2$ (\ref{sec:heco2}) & 4.53 & 0.13 & n/a  & 0.16 \\
   
   %D$^{3}$M2 & 2 THGEM & $^4$He:CO$_2$ (\ref{sec:heco2}) &4.72 & 0.09 & n/a  & 0.13 \\
   2-THGEM & 2 THGEMs & $^4$He:CO$_2$ (\ref{sec:heco2}, \ref{sec:heco2_analysis}) & \SI{4.72}{cm} & \SI{0.09}{cm} & n/a  & \SI{0.13}{cm} \\
  \bottomrule
  \end{tabular}
  \setlength{\abovecaptionskip}{6pt}
  \caption{GEM setups and gas mixtures used.  The proportion of the $^4$He:CO$_2$ mixture is 70:30, while SF$_6$ is pure.  Dimensions along the drift direction of the drift, transfer, and collection regions are listed.  GEM specifications are provided in Table~\ref{tab:GEM_dimensions}.}
  \setlength{\belowcaptionskip}{0pt}
  \label{tab:dcube_dimensions}
\end{table*}

Gas Electron Multipliers (GEMs)~\cite{Sauli:1997qp}, a form of Micro-Patterned Gaseous Detectors (MPGD), are now widely used in the detection of elementary particles.  Their higher rate capabilities and finer segmentation than traditional multi-wire proportional chambers (MWPCs) have led to their use in modern experiments utilizing gaseous time projection chambers (TPCs)~\cite{Nygren:1978rx} with GEM-based charge readout~\cite{ALICETPC:2020ann}, and have application in particle tracking systems \cite{Colaleo:2015vsq}.  When coupled with highly segmented detection elements, GEM-based detectors can produce high-definition 3d images of individual ionization events~\cite{Vahsen:2014fba} where the avalanche gain and its effect on energy resolution play an important role.  The measurements presented here were performed during gas component and gain optimization studies for directional neutron detection~\cite{Jaegle:2019jpx, Lewis:2018ayu} and directional dark matter detection at the University of Hawaii~\cite{Vahsen:2014fba, Jaegle:2019jpx, Lewis:2018ayu, Vahsen:2011qx, Kim:2008zzi, Vahsen:2014mca, Lewis:2014poa, Vahsen:2020pzb}.  More generally, detectors capable of achieving recoil directionality by imaging ionization distributions with high spatial resolution may enable a number of novel, future dark matter (DM), neutrino, and precision recoil experiments~\cite{Vahsen:2021gnb}.

%Within the directional DM detection community, gaseous TPCs remain the most mature technology due to their superior ability in discriminating nuclear (signal) from electron (background) recoil ionization tracks.  Central to this ability are detectors that feature a high signal-to-noise ratio (SNR), low diffusion in the drift region, and the ability to operate with low density gas \cite{Vahsen:2020pzb, Vahsen:2021gnb, Mayet:2016zxu, Battat:2016pap}.  These features allow the subtle differences between the ionization track topologies from nuclear and electron recoils to be measured, especially when coupled with highly segmented readouts.  At low ionization energies, $\mathcal{O}$(1-\SI{10}{keV}), such discrimination becomes difficult and the detection of every primary electron becomes crucial, and this is where a high SNR becomes important.  Operation at low gas pressures is important as it allows for long recoil tracks which better constrain track direction, and low diffusion in the drift region is important for retaining individual ionization track topologies over longer drift lengths.  A good energy resolution is crucial for directional DM searches in order to resolve the so-called `head/tail signal', which drastically reduces the number of detected DM recoils required to claim a discovery via the expected dipole distribution in galactic coordinates \cite{Morgan:2004ys}.  The detection of such a signal, from DM collisions, would definitively prove that DM is not of terrestrial or solar origin \cite{Spergel:1987kx}.  

Within the directional DM detection community, gaseous TPCs remain the most mature technology due to their superior ability to discriminate nuclear (signal) from electron (background) recoil ionization tracks.  Central to this ability are detectors that feature a high signal-to-noise ratio (SNR), low diffusion in the drift region, and the ability to operate with low density gas~\cite{Vahsen:2020pzb, Vahsen:2021gnb, Mayet:2016zxu, Battat:2016pap}.  At low ionization energies, $\mathcal{O}$(1-\SI{10}{keV}), such discrimination becomes difficult and the detection of every primary electron becomes crucial, and high SNR becomes particularly important.  Operation at low gas pressures is important as it allows for long recoil tracks which better constrain track direction, and low diffusion in the drift region is important for retaining individual ionization track topologies over longer drift lengths.  A good energy resolution is crucial for directional DM searches in order to resolve the so-called `head/tail signal', which drastically reduces the number of detected DM recoils required to claim a discovery via the expected dipole distribution in galactic coordinates~\cite{Morgan:2004ys}.  Detection of such a signal from DM collisions would definitively prove that DM is not of terrestrial or solar origin~\cite{Spergel:1987kx}.  

Here we focus on two topics addressing these issues:  First, GEM-based detector operation at high gas (avalanche) gain with a  $^4$He:CO$_2$ (70:30) gas mixture, resulting in high SNR.  Second, measurements with SF$_6$, a Negative Ion (NI) gas, which has been shown to be suitable for low gas pressure operation with low (near thermal) diffusion~\cite{Phan:2016veo}.

GEMs exploit the principle of Townsend avalanching, where free electrons are accelerated by a strong electric field and produce further ionization that can be measured.  The gaseous avalanching process is stochastic in nature and complicated by microscopic effects including gas impurities, non-ionizing collisions, saturation effects, and Penning effects in mixtures.  All of these effects tend to increase the avalanche variance and will ultimately deteriorate the energy resolution.  There exist computer simulation programs~\cite{garfield} which exploit the Monte Carlo technique and take these effects into account to varying degrees for estimating avalanche gain.  A goal of our work is to compare the detector responses of multi-GEM and multi-THGEM setups operating with a $^4$He:CO$_2$ (70:30) gas mixture, and to estimate the best energy resolution achievable with this technology in a laboratory setting.  We have produced a simple phenomenological model, and fit it to high avalanche gain data obtained with four different experimental setups.

%Using pulse-height data obtained with a $^4$He:CO$_2$ (70:30) gas mixture operating at high avalanche gain in four different experimental setups.
%The amount of ionization produced by a single initial electron is generally referred to as the \textit{gas gain}.  Since the gas gain produced in each single-electron avalanche is not equal, a variance term can be associated with the distribution of single-electron avalanches, i.e. the \textit{avalanche variance}, which is a limiting factor of detector energy resolution.

Our results show good model agreement for the gain and energy resolution over a large parameter space, and the parameterization and analysis methods presented are broadly applicable to GEM-based detectors, or any gas avalanching devices utilizing a moderately uniform electric field with an electron-drift gas.  We summarize three main findings.  1) The robust, nearly-linear relationship between the reduced first Townsend coefficient (RFTC) and the average, reduced electric field in the GEMs.  2) A novel method for extracting the effective ionization potential of the gas mixture, which could be fine-tuned and performed for other gas mixtures in future experiments.  3) After correcting for gain degradation due to material outgassing, the energy resolution is well modeled with the RFTC.  In doing so, we estimate the \textit{minimum} energy resolution achievable with GEM-based detectors for a given gas mixture and ionization energy.

\begin{table*}[!ht]
  \centering
  %\begin{tabular}{l | c | c | c | cr}
  \begin{tabular}{lccccc}
  \toprule
  GEM type & Substrate thickness ($t$) &  Active area & Hole diameter & Pitch & Rim \\
  \hline 
  \noalign{\vskip 2mm}
  %Thin GEM (tGEM)  & \SI{50}{\text{$\mu$}m} & $5 \times \SI{5}{cm^2}$ & \SI{70}{\text{$\mu$}m} & \SI{140}{\text{$\mu$}m} & n/a \rule{0pt}{3ex}\\
  Thin GEM (tGEM)  & \SI{50}{\text{$\mu$}m} & $5 \times \SI{5}{cm^2}$ & \SI{70}{\text{$\mu$}m} & \SI{140}{\text{$\mu$}m} & n/a \\
  Thick GEM (THGEM) &  \SI{400}{\text{$\mu$}m} & $5 \times \SI{5}{cm^2}$ & \SI{300}{\text{$\mu$}m} & \SI{500}{\text{$\mu$}m} & \SI{50}{\text{$\mu$}m}\\
\bottomrule

 \end{tabular}
  \setlength{\abovecaptionskip}{6pt}
  \caption{Specifications of the GEMs used in this work.}
  \setlength{\belowcaptionskip}{0pt}
\label{tab:GEM_dimensions}
\end{table*}

\section{Gases studied}\label{sec:gases_used}

%\subsection{Gas mixtures studied}\label{sec:gases_used}

We will present primary results from two gas mixtures: $^4$He:CO$_2$ (70:30) and SF$_6$ (sulfur hexafluoride).  Helium is always understood to be $^4$He.  He:CO$_2$ consists of a relative proportion of 70:30, while SF$_6$ is pure (99.999\%).  In \ref{sec:arco2} we apply the analysis methods developed here to data from an Ar:CO$_2$ (70:30) gas mixture from Ref.~\cite{Vahsen:2014fba}.

\subsection{$^4$He:CO$_2$ (70:30) gas mixture}\label{sec:heco2_intro}

This work developed within an experimental program studying directional neutron detection.  Helium is close in mass to the neutron, and gives relatively long recoils tracks which is the primary motivation for its use here.  Ref.~\cite{Jaegle:2019jpx} provides performance details of the gas mixture as it relates to extracting a directional signal from scattering neutron recoils.  Another advantage of using helium is that $\alpha$-particle calibration sources can be used to study the detector response to helium nuclei.

%\subsection{Introduction to Negative Ion (NI) gases}\label{sec:NI_intro}
\subsection{SF$_6$ - A Negative Ion (NI) gas}\label{sec:sf6_intro}

%\subsection{Introduction}\label{sec:sf6_intro}

For the directional DM detection community, maximizing individual recoil track information is of paramount importance, and the use of so-called Negative Ion (NI) gases is a possible way to go about this.  The idea was first proposed roughly twenty years ago to minimize diffusion in detector drift regions without using magnetic fields~\cite{Martoff:2000wi}.  The central premise is that the primary electrons will attach to the highly electronegative atoms or molecules to form negative ions which then drift to the amplification stage.  The near thermal diffusion of the more massive ions is lower than that of electrons, which should allow ionization distributions to maintain their features over longer drift distances.  The DRIFT collaboration made use of the NI gas CS$_2$ for this purpose~\cite{SnowdenIfft:1999hz}.  The existence of multiple NI species was also discovered when mixing CS$_2$ with O$_2$~\cite{Snowden-Ifft:2014taa}, which enabled a novel, 3d fiducialization of their detector~\cite{Battat:2014van}.  The different NI species drift with different velocities allowing for an absolute position to be assigned along the drift coordinate.  Conversely, electrons all drift at the same speed which only allows for a relative assignment of position along the drift axis by studying the amount of diffusion the ionization distribution has undergone~\cite{Lewis:2014poa}. 

%\subsection{SF$_6$ - A Negative Ion (NI) gas}\label{sec:sf6_intro}

Recently, multiple NI charge carriers have been shown to result from ionization produced in an SF$_6$-based TPC, and Refs.~\cite{Phan:2016veo} and \cite{Phan:2016zvy} lay the foundation for SF$_6$ as a promising directional DM target.  Moderate gas gains have been measured in SF$_6$~\cite{Baracchini:2017ysg, Ishiura:2019ebd, Ikeda:2020pex} and, unlike CS$_2$, it is non-toxic and in gaseous phase at standard room pressure and temperature.  It is also non-corrosive, non-flammable, and the fluorine atoms have an unpaired spin which would allow spin-coupling models of DM to be probed.  However, in addition to the tendency of NI gases to suppress scintillation light, the avalanche process is complicated by the initial removal of electrons from the highly electronegative ions and reattachment processes.  SF$_6$ seems to suffer from large gain fluctuations and appears to be highly sensitive to impurities, including water vapor~\cite{Ikeda:2020pex}.  Because NI gases are usually an additive in gas mixtures, maintaining the correct ratio of partial pressures while filtering impurities presents a particular challenge.  Nonetheless, SF$_6$ looks to be a promising candidate for TPCs aspiring to measure a directional DM signal.

\section{Experimental setups}\label{sec:system}

This section describes the mechanical experimental setups, high voltage bias schemes, the data acquisition (DAQ), and the detector calibration procedures.

%%%
\subsection{Mechanical structures}

The experimental setups were housed in a \SI{30}{liter} vacuum vessel which was evacuated with a Drivac BH2-60HD combination roughing and turbo molecular drag pump.  Four experimental setups were used in this work, see Fig.~\ref{fig:dcube_milli}.  The mechanical structures consist of a white Delrin (acetal) support frame to which the copper and/or aluminum field cage rings, GEMs, and other parts are mounted.  The setups consist of a cathode (either mesh or solid metallic plate), a GEM assembly, and an anode.  In each setup, the anode is a metallic plate (top of each in Fig.~\ref{fig:dcube_milli}).  For a relative size comparison, the same circular aluminum base plate is pictured supporting each setup.  Further details can be found in Ref.~\cite{Thorpe:2018irh}.

Table \ref{tab:dcube_dimensions} specifies the number and types of GEMs used (which we refer to as the GEM assembly), the gas mixtures tested, and important mechanical dimensions for each setup.  Thin GEMs (tGEMs) with substrate thickness of \SI{50}{\text{$\mu$}m} and thick GEMs (THGEMs)~\cite{Chechik:2004wq} with a substrate thickness of \SI{400}{\text{$\mu$}m} were used.  The remaining GEM mechanical specifications are listed in Table \ref{tab:GEM_dimensions}.  The setups are named according to their GEM assemblies: 2-tGEM, 3-tGEM, 1-THGEM, and 2-THGEM, where the number corresponds to the number of that type of GEM that were used.  All GEMs were produced by the CERN PCB workshop.  The THGEMS were part of a custom order, while the tGEMs are a standard type readily available from the CERN Stores.  The tGEMs were considered functional if they drew $\approx$~\SI{10}{nA} or less at \SI{500}{V} in air at STP, and no GEMs were reused between setups.

The capability of turning the $^{55}$Fe radioactive source on/off was added after 3-tGEM was disassembled.  A linear motion feedthrough was used and the collimated source holder was fit onto the end, facing upwards (lower left of Fig.~\ref{fig:dcube_milli}).  The detector setups, after 3-tGEM, were designed such that the radioactive source could be placed under the cathode for the `on' position, and near the vessel wall for the `off' position.

\subsection{High voltage bias schemes}

\begin{figure}[ht!]
  \centering
  \includegraphics[width=\linewidth]{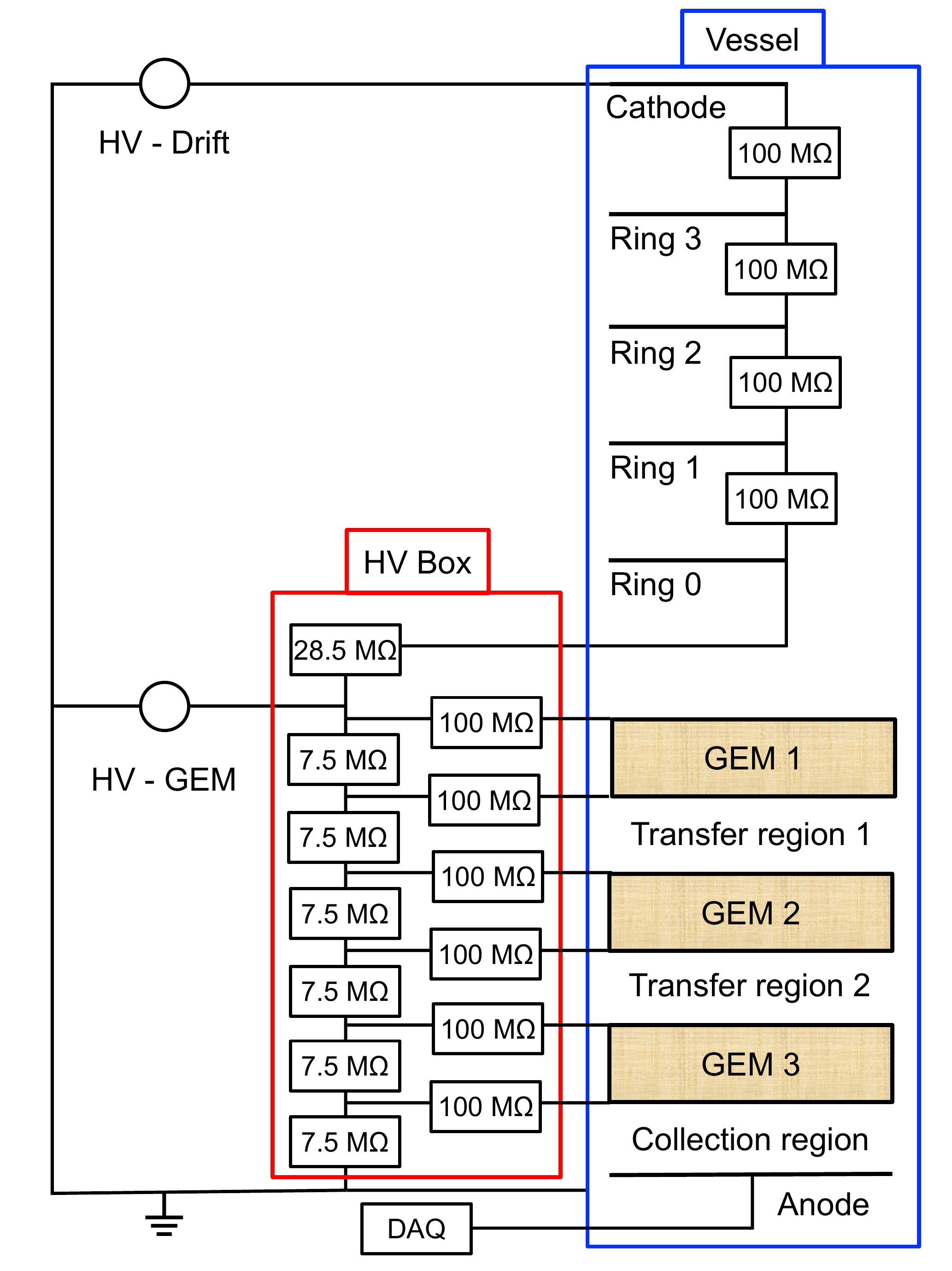}  
  \setlength{\abovecaptionskip}{0pt}
  \caption{Schematic of the high voltage bias scheme employed in 3-tGEM (setup with three \SI{50}{\text{$\mu$}m} GEMs).  Outlined, respectively in blue and red online, are the vacuum vessel and the external resistive voltage divider for the GEMs.  The electrostatic drift field and GEMs are powered by independent HV supplies (HV - Drift and HV - GEM, respectively).  The \SI{100}{M\Omega} resistors before the GEMs are to protect against any discharges.  In setups with fewer GEMs, there are correspondingly fewer transfer regions and the resistor values on the left side of HV Box differ; however, a similar electrical scheme used in all experimental setups.  Mechanical dimensions are listed Table~\ref{tab:dcube_dimensions} (diagram not to scale).}
 \setlength{\belowcaptionskip}{0pt}
 \label{fig:dcube_sketch}
 \end{figure}

Multiple high voltage (HV) power supplies were chosen for their low jitter, including a Keithley model 248 (\SI{5}{kV}), a Stanford Research Systems model PS370 (\SI{20}{kV}), and a CAEN NDT1470 (\SI{8}{kV}).  Figure~\ref{fig:dcube_sketch} depicts the HV bias scheme used in 3-tGEM.  With the exception of 2-tGEM, the field cage rings are spaced \SI{1}{cm} apart and powered by a resistive divider contained within the vacuum vessel (outlined in blue).  The field cage rings for 2-tGEM are spaced \SI{2}{cm} and each is powered by a separate HV channel.  In all setups, the GEMs are biased by an independent HV supply (HV - GEM) with an external resistive divider (HV Box, outlined in red).  This is done to free up HV channels and to provide \SI{100}{M\Omega} resistors to protect the GEMs against discharges.  The resistor values were changed as required by the particular setup; however, Fig.~\ref{fig:dcube_sketch} is a good representation of the HV bias scheme used in each setup.

Common to all of the GEM resistive dividers is that the electric fields within the transfer and collection regions are proportional to the total voltage across the GEMs, $V_G$.  This means that the transfer and collection fields change proportionally to $V_G$, with the constant of proportionality specific to the setup.  For reference, Tables~\ref{tab:heco2_fields} and \ref{tab:sf6} list the electrical field values corresponding to the highest $V_G$ (highest gain) value for each setup.

\subsection{Data acquisition}

For all of the measurements presented here, the anode collects the avalanched charge from the GEMs into a single-channel analog readout.  The box labeled `DAQ' in Fig.~\ref{fig:dcube_sketch} represents the following in order from the anode: 1) A charge sensitive preamplifier (either an Endicott eV-5093 or a Cremat CR-110 with respective gains of \SI{3.6}{V/pC} and \SI{1.4}{V/pC}).  2) A shaping amplifier (Canberra AFT 2025 spectroscopy amplifier).  3) A Pulse Height Analyzer (PHA) (Ortec EASY-MCA) with computer software.  4) A computer with data storage.  The preamplifier is housed inside of an Endicott eV-550 module which greatly reduces the RF-noise and provides AC-coupling to the detector.  Individual waveforms were continuously monitored with an oscilloscope, but only the PH data is presented in this work.  Issues concerning PH measurements with NI gases are discussed in Section~\ref{sec:sf6}. 

\subsection{Detector calibration}\label{sec:calibration}

The detectors are calibrated by the standard method of injecting voltage pulses into a \SI{1}{pF} test capacitor and measuring either the output of the shaping amplifier, or the corresponding PH bin number.  The ratio of the PHA bin number to the magnitude of the injected voltage pulse is recorded as the system response, $\SI{2.86}{PHA \, bins/fC}$.  3-tGEM was calibrated in a similar manner; however, an additional filter was present in the readout chain to reduce the electronic noise.  This resulted in a response value, taken as the magnitude of the output voltage pulse from the shaping amplifier to the magnitude of the injected voltage pulse, of $\SI{0.91}{V/pC}$ with an uncertainty of $\approx$~14\% due to the uncertainty of the added capacitance.  In all cases, the response value is directly proportional to, and used to determine the gain values from the PH spectra.  The PHA has an operating range of 0 - \SI{10}{V}, which is divided into 8196 bins, and its linearity and repeatability were checked and monitored.  %Using the calibrated PH scale, we can estimate the magnitude of the aforementioned noise floor.  We find that the level below which measurements are not possible is in the range of $\mathcal{O}(10^4 - 10^5)$ electrons.  If we assume that this is Johnson–Nyquist noise then this gives a capacitance of $\mathcal{O}(100)$ \SI{}{nF} for the noisiest setups.  This is not unreasonable considering the physical dimensions of the Readout plate. 

\section{Definitions and relationships of quantities}\label{sec:formalism}
 
This section formally develops the parameters which quantitatively define detector operation at high gas gain.  In doing so, relationships that describe the individual detector data sets are explained.

\subsection{Townsend's equation with GEMs}\label{sec:townsend}

The first Townsend coefficient, $\alpha$, quantifies the progression of a charge avalanche, has units of inverse length, and is a function of $E/p$, where $E$ is the electric field strength in the GEM and $p$ is the gas pressure.  Townsend's equation uses $\alpha$ to describe the exponential growth of a charge avalanche over a path length, $r$.  In general, the electric field is a function of $r$, and $\alpha$ needs to be integrated over the entire path of the avalanche; however, this is simplified by using a uniform field assumption, see Section~\ref{sec:uniform_field}.  We consider a system with a number of GEMs $n_g$ all with thickness $t$.  This system will produce a gain $G$ for a voltage $V_G$ applied evenly across all of the GEMs, i.e the single GEM voltage is $V_{GEM} = V_G/n_g$.  Under these assumptions,
\begin{equation}
  \textrm{ln}(G) = \alpha n_g t.
  \label{eqn:gain_gem}
\end{equation}
The gain is an exponential function of $\alpha$.  Additionally, if $\alpha$ is proportional to the electric field strength $E = V_G/n_gt$, then the gain will be an exponential function of $V_G$.

%%
%\begin{equation}
%  \frac{dn}{n} = \alpha dr,
%  \label{eqn:townsend}
%\end{equation}
%%
%\\
%where $n$ is the total number of electrons produced in the avalanche.  The gas gain can be defined as the ratio of the total number of electrons produced in the avalanche to the number of electrons in the primary ionization cloud, $n_0$,  

%%
%\begin{equation}
%  G \coloneqq \frac{n}{n_0}.
%  \label{eqn:gain_def}
%\end{equation}
%%
%\\
%We note that if all avalanches are presumed to be identical then this is equivalent to defining the gain as the number of electrons produced by a single electron avalanche.

%%%
%\subsection{Reduced first Townsend coefficient (RFTC)}\label{sec:townsend}

%By defining $\alpha$ as the product of the number of mean free paths between collisions along the field direction that an ionized electron can take and the probability that a mean free path is larger than the mean free path for ionization, Ref.~\cite{AOYAMA1985125} provides a generalization of the \textit{reduced} first Townsend coefficient.  Assuming an ideal gas, it can be written in terms of the gas pressure, $p$, as

Reference~\cite{AOYAMA1985125} provides a generalization of the \textit{reduced} first Townsend coefficient, $\alpha/p$.  Assuming an ideal gas, the generalization can be written in terms of the gas pressure as

%%
%\begin{equation}
%  \frac{\alpha}{N} = K~\Bigg (\frac{E}{N} \Bigg )^{m} \textrm{exp}~ \Bigg(-L~\Bigg(\frac{N}{E}\Bigg)^{1-m}\Bigg),
%  \label{eqn:general_townsend}
%\end{equation}
%%
%\\

%
\begin{equation}
  \frac{\alpha}{p} = A~\Bigg (\frac{E}{p} \Bigg )^{m} \textrm{exp}~ \Bigg(-B~\Bigg[\frac{p}{E}\Bigg]^{1-m}\Bigg).
  \label{eqn:general_townsend_2}
\end{equation}
$A$, $B$, and $m$ are experimentally fitted parameters related to the gas mixture.  An important assumption is that the cross section of interaction between the accelerating electrons and atomically bound electrons is proportional to $(E/p) \, ^m$, with $0 \leq m \leq 1$.  Many of the historical, empirical relationships describing gas multiplication data are simple manifestations of Eq.~\ref{eqn:general_townsend_2}.  For example, Diethorn used a form where ${\alpha} \sim E$, or $m=1$ \cite{osti_4345702}.  Williams and Sara used a form with $m=0$ \cite{WILLIAMS1962229}.  We can combine Eq.~\ref{eqn:gain_gem} with Eq.~\ref{eqn:general_townsend_2} to obtain

\begin{equation}
\frac{\textrm{ln}(G)}{n_gpt} = A~ \Bigg(\frac{V_G}{n_gpt}\Bigg)^{m} \textrm{exp}~ \Bigg(-B~ \Bigg[ \frac{n_gpt}{V_G} \Bigg]^{1-m} \Bigg).
\label{eqn:reduced_townsend}
\end{equation}
Equation~\ref{eqn:reduced_townsend} contains the gas gain and other experimental parameters relevant for a GEM-based detector.

%\subsection{Reduced quantities}\label{sec:reduced_quantities}

We introduce an abbreviated notation: 

\begin{equation}
 %\Gamma = \frac{\textrm{ln}(G)}{n_gpt} \; \; \; \; \textrm{and} \; \; \; \; \Sigma = \frac{V_G}{n_gpt}, 
 \Sigma \equiv \frac{V_G}{n_gpt}  \; \; \; \; \textrm{and} \; \; \; \; \Gamma \equiv \frac{\alpha}{p} = \frac{\textrm{ln}(G)}{n_gpt}.
\label{eqn:reduced_quantities}
\end{equation}  
$\Sigma$ is the reduced avalanching field strength in the gain stage averaged over the number of GEMs and, when there is no ambiguity, this will be referred to as the reduced field.  $\Gamma$ is the RFTC.  Equation~\ref{eqn:reduced_townsend} can be expressed as

\begin{equation}
\Gamma = A~ \Sigma ~^{m} e^{-B~ \Sigma~^{m-1}}.
\label{eqn:reduced_gain}
\end{equation}
Equation~\ref{eqn:reduced_gain} can be used to constrain the parameters $A$, $B$, and $m$.

%%%
\subsection{Gain versus total GEM voltage}\label{sec:gain_curve}

Setting $m = 1$ in Eq.~\ref{eqn:reduced_gain} is equivalent to $\alpha / p \sim E / p$, which is equivalent to $\textrm{ln}(G) / n_gpt \sim V_G / n_gpt$, for a GEM-based detector.  We can add a voltage offset to reflect the real electric field strength required to initiate the gas avalanche process, and write

\begin{equation}
G = 10^{(V_G - V_1)/V_2}.
\label{eqn:gain_voltage}
\end{equation}
$V_1$ and $V_2$ are returned fit parameters which provide information about the behavior of a given detector setup at a given gas pressure. Specifically, $V_1$ is the voltage at which the gain is unity and $V_2$ is the voltage required to increase the gain by an order of magnitude.  Equation~\ref{eqn:gain_voltage} will be fit to the individual gain versus total GEM voltage data sets.

The reduced quantities allow for variable values of $n_g$, $p$, and $t$.  We use them to write an expression analogous to Eq.~\ref{eqn:gain_voltage}

\begin{equation}
%\Gamma = A'~ \Sigma + B',
%\Gamma = (\Sigma - \Sigma_1) / \Sigma_2,
\Gamma = (\Sigma - \Sigma_1) / V_3,
\label{eqn:red_gain_field}
\end{equation}
where $\Sigma_1$ and $V_3$ are fit parameters that play the same role as $V_1$ and $V_2$ in Eq.~\ref{eqn:gain_voltage}.  $\Sigma_1$ is the reduced field required to achieve zero RFTC and $V_3$ is the voltage required to increase the RFTC by unity (RFTC units are \SI{}{1/cm/torr}).  A similar relationship was used by Zastawny to describe gas multiplication data with a uniform field in CO$_2$~\cite{Zastawny_1966}.   The added voltage offset implies a gain of unity at some non-zero field, which acts as a threshold.  We therefore refer to Eq.~\ref{eqn:red_gain_field} as the \textit{threshold model}.  The drawback of the threshold model is that it provides minimal information about the gas mixture, and returns operational parameters.  

%However, we will see that setting $m = 0$ in Eq.~\ref{eqn:reduced_gain} provides an equally valid description without the need for the voltage offset, while also giving access to more information about the gas itself in the form of the parameters $A$ and $B$.  

%A good rule of thumb summarizing this discussion can be stated as follows:  The RFTC is proportional to the reduced field in the GEM(s).  When a single gas pressure, GEM thickness, and specific number of GEMs are considered this implies that $\textrm{ln}(G)$ is proportional to $V_G$.  However, this statement is only strictly true when $m = 1$, which is not required. 

%
\begin{figure*}[!h]
  \centering
 	
	%\subfigure {\includegraphics[width=7cm]{gain/images/dcube_milli_2_crop2.png}}
	\subfigure [$^4$He:CO$_2$ (70:30) (\SI{760}{torr}) measured with 2-tGEM.]{\includegraphics[width=0.47\linewidth]{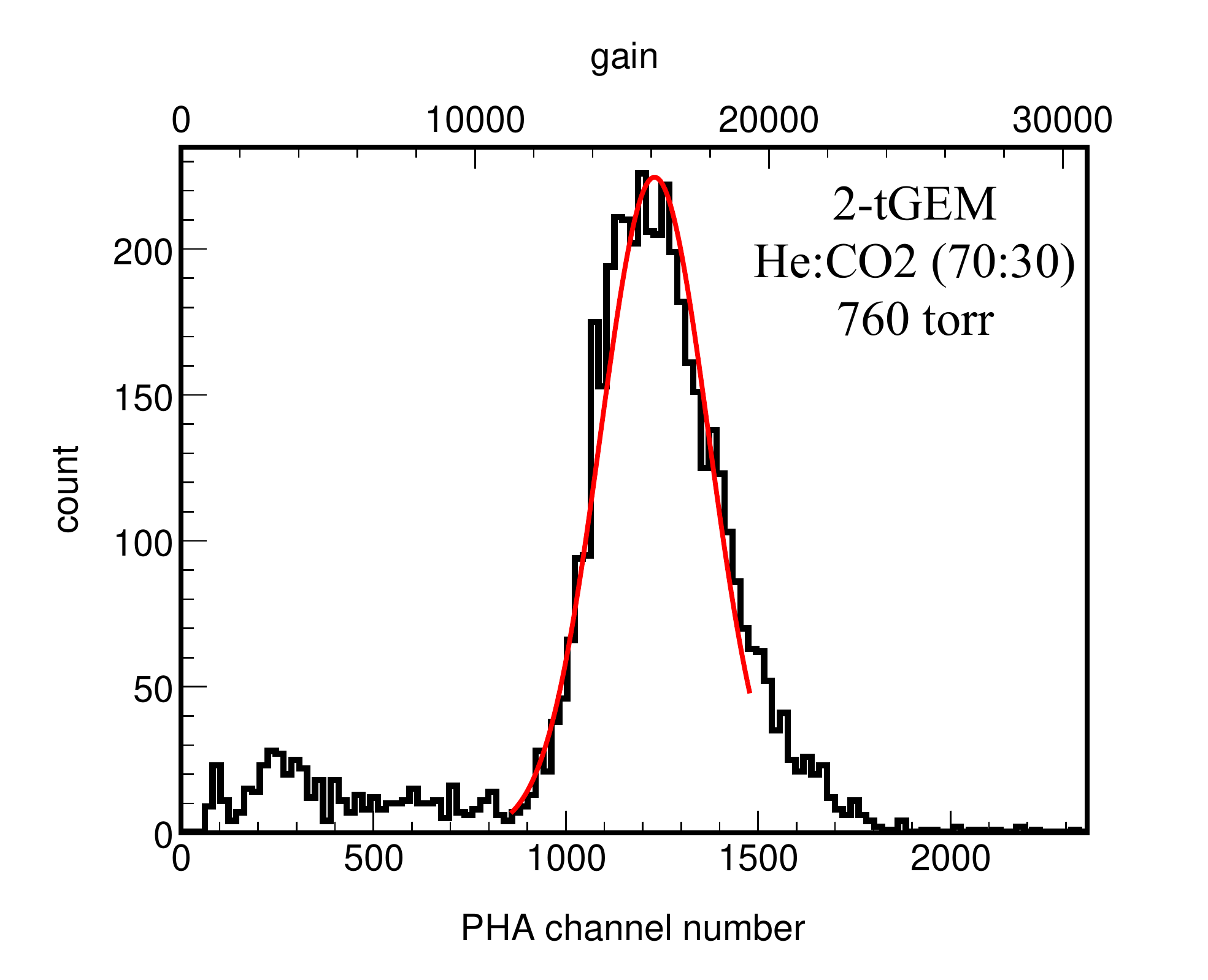}}
	\subfigure [$^4$He:CO$_2$ (70:30) (\SI{760}{torr}) measured with 3-tGEM.]{\includegraphics[width=0.47\linewidth]{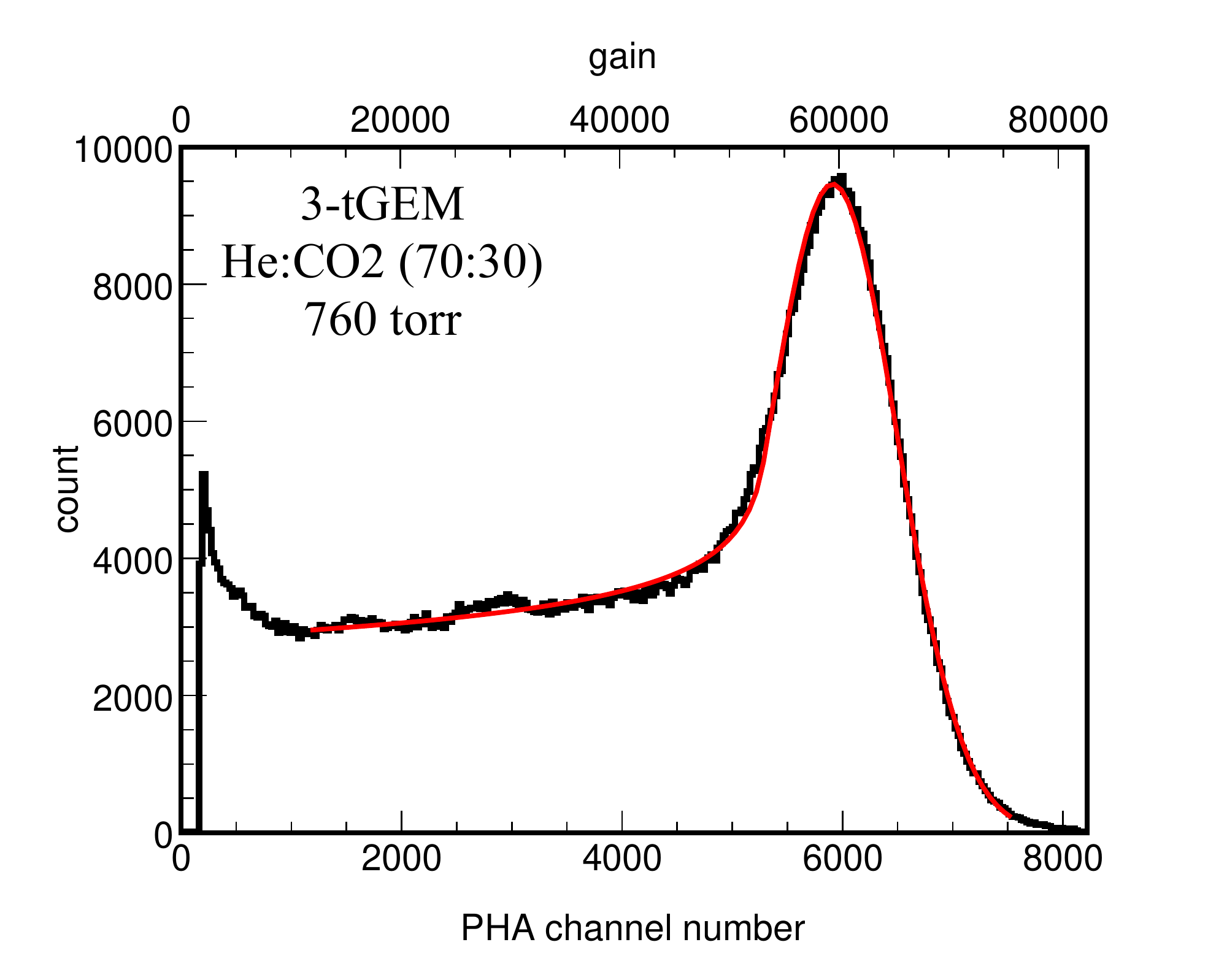}}
	\subfigure [$^4$He:CO$_2$ (70:30) (\SI{380}{torr}) measured with 1-THGEM.]{\includegraphics[width=0.47\linewidth]{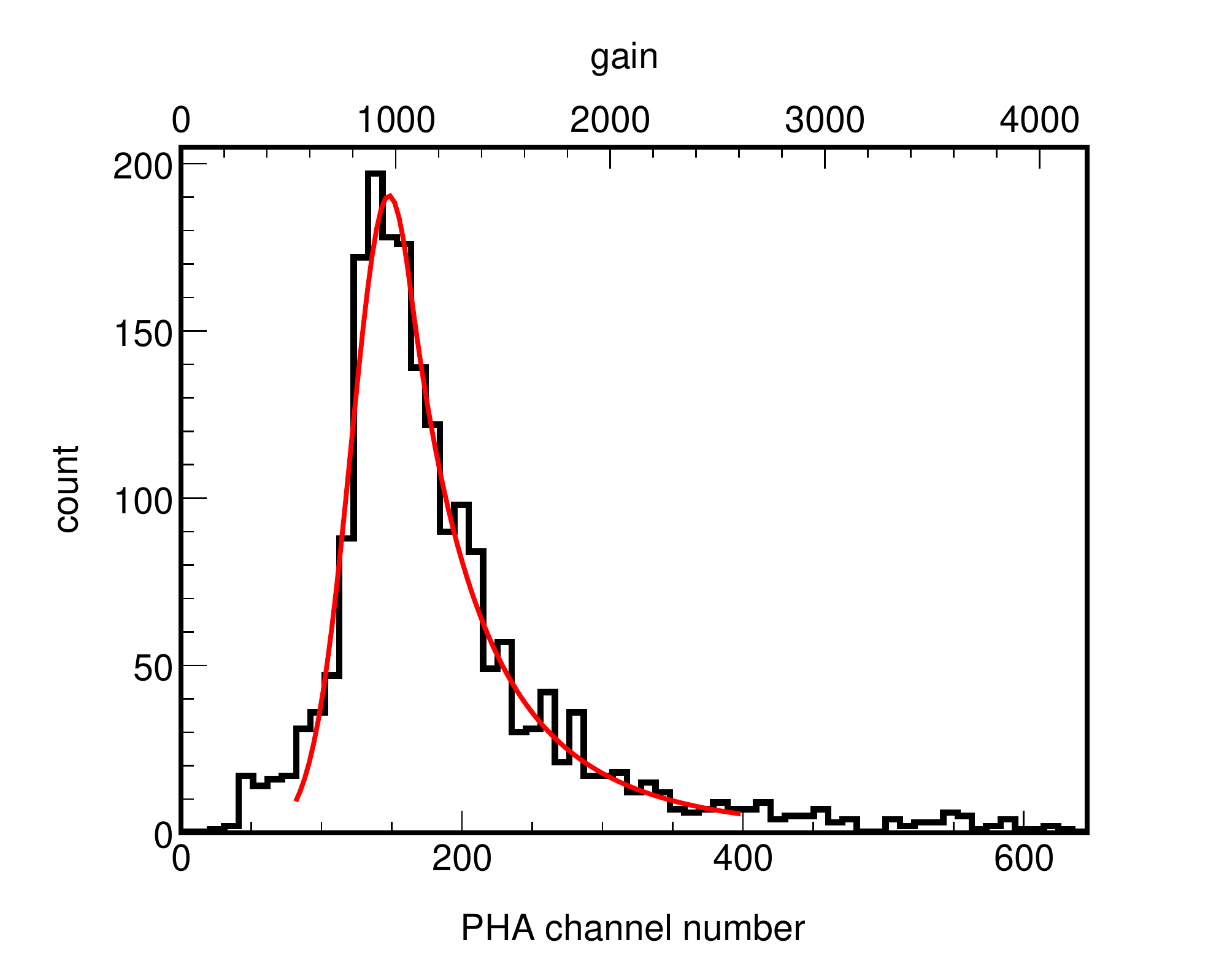}}
	\subfigure [$^4$He:CO$_2$ (70:30) (\SI{760}{torr}) measured with 1-THGEM.]{\includegraphics[width=0.47\linewidth]{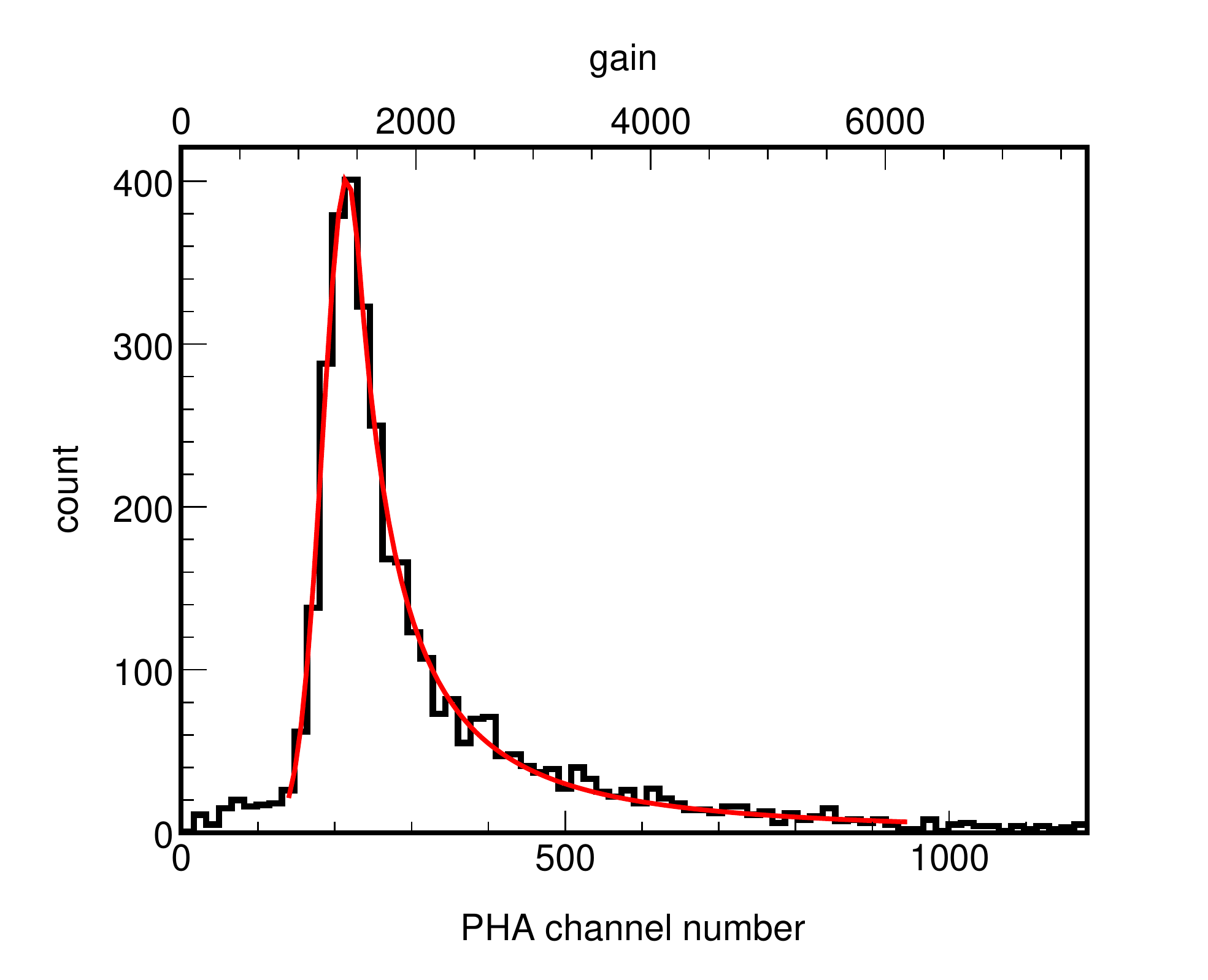}}
	\subfigure [$^4$He:CO$_2$ (70:30) (\SI{760}{torr}) measured with 2-THGEM.]{\includegraphics[width=0.47\linewidth]{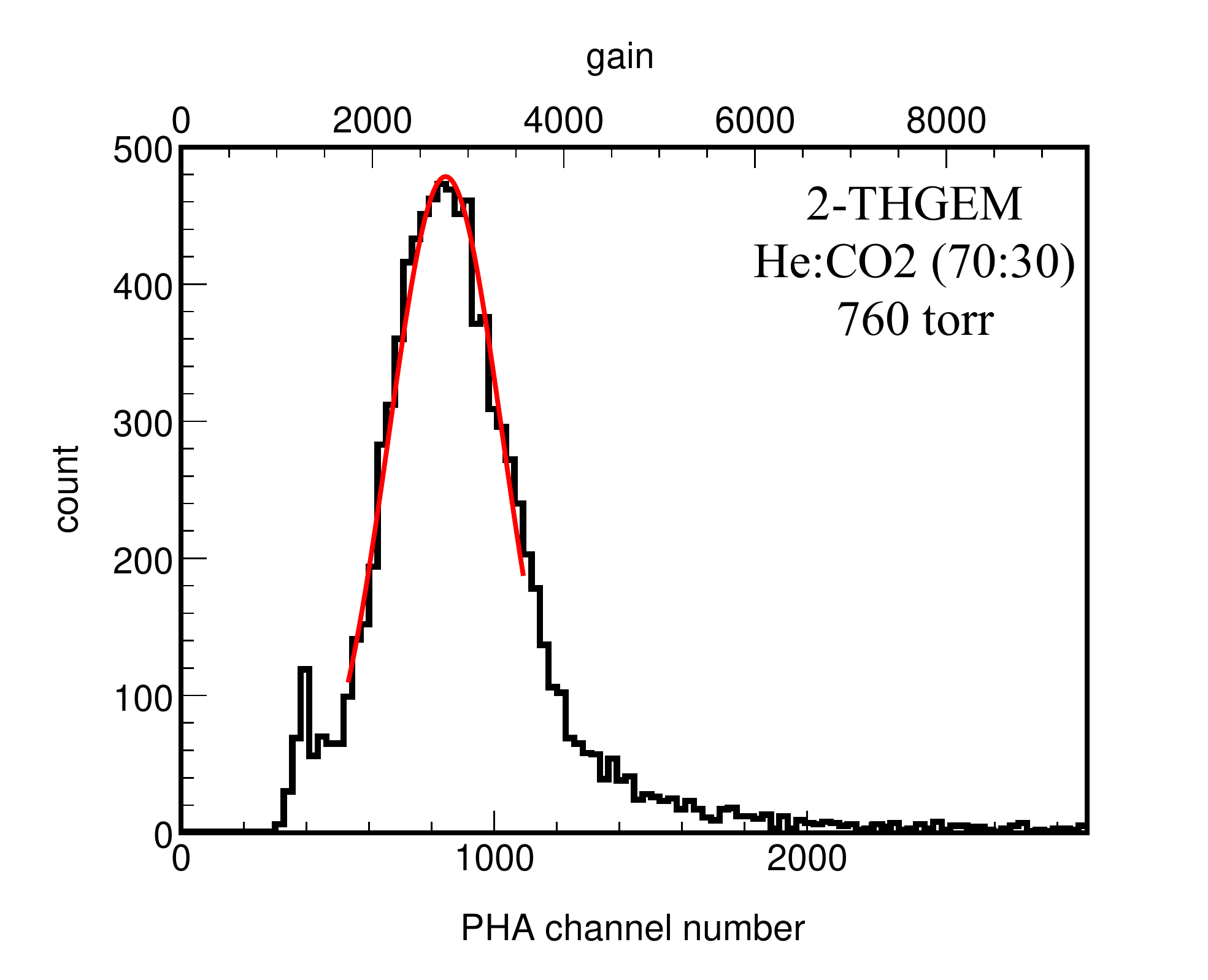}}
	%\vspace{0pt}	
	%\subfigure {\includegraphics[width=10.5cm]{gain/images/milli2_sources_crop2.png}}
	%\subfigure {\includegraphics[width=0.33\linewidth]{images/sf6_spec_THGEM_20_torr.pdf}}
	\subfigure [SF$_6$ (\SI{40}{torr}) measured with 1-THGEM.]{\includegraphics[width=0.47\linewidth]{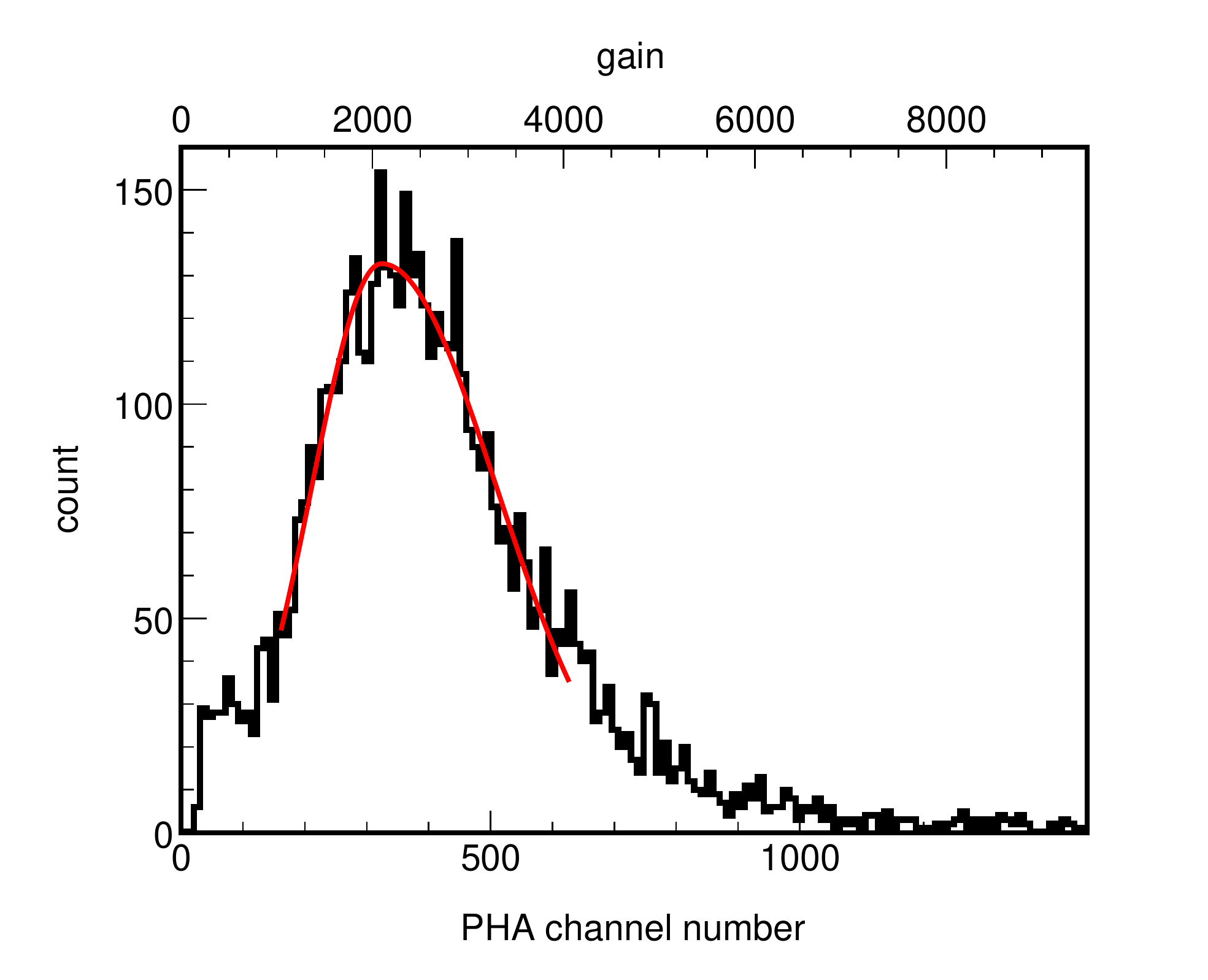}}
	
  \setlength{\abovecaptionskip}{2pt}
  \caption{Examples of pulse-height spectra from which gain and PH resolution values are extracted.  The solid lines (red online) are fits to the distributions.}
 \setlength{\belowcaptionskip}{0pt}
 \label{fig:spectra}
 \end{figure*}
\subsection{Pulse-height (PH) and energy resolution}\label{sec:gain_resolution}

%There are many effects that contribute to fluctuations in the values of the measured charge.  We briefly discuss the main contributions relevant to our study, and how they manifest in the measured PH resolution.}  

We define the \textit{fractional PH resolution} as

\begin{equation}
 R \equiv \frac{\sigma_{PH}}{\mu_{PH}}, 
\label{eqn:res_def}
\end{equation}     
where $\sigma_{PH}$ and $\mu_{PH}$ are the returned parameters from fitting the PH spectra.  All of the resolution measurements shown here are of the PH resolution.

%The energy resolution of a complete GEM-based detector will typically be determined by a number of effects, including those from detection electronics and downstream digitization. Nevertheless, the fundamental limit on achievable energy resolution is when only variation in the avalanche charge contributes to the energy resolution. (Note that this includes fluctuations both in primary ionization and in gain.) We can estimate resolution in this fundamental limit by considering the PH resolution at high gas gain, where the anode noise term becomes negligible. In that limit, we will therefore refer to the fractional PH resolution obtained as the `energy resolution'.

The PH resolution can be described by three terms:  The first is the Fano factor, which quantifies the variance in the initial ionization distribution.  Unless a known number of primary electron can be produced, which is typically not the case, the Fano factor will be an inherent part of the measured PH resolution.  The second term is the avalanche variance, which has a complicated dependence on $E/p$ and the gain, but does have a minimum value.  The third term corresponds to the `detector noise' due to the analog amplification electronics used, which appears to be gain independent.  We expect that the quadrature sum of these three terms equals $R$.  The first two terms can be probed by considering the PH resolution at high gain, where the detector noise contribution becomes negligible.  The second term can be further minimized to estimate the best (lowest) energy resolution achievable with this technology.

%Additional effects that can broaden the PH resolution include the variance in gain between the GEM holes and the gain stability.  The latter is explicitly addressed in Section~\ref{sec:stability} and in Ref.~\cite{Jaegle:2019jpx}, where a double-tGEM setup is combined with a highly segmented pixel ASIC readout that shows week-long, percent-level gain stability with a radioactive $\alpha$-particle source.  As for the former, we have used highly stable HV supplies and a collimated $^{55}$Fe source in this work, but the degree of the transverse diffusion in the charge cloud in the drift gap of similar detectors is addressed in Refs. \cite{Vahsen:2014fba}, \cite{Lewis:2014poa}, and \cite{Lewis:2021}.  We can estimate that the charge clouds in the atmospheric pressure He:CO$_2$ measurements presented here could be spread over three GEM holes in the case of the tGEM setups, and possibly more for lower gas pressures.  However, this wouldn't be true for events drifting only a portion of the drift length, and the effect on the resolution is minimal given the number of pulses in each PH spectrum.}

%To obtain accurate values for the energy resolution, the detector noise term must be minimized.  This is done quantitatively defining values associated with detector operation at high gas gain, where the detector noise term is overwhelmed.  Following this, the avalanche variance will be the only remaining term dependent on the gain and reduced field.}

\subsection{PH resolution versus gain and asymptotic quantities}\label{sec:gain_res_curve}

$R$ has a dependence on the gain itself, which is well described by

\begin{equation}
R = \sqrt  { \; \Bigg( \frac {a}{G} \Bigg)^2 + b \, ^2},
%\frac{\sigma_E}{G} = \sqrt  { \; \Bigg( \frac {a}{G} \Bigg)^2 + b \, ^2},
%\sigma_G/G = \sqrt{(a/G)^2 + b^2}
\label{eqn:gain_res}
\end{equation}     
where $a$ and $b$ are returned fit parameters.  $a$ can be interpreted as a constant-value, gain-independent detector noise term.  $b$ is the asymptotic, fractional PH resolution achieved with high avalanche gain operation.  For $R$ to approach $b$, the gain must be much larger than the detector noise.  To this end, we define the \textit{asymptotic gain},

\begin{equation}
 G_{\infty} \equiv 100 \times a, 
\label{eqn:asy_gain}
\end{equation}     
as the gain at which the \textit{asymptotic resolution} $b$ can be ensured.  Eq.~\ref{eqn:gain_res} will be used to fit the individual PH resolution versus gain data sets.

%Without the factor of 100, one could regard $a$ as the constant value, gain-independent, detector noise term which adds in quadrature with the gain-dependent asymptotic resolution term, and is being suppressed at high-gain.  However, the factor of 100 is introduced and $a$ is defined as the gain at which $b$ is ensured to be near its minimum, or asymptotic, value.  This will also return $b$ as a percentage and it is presented this way throughout the paper.  Eq.~\ref{eqn:gain_res} will be employed throughout Section~\ref{sec:measurements} and is sometimes referred to as a given setup's `PH resolution curve'.

%%
\subsection{Asymptotic, reduced quantities}\label{sec:asy_red_quant}

We consider a particular detector operating at its asymptotic gain and define the asymptotic, reduced, GEM-averaged quantities:

\begin{equation}
 %\Gamma_{\infty} = \frac{\textrm{ln}(G_{\infty})}{n_gpt} \; \; \; \; \textrm{and} \; \; \; \; \Sigma_{\infty} = \frac{V_{G_{\infty}}}{n_gpt}, 
 \Sigma_{\infty} \equiv \frac{V_{G_{\infty}}}{n_gpt}  \; \; \; \; \textrm{and} \; \; \; \;  \Gamma_{\infty} \equiv \frac{\textrm{ln}(G_{\infty})}{n_gpt}.
\label{eqn:asy_reduced_quantities}
\end{equation}  
$\Sigma_{\infty}$ is the asymptotic, reduced, avalanching field strength in the gain stage averaged over the GEMs.  $\Gamma_{\infty}$ is the RFTC evaluated at the detector's asymptotic gain value.  $G_{\infty}$ is defined as above, where $a$ is extracted from Eq.~\ref{eqn:gain_res} after fitting to a particular setup's resolution versus gain data.  $V_{G_{\infty}} = \textrm{log}(G_{\infty})V_2 + V_1$ comes from Eq.~\ref{eqn:gain_voltage}, where $V_1$ and $V_2$ are extracted after fitting to the setup's gain versus total GEM voltage data.  

$\Sigma_{\infty}$ and $\Gamma_{\infty}$ will be referred to as the \textit{asymptotic reduced field} and the \textit{asymptotic RFTC}, respectively.  These quantities help reduce the effect of systematic differences between the experimental setups because they describe each detector in a regime where the effect of the detector noise is negligible.  %This is a reasonable choice because high gas gain operation is of interest for many communities including those involved in low-energy, rare-event searches.  

\section{Measurements and results with $^4$He:CO$_2$ (70:30)}\label{sec:measurements}

This section presents the He:CO$_2$ gain and resolution results.  All measurements are performed with a collimated $^{55}$Fe source (\SI{5.9}{keV}) in room temperature gas.  All gain measurements here are `effective' and not absolute (discussed in Section~\ref{sec:systematics_summary}).  The radioactive source was less collimated for the 3-tGEM measurements than for other setups.

%Furthermore, the gain measurements are `effective' as we do not account for charge losses due to transfer efficiencies or other mechanisms.  We choose the maximum transfer and collection field values that allow for stable operation, but optimize no further.  From Fig.~29 in Ref.~\cite{Bachmann:1999xc}, we can see that the different collection field values used in this work will have the largest impact.  As shown in Table \ref{tab:heco2_fields}, the collection field value used in 2-tGEM is higher compared to the other setups, and the largest difference is between 2-tGEM and 3-tGEM.  If this were accounted for, we estimate an $\approx$ \SI{20}{\%} difference in measured gain would result, while the remaining differences, including in the drift and transfer fields, would have less impact.}

%FIXME
%However, in the end, we will be interested in the RFTC, which is the logarithm of the gain, and these differences will matter less.
%
%Also, this issue should not effect the fractional PH resolution as the mean and the sigma of the PH distribution should be effected equally.  ???

%In this Section~we discuss gain and PH resolution measurements.  Section~\ref{sec:gain_spec} explains how gain and PH resolution are reconstructed from pulse-height data.  In Sections \ref{sec:sf6} (SF$_6$) and \ref{sec:heco2} (He:CO$_2$) we present the measurements performed with the experimental structures pictured in Fig.~\ref{fig:dcube_milli} and the various GEM setups listed in Table \ref{tab:dcube_dimensions}.  A discussion of the systematic issues can be found in Section~\ref{sec:systematics}.

%%%
\subsection{Measurement technique and procedure}

The measurement procedure consists of pumping the vacuum vessel down to approximately 10$^{-5}$ \SI{}{torr} for at least a few hours, or a series of pump-and-purge cycles before filling.  As the vacuum vessel reaches the desired pressure, the HV supplies are switched on and the stability of the gas gain is monitored.  Once suitable stability (i.e. no noticeable changes occurring at a specific GEM voltage) is achieved, the GEM voltages are varied and other measurements are performed.  Gain stability was typically monitored carefully and for long time periods surrounding other measurements (see Section~\ref{sec:stability}).  As opposed to the lower pressure measurements with SF$_6$, the He:CO$_2$ measurements were performed without gas flowing through the vessel. 

The $^{55}$Fe source is always placed outside the sensitive volume, near the cathode, and pointing along the drift axis towards the GEM(s).  Except for the 3-tGEM measurements, PH spectra have been background subtracted by completing a time normalized non-source run.  Each spectrum is recorded for five minutes to account for any rapid time variation of the gain, and to achieve a large statistical sample.  The spectra are quality checked for similar event rates, and each spectrum is fit to extract the measured effective gain and PH resolution values.

%Unless otherwise specified, pulse-height spectra are recorded over five minutes to account for any acute time variation of the gain, and to achieve a large statistical sample.  

%The integrals of the five minute spectra are checked for similar values to ensure that the full distribution is being measured, and each spectrum is used to extract a single gain value.  

%The $^{55}$Fe source is always placed near the cathode pointing along the drift axis into the drift region centered on the GEM(s).   

%All of the He:CO$_2$ and SF$_6$ spectra, with the exception the 3-tGEM data set, have been background subtracted by completing a time normalized non-source run.  This feature was added at some point later in time, and not available for the earliest measurements.

%%%%%%%%%%%%%%.   Tables: Data sets first and then fit parameters HeCO2
\begin{table*}[ht!]
\centering
  %\begin{tabular}{l*{6}{c}r}
  %\begin{tabular}{l | l | >{\centering\arraybackslash}p{1.8cm} | >{\centering\arraybackslash}p{1.8cm} | >{\centering\arraybackslash}p{2.6cm} | >{\centering\arraybackslash}p{2.7cm} | >{\centering\arraybackslash}p{2.5cm} }
  \begin{tabular}{l  >{\centering\arraybackslash}p{1.8cm}  >{\centering\arraybackslash}p{1.8cm}  >{\centering\arraybackslash}p{2.6cm}  >{\centering\arraybackslash}p{2.7cm}  >{\centering\arraybackslash}p{2.5cm} }
  \toprule
  Setup & Pressure (\SI{}{torr}) & Drift field (\SI{}{V/cm}) & GEM field(s) 1:2:3 (\SI{}{kV/cm}) & Transfer field(s) 1:2 (\SI{}{V/cm}) & Collection field (\SI{}{V/cm}) \\
  \hline
\noalign{\vskip 2mm}
   %2-tGEM & 760 & 500 & 99.7 : 97.9 & 3853 & 3588  \rule{0pt}{3ex} \\ %\hline
   2-tGEM & 760 & 500 & 99.7 : 97.9 & 3853 & 3588  \\ %\hline
   3-tGEM & 760 & 509 & 74.0 : 75.1 : 74.9 & 1758 : 1706 & 2164 \\ %\hline
   1-THGEM & 380 & 239 & 26.5  & n/a & 2161 \\ %\hline
   1-THGEM & 570 & 291 & 33.0 & n/a & 2692 \\ %\hline   
   1-THGEM & 760 & 469 & 37.3 &  n/a & 3047 \\ %\hline
   2-THGEM & 760 &  500 & 32.6 : 31.0 & 2465 & 2211 \\ %\hline
  \bottomrule
   \end{tabular}
   
   \setlength{\abovecaptionskip}{6pt}
  \caption{Detector settings for $^{55}$Fe measurements with $^4$He:CO$_2$ (70:30).  The GEM, transfer, and collection field(s) are proportional to the total GEM voltage, $V_G$.  The values shown are for the highest $V_G$.  The drift field is held constant.}
  \setlength{\belowcaptionskip}{0pt}
  \label{tab:heco2_fields}
\end{table*}
%

%%%
\subsection{Extracting the gain and PH resolution from data}\label{sec:gain_spec}

Figure~\ref{fig:spectra} shows examples of PH spectra from all of the experimental setups, each recorded over five minutes.  We perform a $\chi^2$-minimization fit to each PH spectra using a fit function consisting of a gaussian signal with floating mean $\mu$ and standard deviation $\sigma$.  Some of the fit functions also contain exponential tails as shown in Fig~\ref{fig:spectra}.  The fitted value of $\mu$ provides the gain, and $\sigma / \mu$ provides the fractional PH resolution $R$.

%The gas gain, $G$, is defined as the mean obtained from a $\chi^2$-minimization of a fit function to the PH spectrum using the gain calibrated PH scale and, as mentioned, the PH resolution is defined as the ratio of the standard deviation to the mean of the distribution.  
%
%Gain scales are provided at the top of each spectrum in Fig.~\ref{fig:spectra}.  The PH spectra are fit with a Gaussian, or modified in the case of SF$_6$, distribution to the peak and exponential tails are used to amend any modified signal shape as needed, as not all of the distributions are symmetric
%
%\textcolor{blue}{Aside from 3-tGEM, the lowest gain spectra tend to be the least symmetric.  It could be that the low side of the spectra is partially falling below the noise floor, but the effect is more pronounced for 1-THGEM at similar gain values.  The "high-side" tail is a typical feature for 1-THGEM using He:CO$_2$, but is substantially diminished using SF$_6$.  The gain variation is also highest for 1-THGEM when considering gain stability over time.  It is likely the lower field in a single THGEM gives a higher variance in avalanche starting positions, and those starting past the mean produce fewer measurable signals.  Adding another THGEM, as with 2-THGEM, would help wash this effect out.}

%%

%
\begin{figure*}[h!]
  \centering
  \includegraphics[width=7cm]{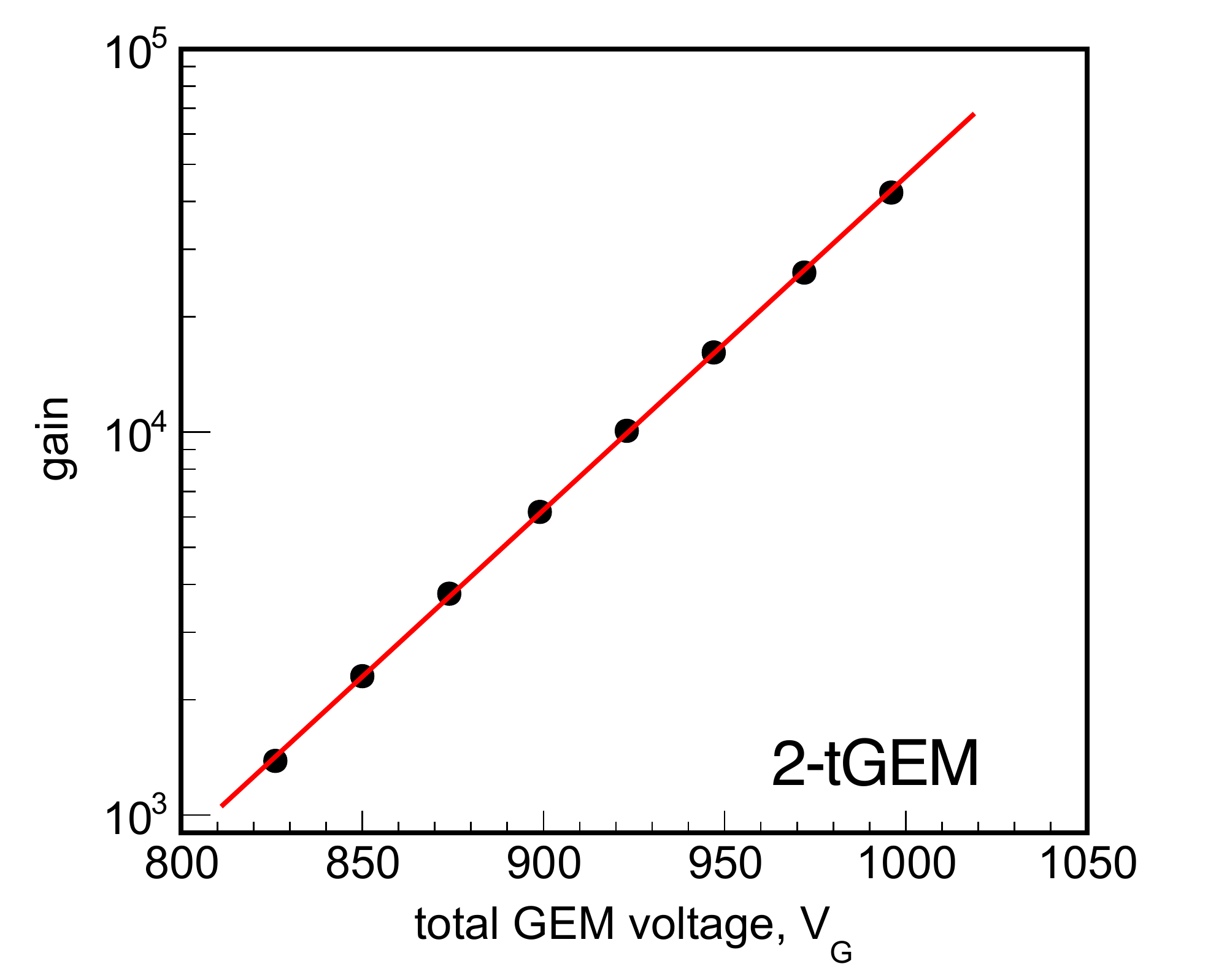}
  \includegraphics[width=7cm]{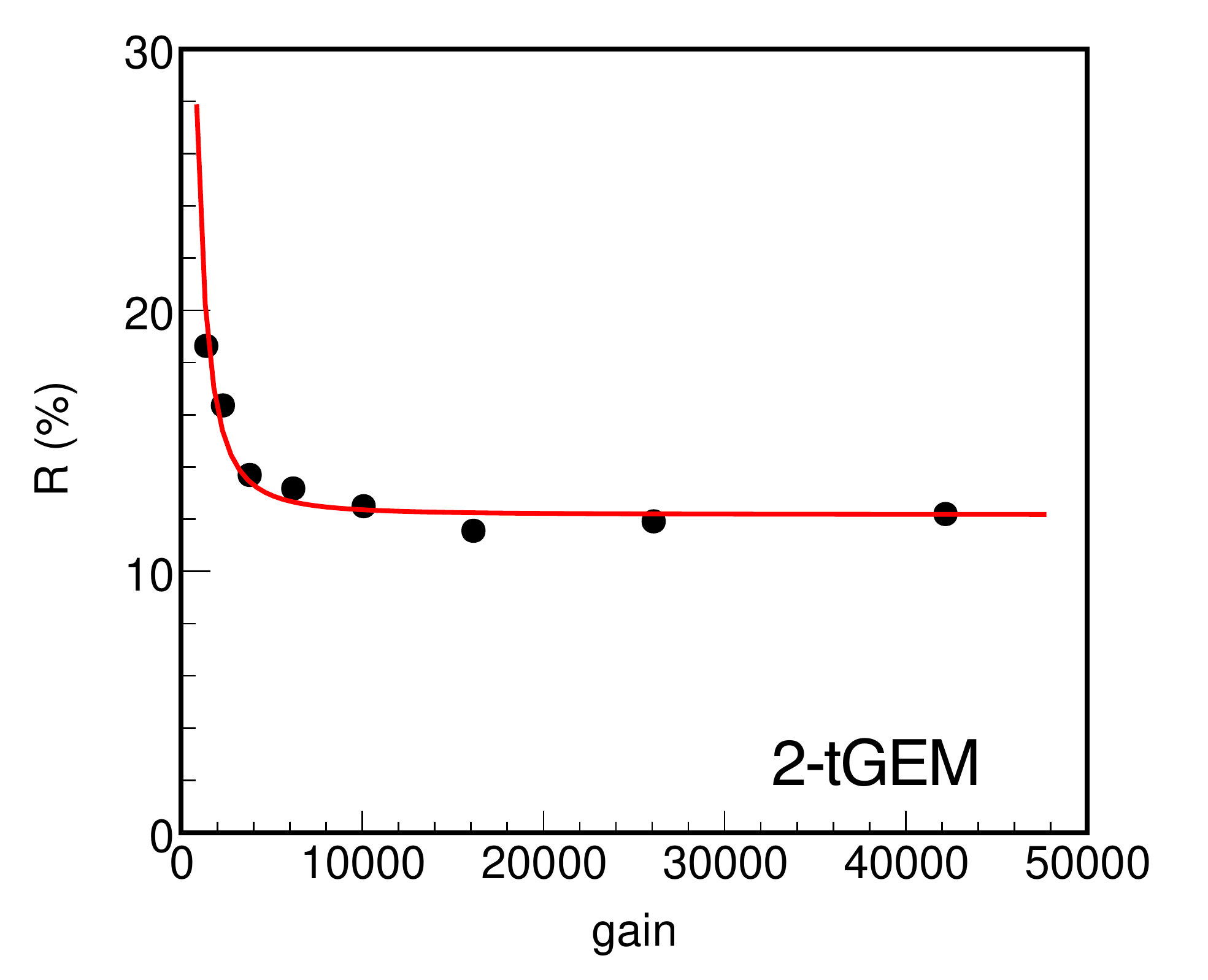}
  \includegraphics[width=7cm]{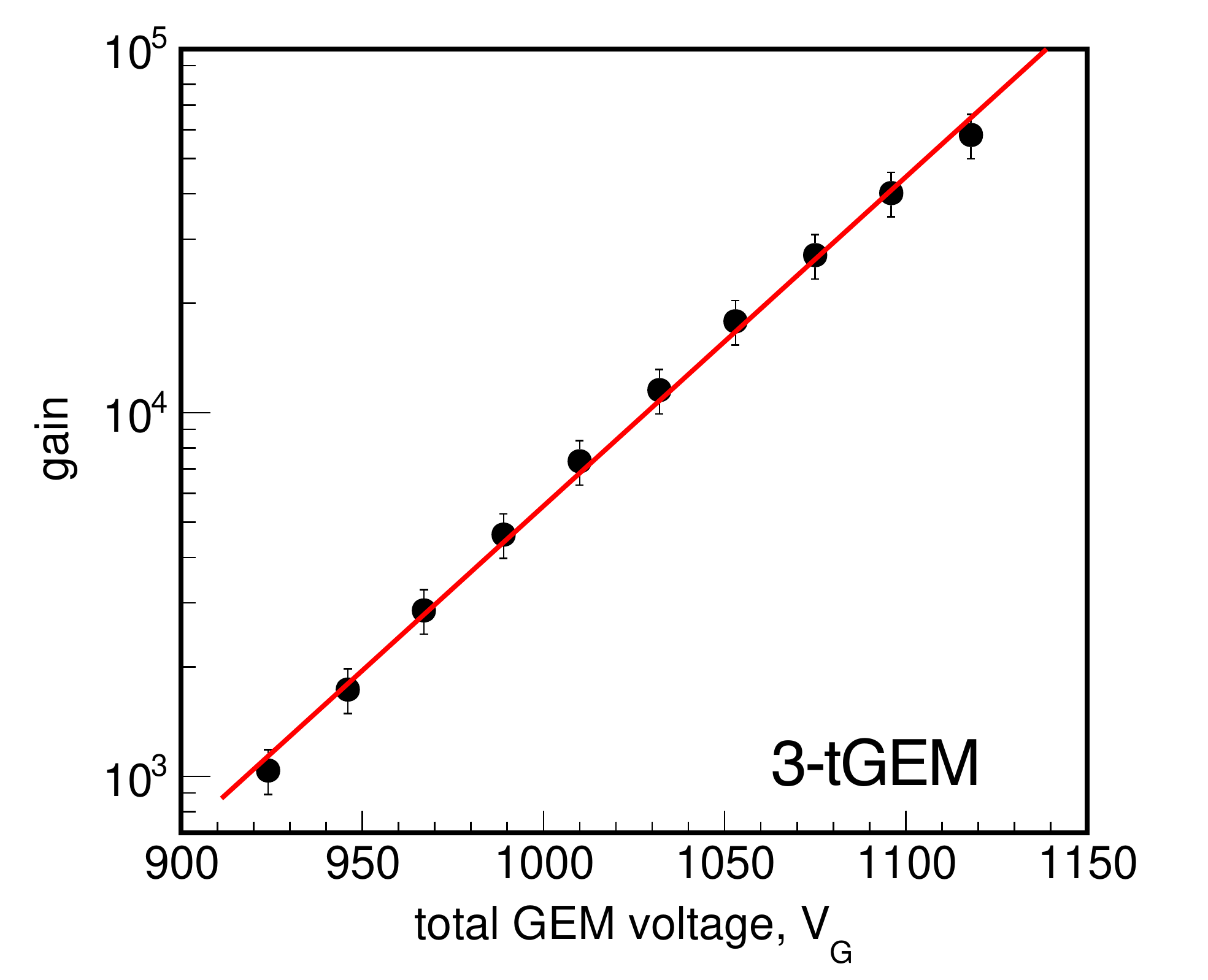}
  \includegraphics[width=7cm]{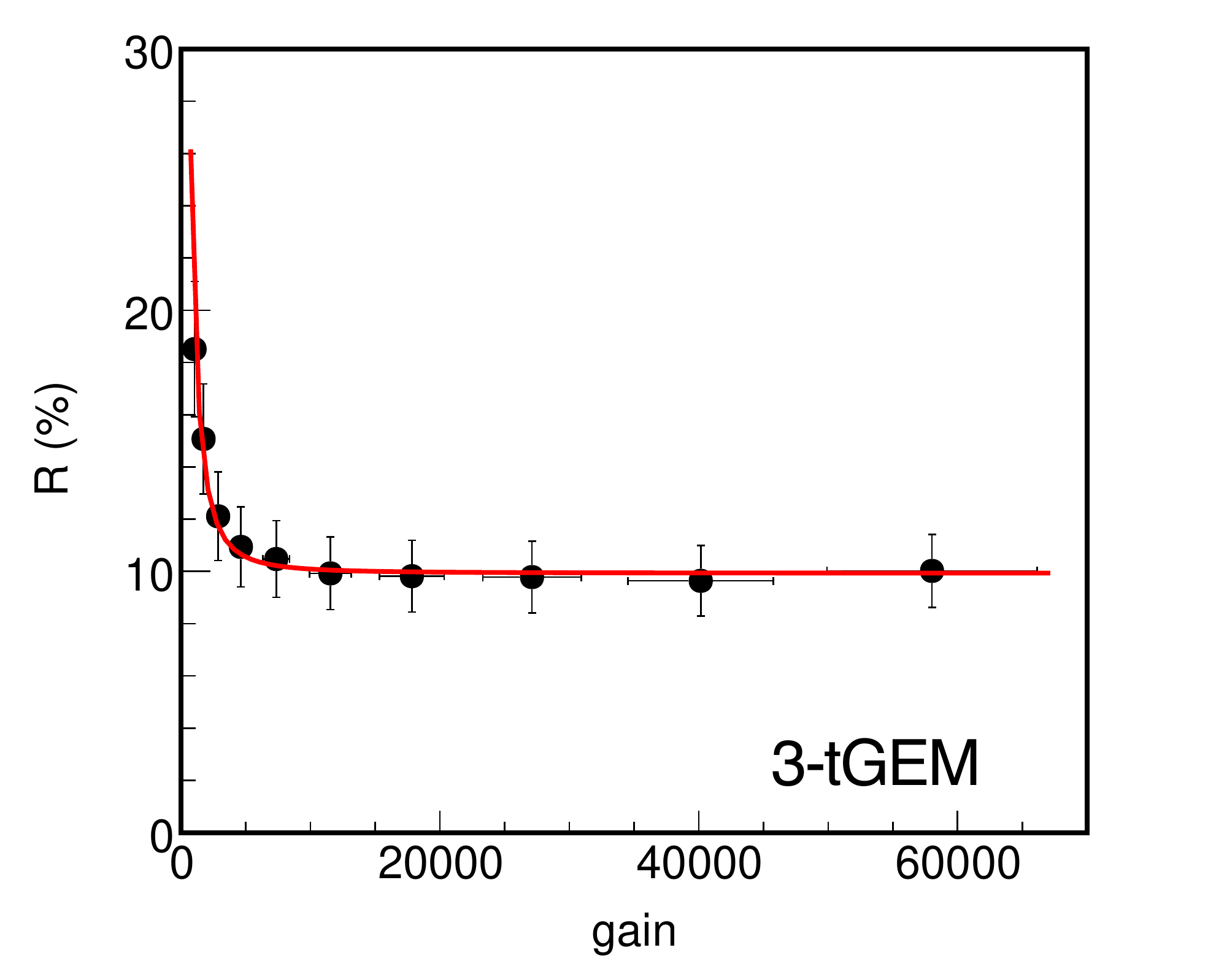}	
  \includegraphics[width=7cm]{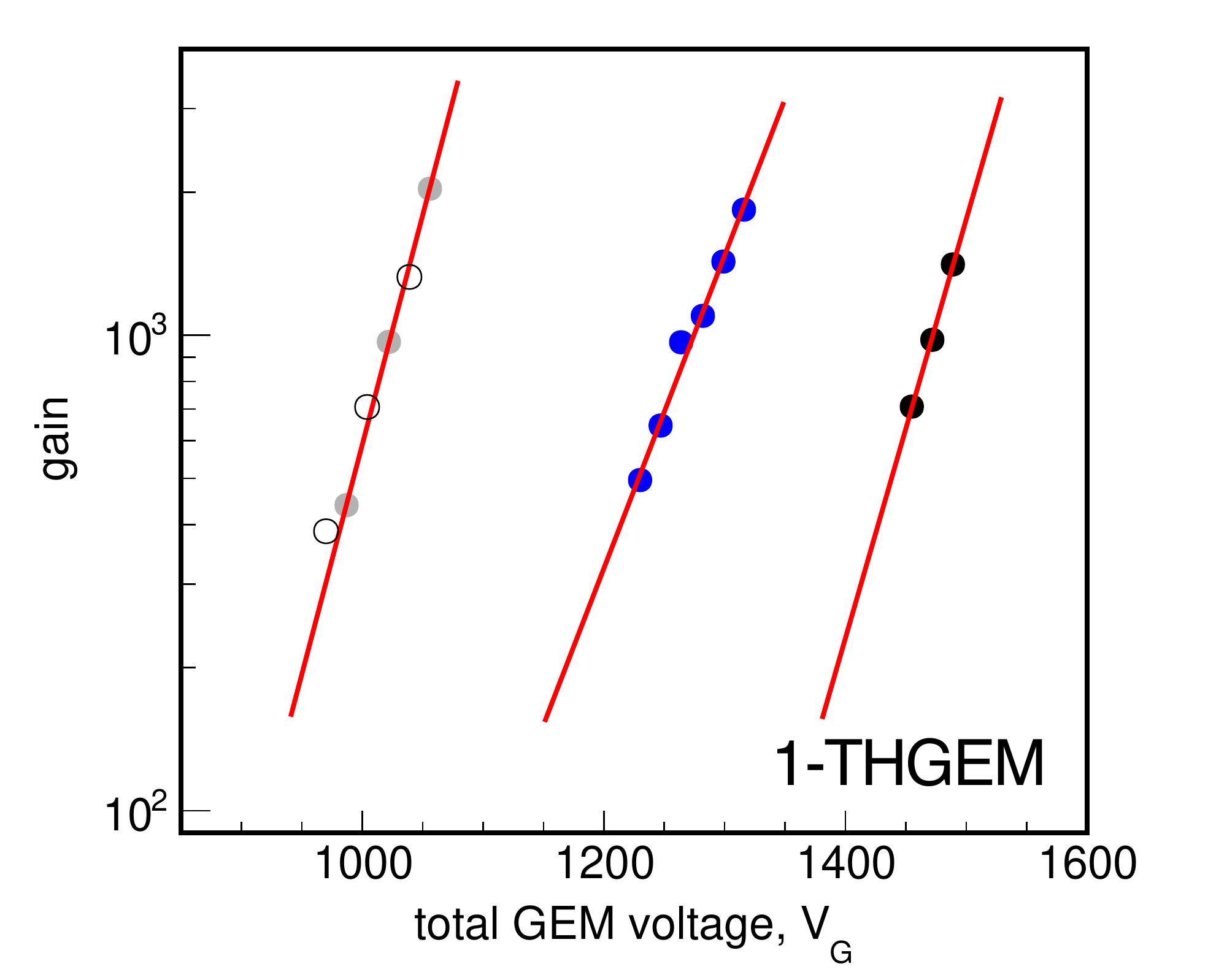}
  \includegraphics[width=7cm]{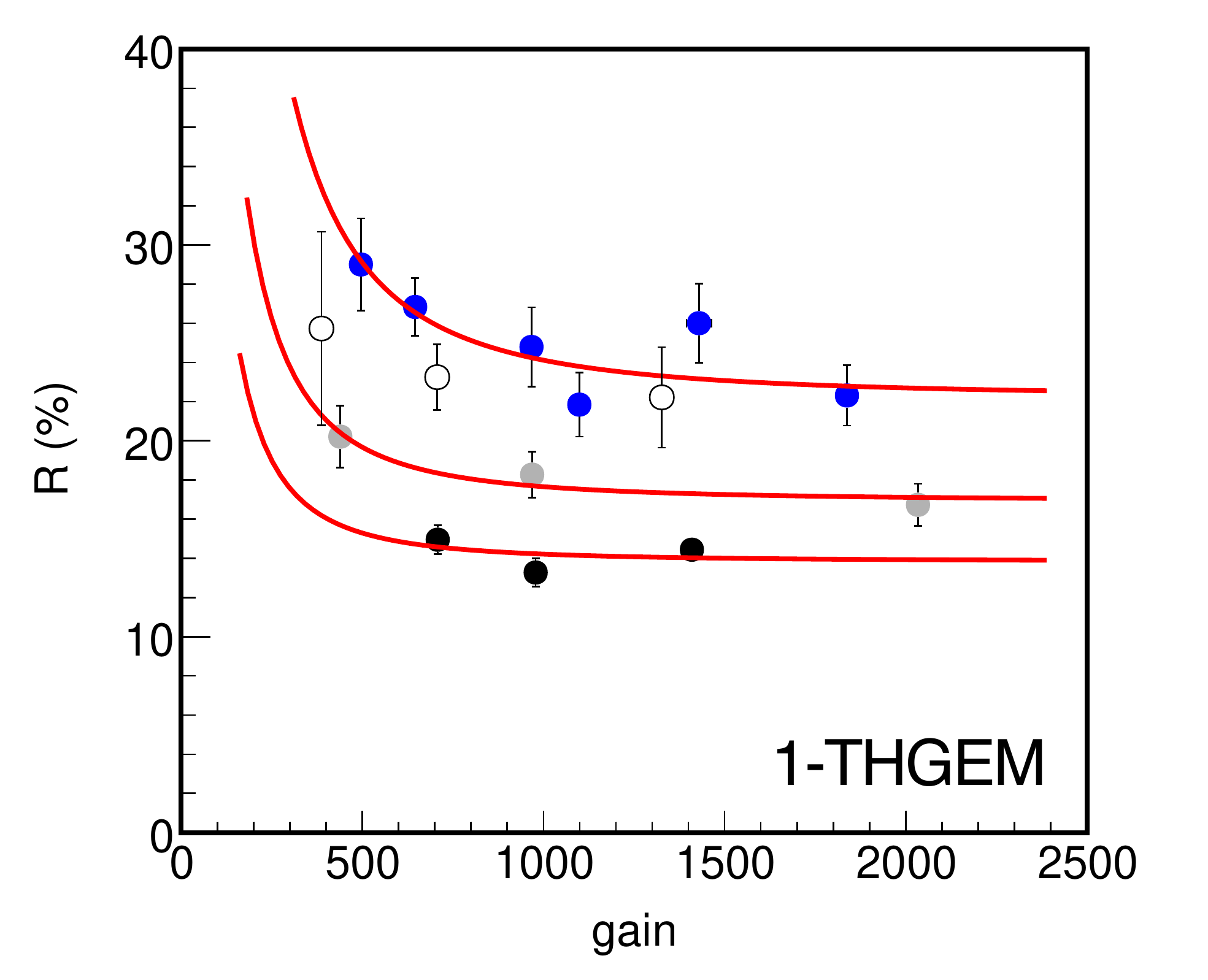}
  \includegraphics[width=7cm]{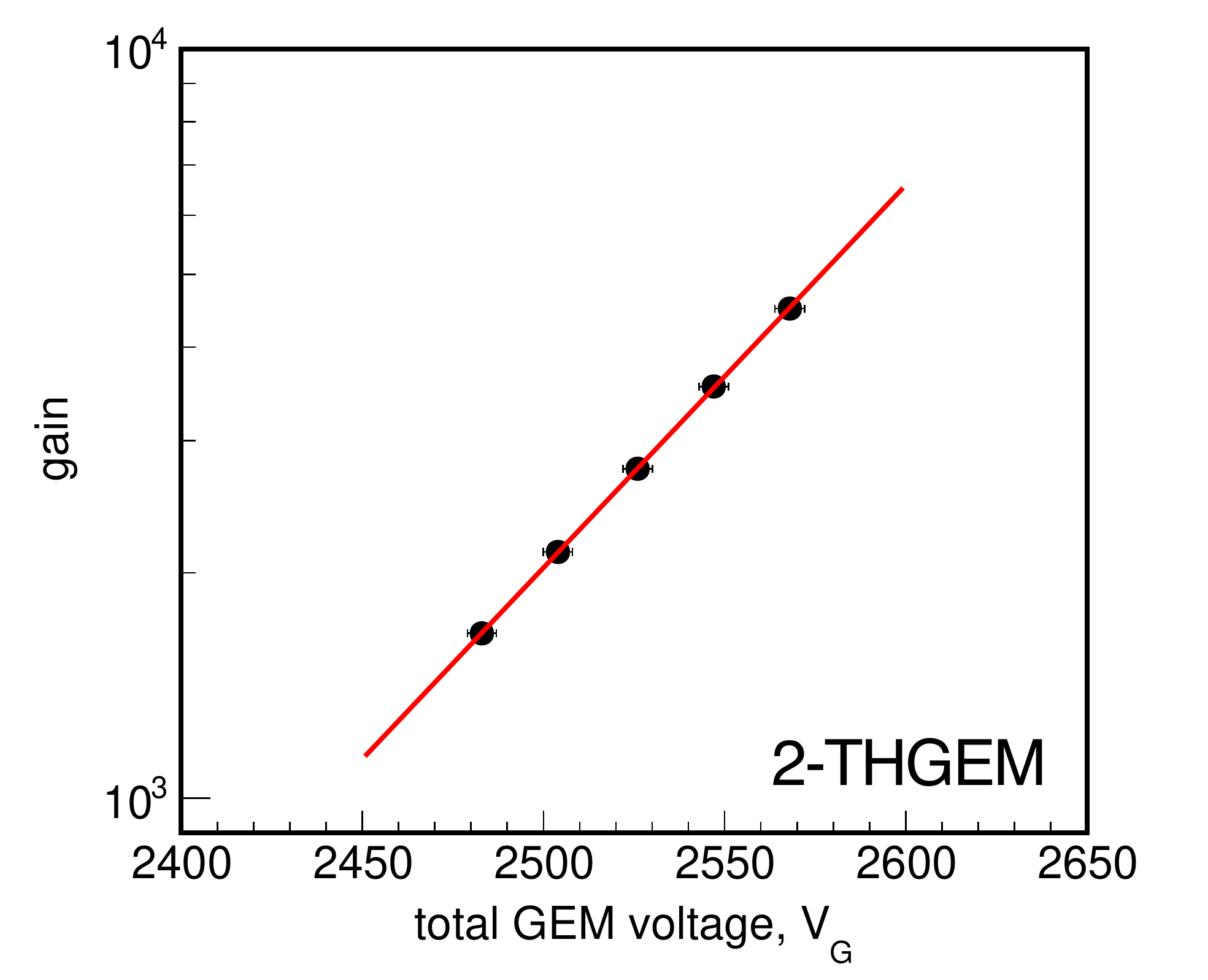}
  \includegraphics[width=7cm]{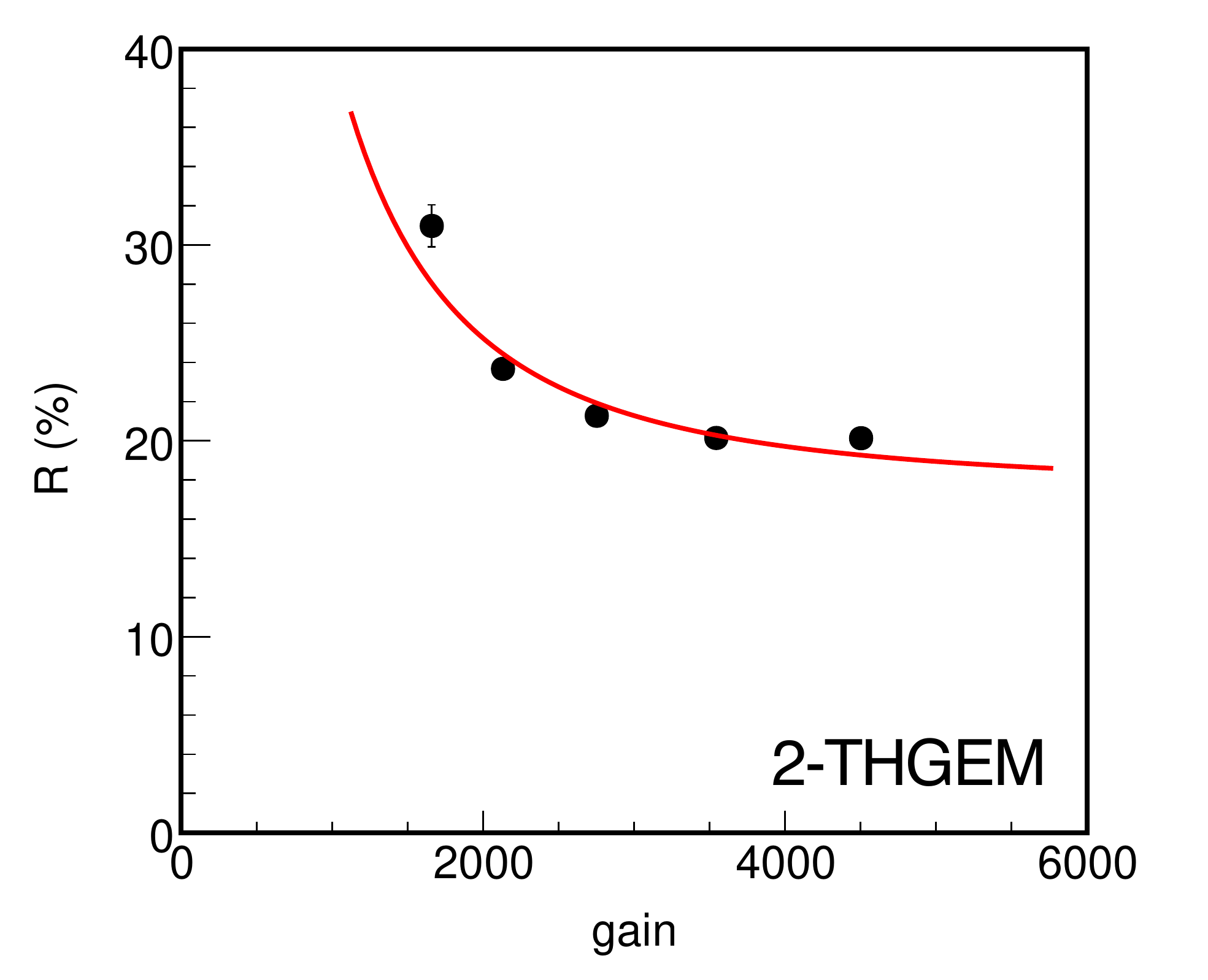}
  \setlength{\abovecaptionskip}{0pt}
  \caption{Results of $^{55}$Fe measurements with $^4$He:CO$_2$ (70:30).  Top row: 2-tGEM setup.  Second row: 3-tGEM setup.  Third row: 1-THGEM setup, from left to right: \SI{380}{torr} (gray, open data points are not included in fits shown), \SI{570}{torr} (blue points), and \SI{760}{torr} (black points).  Last row: 2-THGEM setup.  The solid lines (red online) are fits of Eqs. \ref{eqn:gain_voltage} (left) and \ref{eqn:gain_res} (right) to the data.}
  \setlength{\belowcaptionskip}{0pt}
  \label{fig:heco2_gain_res}
\end{figure*}
\subsection{Analysis of gain and PH resolution data}\label{sec:gain_spec}

Once the PH spectra are fit and values for the gain and resolution have been extracted, we make use of Eqs.~\ref{eqn:gain_voltage} and \ref{eqn:gain_res}.  We summarize the procedure applied to all measurements, including the SF$_6$ measurements discussed in Section \ref{sec:sf6}:

\begin{itemize}

\item{Check for similar number of counts (within $\approx~10\%$) in each spectrum; spectra with a low number of counts are discarded.}

\item{Perform bin-by-bin background subtraction with time normalized non-source spectra (not applicable to 3-tGEM).}

\item{Perform $\chi^2$-minimization of a fit function on the spectra; extract mean and sigma values.}

\item{Calculate fractional PH resolution as the sigma divided by the mean.}

\item{Determine gain from the mean and calibrated PH scale.}

\item{Determine precise voltage values by applying voltage correction factors.}

\item{Plot gain versus total GEM voltage, $V_g$, for each detector setup; fit with Eq.~\ref{eqn:gain_voltage} and extract parameters $V_1$ and $V_2$.  See Figs.~\ref{fig:heco2_gain_res} (left) and \ref{fig:sf6_gain_res} (top) for He:CO$_2$ and SF$_6$ data, respectively.}

\item{Plot PH resolution versus gain for each detector setup; fit with Eq.~\ref{eqn:gain_res} and extract parameters $a$ and $b$.  See Figs.~\ref{fig:heco2_gain_res} (right) and \ref{fig:sf6_gain_res} (bottom) for He:CO$_2$ and SF$_6$ data, respectively.}

\end{itemize}

\subsection{Measurements in $^4$He:CO$_2$ (70:30)}\label{sec:heco2}

We will discuss details and peculiarities of six He:CO$_2$ data sets obtained the with four setups pictured in Fig.~\ref{fig:dcube_milli}.  Given the time span, these measurements were performed with two different bottles of gas, which were from the same vendor with identical specifications.  The operational parameters are summarized in Table~\ref{tab:heco2_fields}.  Overall, the tGEM setups readily produce gas gains up to $6 \, \times 10^4$ and PH resolutions between 10\% and 20\%, while the THGEM setups produce gas gains of a few $\mathcal{O}$(10$^3$) and PH resolutions between 14\% and 30\%.  PH spectra are shown in Fig \ref{fig:spectra}.  The results are shown in Fig.~\ref{fig:heco2_gain_res}, with the gain versus total GEM voltage (left) and the PH resolution versus gain (right).  The returned fit parameters are summarized in Table~\ref{tab:fit_par_he}.

%Thin GEMs are more fragile and more challenging to work with than THGEMs, but their performance is generally better.  They are able to stably produce reduced fields above $\SI{100}{V/cm/torr}$ resulting in high avalanche gain and good resolution.  Here we present two data sets, one recorded with the 2-tGEM setup and the other recorded with the 3-tGEM setup.  

%THGEMs are more mechanically robust than thin GEMs, however they can not achieve the same field strength as thin GEMs and generally produce worse resolution.  The primary motivation for their use here was the study of SF$_6$.  It was thought that the THGEMs would perform better in the  required low pressure regime.  He:CO$_2$ was also studied and four sets of measurements are presented here, three using the 1-THGEM setup and one using the 2-THGEM setup. 

 %
\begin{table*}[h!] %\label{tab:fit_par}gray
  \centering
  %\begin{tabular}{l | l |  >{\centering\arraybackslash}p{1.8cm} | c | c | c | c r}
  %\begin{tabular}{l  >{\centering\arraybackslash}p{1.8cm} cccc}
  \begin{tabular}{l >{\centering\arraybackslash}p{1.1cm} >{\centering\arraybackslash}p{3.2cm} cccc}
 \toprule
  Setup & Pressure (torr) & Reduced GEM field(s) 1:2:3 (\SI{}{V/cm/torr})  & $V_1$ (\SI{}{V})& $V_2$ (\SI{}{V}) & $G_{\infty} = 100 \times a$ & $b$ (\SI{}{\%})  \\
 
  \hline
   \noalign{\vskip 2mm}
   2-tGEM & 760 & 131 : 129  & $463 \pm 4$ & $115 \pm 1$ & $(2.15 \pm 0.05) \times10^4$ & $12.17 \pm 0.09$ \\ %%   
   3-tGEM & 760 & 97 : 99 : 99 & $(5.9 \pm 0.1) \times 10^2$ & $110 \pm 4$ & $(1.8 \pm 0.3) \times 10^4$ & $ 9.9 \pm 0.5$ \\ %
   %1-THGEM & 380 & 70 &  $677 \pm 9$ & $115 \pm 3$ & $(6 \pm 1) \times 10^3$ & $17.8 \pm 0.9$ \\
   1-THGEM & 380 & 70 &  $ (7.1 \pm 0.1) \times 10^2$ & $104 \pm 4$ & $(5 \pm 2) \times 10^3$ & $16.9 \pm 0.9$ \\
   1-THGEM & 570 & 58 & $(8.2 \pm 0.2) \times 10^2$ & $152 \pm 5$ & $(9 \pm 2) \times 10^3$ & $22 \pm 1$ \\ %
   1-THGEM & 760 & 49 & $(1.13 \pm 0.3) \times 10^3$ & $(1.1 \pm 0.1) \times 10^2$ & $(3 \pm 3) \times 10^3$ & $13.8 \pm 0.8$ \\ %
   2-THGEM & 760 & 43 : 41 & $(1.86 \pm 0.4) \times 10^3$ & $(2.0 \pm 0.1) \times 10^2$ & $(3.6 \pm 0.2) \times 10^4$ & $17.5 \pm 0.5$ \\
   
%   2-tGEM & 760 & $463 \pm 2$ & $114 \pm 1$ & $(2.01 \pm 0.05) \times10^4$ & $10.21 \pm 0.09$ & 131:129 \rule{0pt}{3ex} \\ %%   
%   3-tGEM & 760 & $(5.9 \pm 0.1) \times 10^2$ & $110 \pm 4$ & $(1.8 \pm 0.3) \times 10^4$ & $ 9.9 \pm 0.5$ & 1\\ %
%   1-THGEM & 380 &  $677 \pm 9$ & $115 \pm 3$ & $(6 \pm 1) \times 10^3$ & $17.8 \pm 0.9$ & 1\\
%   1-THGEM & 570 & $(8.2 \pm 0.2) \times 10^2$ & $152 \pm 5$ & $(9 \pm 2) \times 10^3$ & $22 \pm 1$ & 1\\ %
%   1-THGEM & 760 & $(1.13 \pm 0.3) \times 10^3$ & $(1.1 \pm 0.1) \times 10^2$ & $(3 \pm 3) \times 10^3$ & $13.9 \pm 0.8$ & 1\\ %
%   2-THGEM & 760 & $(1.85 \pm 0.4) \times 10^3$ & $(2.0 \pm 0.1) \times 10^2$ & $(3.6 \pm 0.2) \times 10^4$ & $17.5 \pm 0.5$ & 1\\
   
   \bottomrule
  \end{tabular}
  \setlength{\abovecaptionskip}{6pt}
  \caption{Returned fit parameters for the $^4$He:CO$_2$ (70:30) data shown in Fig. \ref{fig:heco2_gain_res}.  The reduced field values are shown at the highest $V_G$ (highest gain) value for each data set.  Note that no corrections have been applied to these parameters.}
  \setlength{\belowcaptionskip}{0pt}
\label{tab:fit_par_he}
\end{table*}

\subsubsection{2-tGEM}\label{sec:2tGEM}

2-tGEM produced gain and asymptotic resolution values of $\mathcal{O}$(10$^4$) and $\approx$ 12\%.  Owing to experience obtained and low material outgassing, these are among the most stable and least noisy measurements recorded.  This is explicitly highlighted in Section~\ref{sec:stability}, see Fig.~\ref{fig:stability}.  The data being discussed here (the top row of Fig.~\ref{fig:heco2_gain_res}) were recorded near the end of the stability run (the gap on the right side of Fig.~\ref{fig:stability}).  After substantial time, material outgassing had degraded the PH resolution by a couple of percent, which is the reason the asymptotic resolution is slightly higher than with 3-tGEM.  After applying corrections for this degradation, the asymptotic resolutions from the thin GEM setups are nearly identical. 

\subsubsection{3-tGEM}\label{sec:3tGEM}

Similar to 2-tGEM, gains of $\mathcal{O}$(10$^4$) are typical with this setup and this data set (second row of Fig.~\ref{fig:heco2_gain_res}) were recorded at the end of a long stability run.  However, this setup contains more material which leads to higher outgassing resulting in a larger degradation of the gain and resolution over time.  One way to reverse this degradation is to flow gas through the vessel until the gain recovers, and this was done preceding the recording of this data set.  The effect is evident in the slight `bend' in the gain versus $V_G$ data, i.e. the highest and lowest gain data points both fall slightly below the solid line of best fit.  Although gas flow had been stopped, the gain increased by $\approx~\SI{1.5}{\%}$ during this series of measurements while the PH resolution was essentially constant.  

Another issue with 3-tGEM is the lack of background subtraction capability and a higher radioactive source rate.  Both of these issues contribute to the odd shape of the PH spectrum and the large `shelf' to the left of the source peak.  While the asymptotic resolution is slightly better here compared with 2-tGEM, it's important to remember that this was recorded immediately after substantial gas flow and material outgassing can explain the difference.  This suggests that the number of thin GEMs producing the gain, hence the number of transfer regions, has little effect on the PH resolution.  However, this is not observed in the THGEM setups.

\subsubsection{1-THGEM}\label{sec:1THGEM}

The third row from the top in Fig.~\ref{fig:heco2_gain_res} shows the data recorded with 1-THGEM.  Measurements were performed at three different gas pressures, 380 (gray, open data points are not included in fits shown), 570 (blue points), and \SI{760}{torr} (black points).  Overall, THGEMs produce lower gain and worse resolution compared with tGEMs, which can be expected as THGEMs typically operate at lower reduced fields.  Unfortunately, the data set recorded at a pressure of \SI{570}{torr} was recorded following changes in the gas pressure and flow.  The same is true for half of the data set recoded at a pressure of \SI{380}{torr} (open gray points).  We report them for completeness and detail the procedure for clarification.  

A long stability run was performed followed by the recording of the first three \SI{380}{torr} data points in Fig.~\ref{fig:heco2_gain_res} (gray points).  The pressure was then increased and the entire \SI{570}{torr} data set was recorded.  After this, the pressure was lowered to \SI{380}{torr} and the remaining three data points (open gray points) were recorded.  This is the reason the \SI{380}{torr} data, especially the PH resolution versus gain, is split into two different groupings.   Following the measurements at 380 and \SI{570}{torr}, the chamber was evacuated and filled to atmospheric pressure.  After a period of stabilization, the \SI{760}{torr} data set was recorded.  The \SI{570}{torr} (blue points) and the second half of the \SI{380}{torr} (open gray points) data clearly exhibit different gain and resolution trends compared to the other measurements, suggesting the system had not yet stabilized following the pressure change.  Therefore we exclude the second half of the \SI{380}{torr} data from all analysis and exclude the \SI{570}{torr} data from further analysis in Section~\ref{sec:heco2_analysis}.

Finally, the most notable issue with 1-THGEM was that it produced the lowest overall SNR, achieving maximum gains of $\mathcal{O}$(1000).  After the 3-tGEM measurements, the grounding scheme had to be improved before any gain from a THGEM could be measured, and being closer to the noise floor resulted in some non-Gaussian spectra features.  Exponential tails on the high-gain side can be seen in both the 380 and \SI{760}{torr} spectra shown in the middle row of Fig. \ref{fig:spectra}, and were included in the fit functions.  Collecting and analyzing the full waveforms could be beneficial in future measurements with similar setups.

\subsubsection{2-THGEM}\label{sec:2THGEM}

The bottom row in Fig.~\ref{fig:heco2_gain_res} shows the data recorded with 2-THGEM, and there are two things to note.  First, although stable gain was achieved, large voltages were required and sparking was a major issue.  Compared to 1-THGEM, the gain was more stable and uniform with a higher SNR resulting in gains of several thousand.  A redesign of this system to reduce sparking would likely prove fruitful.  

Second, the asymptotic PH resolution is substantially worse compared with 1-THGEM at the same gas pressure.  It is not immediately obvious why this is true for THGEMs, but not tGEMs.  After accounting for systematics, particularly gain and resolution degradation due to material outgassing, an explanation is provided in Section~\ref{sec:heco2_fund_res}.

\section{Systematics and assumptions}\label{sec:systematics}

Given the timeframe and number of experimental setups being discussed, a concerted effort was taken to account for, and minimize systematic issues.  Many systematic differences result in the gain axis being shifted, i.e. the gain versus $V_G$ relationships shifting vertically.  However, depending on the magnitude and time scale, varying temperature and gas purity can affect the relative gain difference within a given data set, i.e. the \textit{slope} of the gain versus $V_G$ relationship.  The temperature was monitored in our environmentally controlled laboratory, and fluctuations were correlated with $\mathcal{O}(\%)$-level gain fluctuations.  Electronegative impurities can absorb charge, and the concentration of impurities can change over time, possibly enough to affect individual measurements.

%For completeness, and in anticipation of the results presented in Section~\ref{sec:heco2_analysis}, we summarize the various systematic uncertainties.  We mention again the approximately 14\% uncertainty on the system response, and the gain as a result, for the 3-tGEM setup.  Given the calibration procedure for the resistive dividers used, there is an approximately 0.16\% uncertainty on all voltage values.  Pressure values have an uncertainty of 1\%, with the exception of those used in SF$_6$, which are estimated as 2\% due to the required gas flow.  The uncertainty in the GEM thickness is estimated to be 1\%, and all of these values are included within the error bars and quoted uncertainties throughout.  Due to the differences in the collection fields, there is an $\approx$ \SI{20}{\%} (maximum) difference in measured gain for 2-tGEM compared to the other setups.  Finally, there is the $\approx$ \SI{28}{\%} under-estimate of the field strength when compared to the uniform field assumption, which is common to all of the measurements.  The last two are not included in subsequent analysis or in the quoted errors.}. What follows are dedicated measurements as a function of the drift field and time.  For both studies, 2-tGEM was used with the $^{55}$Fe radioactive source in He:CO$_2$ at atmospheric pressure. 

For the analysis methods used in Section~\ref{sec:heco2_analysis}, we assume the following:  1) The different detector setups have roughly similar electric fields (within a factor of two of each other) in the transfer and collection regions.  2) The drift field values are similar.  3) The distribution of the voltages in multi-GEM setups is similar for all setups being analyzed.  See Fig.~29 in Ref.~\cite{Bachmann:1999xc} for a comparison of the effects on the gain of 1), 2), and 3).  4) The electric field in the GEM holes is uniform.

%%%%%

%Additional effects that can broaden the PH resolution include the variance in gain between the GEM holes \cite{Hilden:2018isz} and the gain stability.  

%gain differences of up to 2\% per \text{$\mu$}m of hole misalignment  gain differences can occur in multi-GEM setups  \cite{Hilden:2015qpi}

%%%%

\subsection{Electric field uniformity}\label{sec:uniform_field}

The electric field strength inside a GEM hole depends not only on the voltage applied to the GEM, but also on the geometry of the hole itself.  The electric fields above and below the GEM can also have an effect, and the general result is a dipole field within the GEM holes~\cite{Shalem:2006iw}.  Reference~\cite{Hallermann:2010zz} studies the electric field strength dependence on the hole diameter in thinner GEMs, and introduces a parameterization using the applied GEM voltage along with the electric field strengths above and below the GEM to estimate the GEM-hole electric field strength.  Applying this to the $\SI{70}{\text{$\mu$}m}$ diameter tGEMs holes used in our work gives an $\approx$~28\% lower estimate of the electric field strength than the uniform electric field assumption.  Reference~\cite{Hallermann:2010zz} also includes simulation results for the electric field uniformity for a 50 and a \SI{100}{\text{$\mu$} m} GEM, and the results are similar.  We conclude that the field uniformity for the THGEMs is not substantially worse than for the tGEMs used here.

%\textcolor{cyan}{The ratio of the hole diameter to GEM thickness is also similar and, for a given GEM voltage, the peak field strength does decrease with increasing hole diameter, but the avalanche is also continuing over the larger substrate thickness.  As long as the field strength is large enough, the overall effect should be similar given similar ratios.  The uncertainty would then be roughly equivalent for all of the gain values, and act as an offset to the reduced field axis.}

%%%%
\subsection{Detector noise}\label{sec:noise}

More sophisticated, highly-segmented readouts can introduce additional geometric or threshold effects, which contribute to the energy resolution at all gain values.  These effects can be difficult to account for~\cite{Lewis:2021mgp} because they usually result in charge that is not collected, the amount of which is hard to estimate.  The large anode used here is not segmented, hence we avoid these issues.  However, the anode introduces a gain-independent noise term, which acts as a `noise floor'.  This hinders measurements with low SNR which, with low-energy sources, are low-gain signals.  The lowest measurable gain values (with fully resolvable PH spectra peaks) for each setup are reported in Fig.~\ref{fig:heco2_gain_res} as the lowest gain data points.

Using the calibrated PH scale, we can estimate the magnitude of the aforementioned noise floor.  We find that the level below which measurements are not possible is in the range of $\mathcal{O}(10^4 - 10^5)$ electrons.  If we assume that this is Johnson–Nyquist noise then this gives a capacitance of $\mathcal{O}(100)$ \SI{}{nF} for the noisiest setups.  This is not unreasonable considering the physical dimensions of the anode. 

\subsection{Measurements of gain and PH resolution versus drift field}\label{sec:drift}

The drift field strength can have a large effect on the gain and PH resolution.  Essentially, when the transport of charge into the GEM holes becomes most efficient the gain and PH resolution will reach a plateau.  Depending on the gas mixture and voltage distribution over the GEMs, this plateau occurs roughly in the range: 0.5 - \SI{2}{kV/cm}.  As the drift field increases further, the transport efficiency starts to decrease~\cite{Bachmann:1999xc}.

\begin{figure}[h!]
  \centering
  \includegraphics[width=\linewidth]{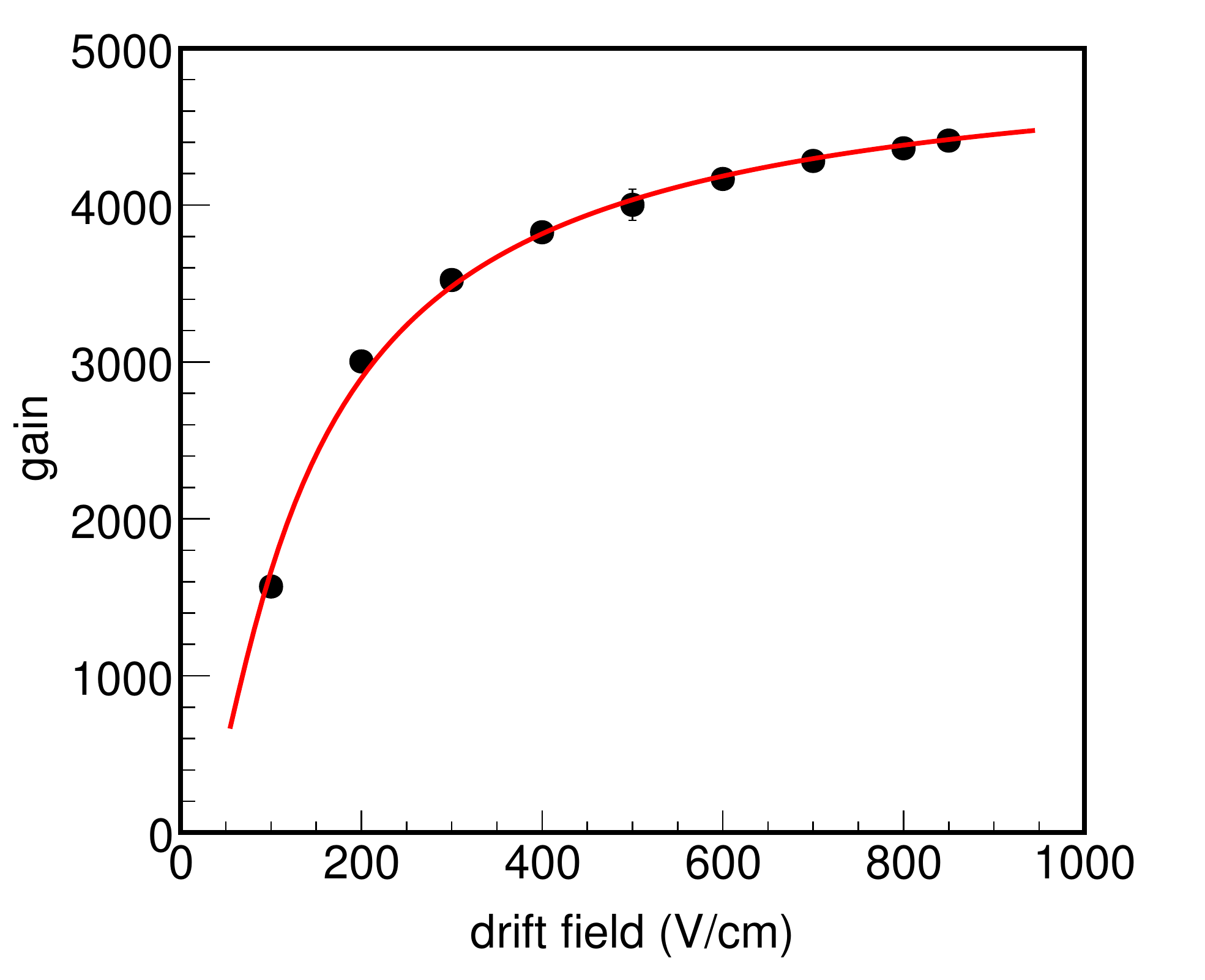}
  \includegraphics[width=\linewidth]{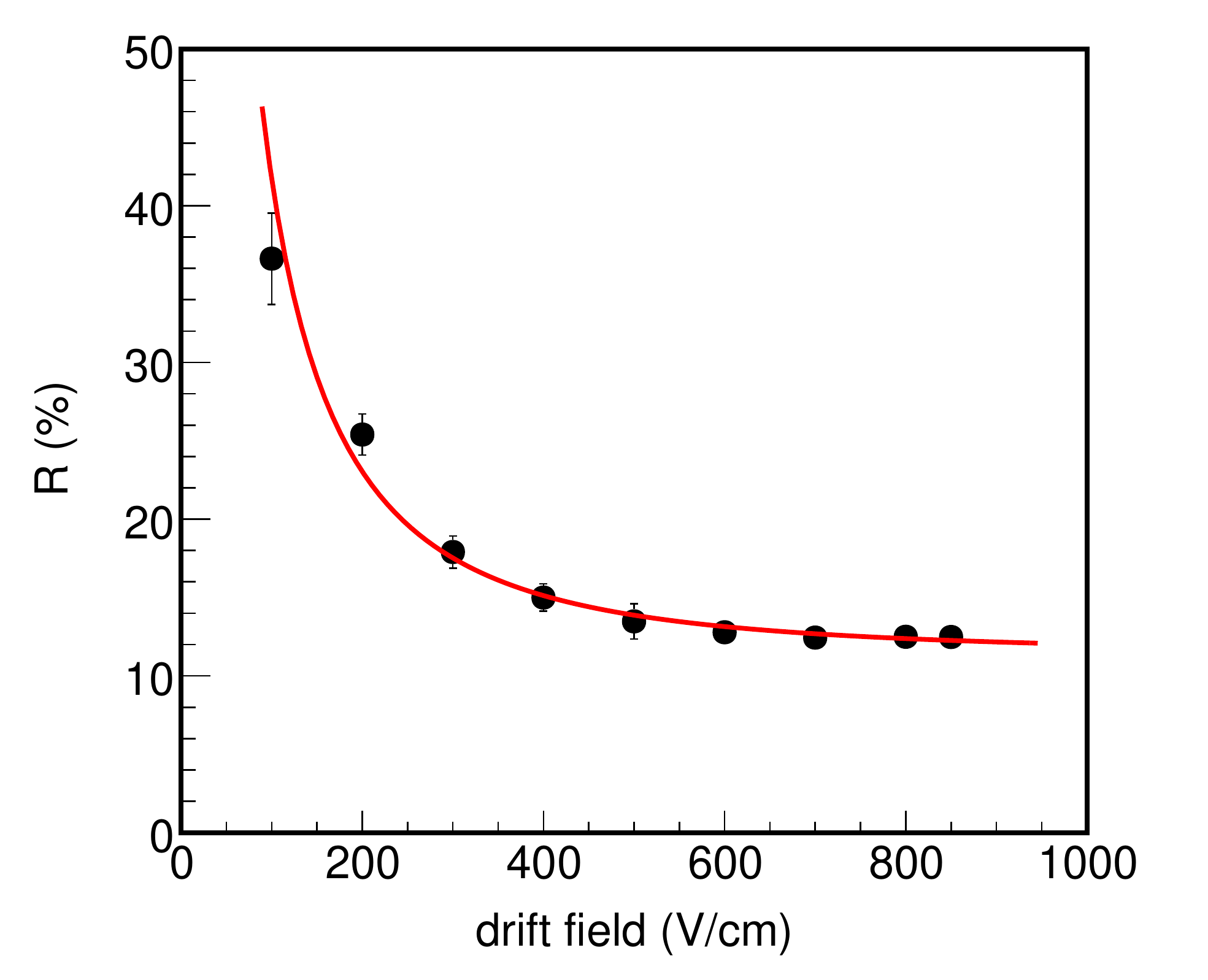}
  \caption{Results of $^{55}$Fe measurements with $^4$He:CO$_2$ (70:30) at atmospheric pressure as a function of drift field using 2-tGEM.  Top: Gain versus drift field.  Bottom: PH resolution versus drift field.  The solid lines (red online) are fits of Eqs. \ref{eqn:drift_gain} (top) and \ref{eqn:drift_res} (bottom) to the data.}
  \label{fig:drift}
\end{figure}

For this study, $V_G$ was held constant at \SI{874}{V} which corresponds to a gain of approximately 4000.  For reference, this is the same setting used for the third data point from the left in the top left plot in Fig.~\ref{fig:heco2_gain_res}.  Fig.~\ref{fig:drift} shows the gain (top) and PH resolution (bottom) versus drift field, $E_{drift}$.

We assume that the gain exponentially approaches a maximum value as the drift field increases and the charge transport becomes most efficient.  We write

\begin{equation}
G = 10^{ \; c - E_1/E_{drift}},
\label{eqn:drift_gain}
\end{equation}
where $c$ and $E_1$ are returned fit parameters.  At infinite drift field, $c$ is the log of the gain.  $E_1$ is the drift field value at which the gain is $\approx$~10\% of its maximum value.

For the purpose of finding a minimum value of the PH resolution versus drift field, we can model the transport inefficiency into the GEM holes as a `noise term' which is minimized with increasing drift field.  We can then write

\begin{equation}
R = \sqrt{ \Bigg( \frac{e} {E_{drift}} \Bigg)^2 + d \, ^2},
%\frac{\sigma_G}{G} = \sqrt{ \Bigg( \frac{e} {E_{drift}} \Bigg)^2 + d \, ^2},
\label{eqn:drift_res}
\end{equation}
to describe the PH resolution as a function of $E_{drift}$.  Similar to Eq.~\ref{eqn:gain_res}, $d$ is the asymptotic fractional PH resolution obtained at high drift field.  $e$ is the value of drift field below which the sigma of the PH distribution is certain to be greater than the mean.  

The solid lines in Fig.~\ref{fig:drift} are fits to the data points of Eqs.~\ref{eqn:drift_gain} (top) and \ref{eqn:drift_res} (bottom).  The extracted fit parameters are listed in Table \ref{tab:fit_par_drift}.  With the exception of the middle data points (at \SI{500}{V/cm}) in the plots in Fig.~\ref{fig:drift}, all of the gain and PH resolution values are obtained from a single five minute spectrum.  However, the middle points are the result of averaging multiple spectra.  The error bars on these data points quantify the fluctuations of the gain and PH resolution measurements over the time period that the entire data set was recorded, and indicate a very stable system.   

The He:CO$_2$ measurements at atmospheric pressure in Section~\ref{sec:heco2} were taken with a drift field of approximately \SI{500}{V/cm} and, as evidenced from the current discussion and the plots in Fig.~\ref{fig:drift}, this is not the optimal drift field for maximizing the gain or minimizing the PH resolution.  However, measured at a drift field of \SI{500}{V/cm}, the PH resolution is within a few percent of its asymptotic value and any previous conclusions will not be substantially affected by this.  Nonetheless, operating at a drift field that ensures adequate electron transport efficiency into the GEM holes is an important design consideration.  We will refer back to the drift field optimization in Section~\ref{sec:heco2_fund_res}.

\begin{table}[h!] 
  \centering
  %\begin{tabular}{c | c | c | c}
  \begin{tabular}{>{\centering\arraybackslash}p{1.8cm}  >{\centering\arraybackslash}p{1.8cm}  >{\centering\arraybackslash}p{1.8cm}  >{\centering\arraybackslash}p{1.8cm} }
  \toprule
  $c$ & $E_1 \SI{}{(V/cm)}$ & $e \, \SI{}{(V/cm)}$ & $d$ (\%)  \\
  \hline
  \noalign{\vskip 2mm}
   $3.70 \pm 0.01$ & $48 \pm 2$ & $ 40 \pm 2 $ & $11.3 \pm 0.4$ \\
  \bottomrule
  \end{tabular}
  \setlength{\abovecaptionskip}{6pt}
  \caption{Fit results for $^{55}$Fe measurements with $^4$He:CO$_2$ (70:30) at atmospheric pressure as a function of drift field using 2-tGEM.}
  \setlength{\belowcaptionskip}{0pt}
\label{tab:fit_par_drift}
\end{table}
\subsection{Measurements of gain and PH resolution versus time}\label{sec:stability}

Many factors can affect the gas gain and PH resolution over different time scales including temperature, gas purity, and device charge up.  An intense effort was undertaken to understand the stability for all of the setups and, generally, the system stability was monitored for days before and after our primary measurements of varying V$_G$.  Over short time scales, HV jitter can cause gain variations in time, and gas purity becomes dominant over longer time scales.  The HV supplies were chosen such that HV jitter would be a sub-dominant effect.  A similar double-tGEM setup is  discussed in Ref.~\cite{Jaegle:2019jpx}, where a highly segmented pixel ASIC readout is employed showing week-long, percent-level gain stability with a radioactive $\alpha$-particle source.  We discuss a few general observed features.

%The exact mechanism of the gain increasing after the initial back-fill is not obvious, although it does seem to be related to the gas itself as opposed to an electronics or readout issue.  The rapidly expanding gas is cold and will warm up after filling into the vessel.  However, if the initial gain variation were related to the gas temperature, one would instead expect the gain to decrease as it goes as the inverse of pressure.  Whatever the case, this period lasts for up to \SI{30}{minutes} after which the gain becomes stable.

Immediately following the initial fill of the vacuum vessel there is a `stabilization' period, where the gain increases a few percent before reaching a plateau.  This seems to be consistent with the charging up of the insulating material in the GEMs~\cite{Correia:2014vla}~\cite{Alexeev:2015kda}.  This period lasts for up to \SI{30}{minutes} after which the gain becomes stable.  If there is neither a gas purification system nor gas is flowed, the gain and resolution will degrade over time.  The degree of the degradation and the timeline involved can depend on many factors, but the largest culprit is the material outgassing of the detector components.  In the setups with low material outgassing, including 2-tGEM, the gain with atmospheric-pressure, electron-drift gases can remain stable within a few percent for many days.  The gain is proportional to the mean of the PH distribution and will decrease along with the gas purity, while the sigma of the distribution increases.  A commonly observed feature is that the sigma degrades more dramatically than the mean.  As a result, the PH resolution is usually more sensitive to gas purity issues than the gain itself.

\begin{figure}[h!]
  \centering
  \includegraphics[width=\linewidth]{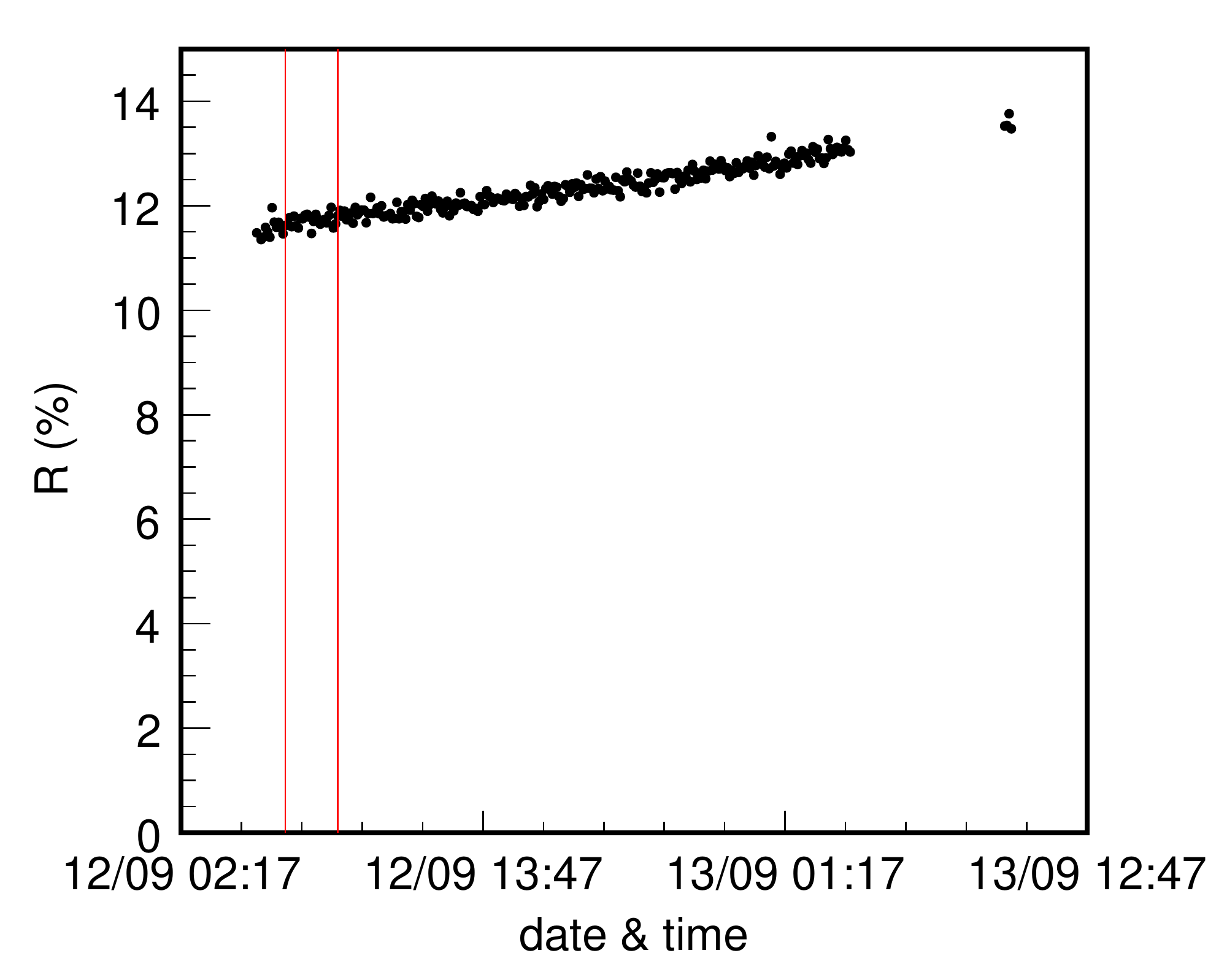}
  \caption{Results of $^{55}$Fe resolution measurements with $^4$He:CO$_2$ (70:30) at atmospheric pressure as a function of time using 2-tGEM.  The solid vertical lines (red online) enclose the two hour time period used for correcting the resolution against degradation over time (see Section~\ref{sec:heco2_fund_res}).}
  \label{fig:stability}
\end{figure}     

Fig.~\ref{fig:stability} shows the PH resolution versus time from 2-tGEM.  Over this time period the sigma of the PH distributions increased by $\approx$~10\%, while the mean decreased by only a few percent.  The gap on the right indicates the time during which the measurements discussed in Section~\ref{sec:2tGEM} were performed.  As a result, the PH resolution during the gain versus $V_g$ measurements is a few percent higher than near the beginning of the stability run.  As the PH resolution changes over time, we can use the stability data to estimate the asymptotic PH resolution if the gain versus $V_g$ measurements were taken at a different point in time.  For instance, in the present 2-tGEM case, if the gain versus $V_g$ measurements were performed during the time bounded by the two vertical red lines in Fig.~\ref{fig:stability}, instead of during the gap on the right, all of the PH resolution values would be $\approx$~\SI{2}{\%} lower than what is shown in Fig.~\ref{fig:heco2_gain_res}.  It follows that the asymptotic PH resolution would be $\approx$~\SI{2}{\%} lower as well.  We will refer back to this topic in Section \ref{sec:heco2_fund_res}.

%\clearpage

\subsection{Summary of systematic uncertainties}\label{sec:systematics_summary}

The following uncertainties are included throughout the analysis:

\begin{itemize}

\item{0.16\% uncertainty on voltages.}

\item{1\% uncertainty on GEM substrate thickness.}

\item{1\% (2\%) uncertainty on He:CO$_2$ (SF$_6$) gas pressures.}

%\item{The effect of material outgassing on the asymptotic resolution is corrected for, see Section~\ref{sec:heco2_fund_res}.}

\item{14\% uncertainty on the calibrated PH scale (system response) for 3-tGEM.}

\end{itemize}

All gain measurements here are `effective' as we do not account for charge losses due to transfer efficiencies or other mechanisms.  We choose the maximum transfer and collection field values that allow for stable operation, but optimize no further.  The following effects and uncertainties are not included in the analysis.  We list them for completeness:

%From Fig.~29 in Ref.~\cite{Bachmann:1999xc}, we can see that the different collection field values used in this work will have the largest impact.  

%As shown in Table \ref{tab:heco2_fields}, the collection field strength in 2-tGEM is higher compared to the other setups, and the largest difference is between 2-tGEM and 3-tGEM.  We estimate an $\approx$ \SI{20}{\%} difference in measured gain due to this effect, while the remaining differences, including in the drift and transfer fields, would have less impact.

\begin{itemize}

\item{The effect of the drift field on the PH resolution of $\approx$~2\% was discussed above and further in Section~\ref{sec:heco2_fund_res}.}

\item{2\% per \text{$\mu$}m of hole misalignment (up to \SI{8}{\text{$\mu$}m}) relative gain differences can occur in multi-GEM setups \cite{Brucken:2021tze}.}

\item{10\% gain variance across a given GEM foil ($10 \times \SI{10}{cm^2}$) due to differences in hole size \cite{Hilden:2018isz}.  However, we estimate a maximum of three GEM holes are producing gain per event here, and this effect is expected to be much less than 10\%.  The degree of transverse diffusion in drift regions of similar detectors is addressed in Refs.~\cite{Vahsen:2014fba}, \cite{Lewis:2014poa}, and \cite{Lewis:2021mgp}.}

\item{Due to the differences in the collection fields, there is an $\approx$~\SI{20}{\%} (maximum) difference in gain between 2-tGEM and 3-tGEM (see Fig.~29 in Ref.~\cite{Bachmann:1999xc}).  All remaining differences due to this effect are less.}

\item{The uniform field assumption over-estimates the field strengths in the tGEM holes by $\approx$~\SI{28}{\%}~\cite{Hallermann:2010zz}.  We consider this a reasonable over-estimate for the THGEMs as well (discussed in Section~\ref{sec:uniform_field}).}

%At the same field, THGEMs produce half the voltage as expected with a uniform field…  Which means you need to double the field to get the same effective voltage, or the same effective gain. 
%
%2-tGEM and 2-THGEM.  Nearly equal gain at a voltage factor of ~3…  All things being equal, if the field in the THGEM had the same uniformity as the tGEM you would expect the same gain at a voltage factor of 8… So these THGEMs are producing less than half the field you would expect.  The relationship is not linear though…  
%
%The uniform field assumption is more drastic for the 100um than the 50um GEM, by roughly a factor of 2.

\item{A lower bound on the gain measured in SF$_6$, by $\approx$ 10\%, resulting from some event charge not being integrated (discussed in Section~\ref{sec:sf6_PH}).}

\end{itemize}

%\section{Further analysis using reduced quantities} \label{sec:analysis}

\section{Analysis and discussion of high gas gain data}\label{sec:heco2_analysis}

This section uses the asymptotic, reduced quantities defined in Section~\ref{sec:asy_red_quant} to further examine the He:CO$_2$ data.  We consider the relationship between $\Sigma_{\infty}$ and $\Gamma_{\infty}$, and discuss the conditions to reach the minimum energy resolution achievable in GEM-based detectors. 

\begin{table*}[h!] %\label{tab:fit_par}gray
  \centering
  %\begin{tabular}{l | l |  >{\centering\arraybackslash}p{1.8cm} | c | c | c | c r}
  \begin{tabular}{l l cccccc }
  %\begin{tabular}{p{1.2cm} p{2cm} >{\centering\arraybackslash}p{1cm} >{\centering\arraybackslash}p{0.7cm} >{\centering\arraybackslash}p{1.2cm} >{\centering\arraybackslash}p{1.3cm} >{\centering\arraybackslash}p{1cm} >{\centering\arraybackslash}p{2.6cm} >{\centering\arraybackslash}p{2cm} }
 \toprule
  Gas (fit to which quantities) & Fit with Eq.~& $\chi^2$ & ndf & $1 - P$ & B/A (\SI{}{V}) & $\Sigma_1$ (\SI{}{V/cm/torr}) & $V_3$ (\SI{}{V}) \\ 
 
  \hline 
\noalign{\vskip 2mm}
%\\[0.0ex]      

%\\[0.0ex]   

%   SF$_6$ (all reduced) & \ref{eqn:red_gain_field} & 2.142 & 4 & 0.290 & n/a & $(2.0 \pm 0.2) \times 10^{2}$ & $80 \pm 4$ \rule{0pt}{3ex} \\ %%   
%   SF$_6$ (all reduced) & \ref{eqn:reduced_gain}  ($m = 0$) & 2.176 & 4 & 0.297  & 37 $\pm$ 3 & n/a & n/a \\ %%

   He:CO$_2$ (asymptotic) & \ref{eqn:red_gain_field} & 0.274 & 3 & 0.035 & n/a & $33 \pm 1$ &  $71 \pm 2$ \\ %%   
   He:CO$_2$ (asymptotic) & \ref{eqn:reduced_gain} ($m = 0) $ & 0.256 & 3 & 0.032 & 35 $\pm$ 2 & n/a & n/a \\ %%   
   \noalign{\vskip 2mm}
    Ar:CO$_2$ (all reduced) & \ref{eqn:red_gain_field} & 0.079 & 11 & $6.5 \times 10^{-11}$ & n/a & $65 \pm 2$ & $54 \pm 2$  \\ %%   
    Ar:CO$_2$ (all reduced) & \ref{eqn:reduced_gain} ($m = 0$) & 0.326 & 11 & $1.4 \times 10^{-7}$ & 27 $\pm$ 3 & n/a & n/a \\ %%
   
   \bottomrule
  \end{tabular}
  \setlength{\abovecaptionskip}{6pt}
  \caption{Summary of parameters extracted from fitting Eqs. \ref{eqn:reduced_gain} and \ref{eqn:red_gain_field} to the asymptotic, reduced quantities for He:CO$_2$ in the top plot of Fig.~\ref{fig:heco2_reduced_gain}, and to the reduced quantities for Ar:CO$_2$ in Fig.~\ref{fig:arco2_reduced_gain} (see \ref{sec:arco2}).  Here, the value $1-P$ is the probability that the $\chi^2$ value from a random model would be less than the quoted result, so a lower value reflects a better description of the data.  Notice that both equations describe the asymptotic, reduced quantities nearly equally well.  We fix $m = 0$ in Eq.~\ref{eqn:reduced_gain}, as there is not enough data to properly constrain it.}
  \setlength{\belowcaptionskip}{0pt}
\label{tab:analysis_par}
\end{table*}
%
 
%%%
%\subsection{Asymptotic, reduced quantities in $^4$He:CO$_2$ (70:30)}\label{sec:heco2_analysis}

%We will now discuss the He:CO$_2$ data within the more formal language of Section~\ref{sec:formalism}.  As mentioned in Section~\ref{sec:1THGEM}, this is a subset of the data discussed in Section~\ref{sec:heco2}.  We exclude some of the \SI{380}{torr} data along with the \SI{570}{torr} data set.  The motivation for this Section~begins by considering all of the plots in Fig.~\ref{fig:heco2_gain_res} together and looking for an obvious relationship between the gain and the energy resolution.  This eventually leads to the consideration of the asymptotic reduced quantities which allow us to understand the gain and energy resolution in GEM-based detectors in a broader context.

%%
\subsection{Asymptotic RFTC versus asymptotic reduced field}\label{sec:heco2_fund_gain}

%We compare two limiting cases of Eq.~\ref{eqn:reduced_gain}.  One is setting $m=0$, which provides a measurement for the effective ionization potential of the gas mixture in the form of $B/A$ \cite{AOYAMA1985125, Engel}.  The other is the threshold model, Eq.~\ref{eqn:red_gain_field}, which is the equivalent of setting $m=1$ in Eq.~\ref{eqn:reduced_gain}  and adding a voltage offset.  As discussed in Section~\ref{sec:1THGEM}, we consider five of the six data sets shown in Fig.~\ref{fig:heco2_gain_res}.

%
\begin{figure}[h!]
  \centering
  \includegraphics[width=\linewidth]{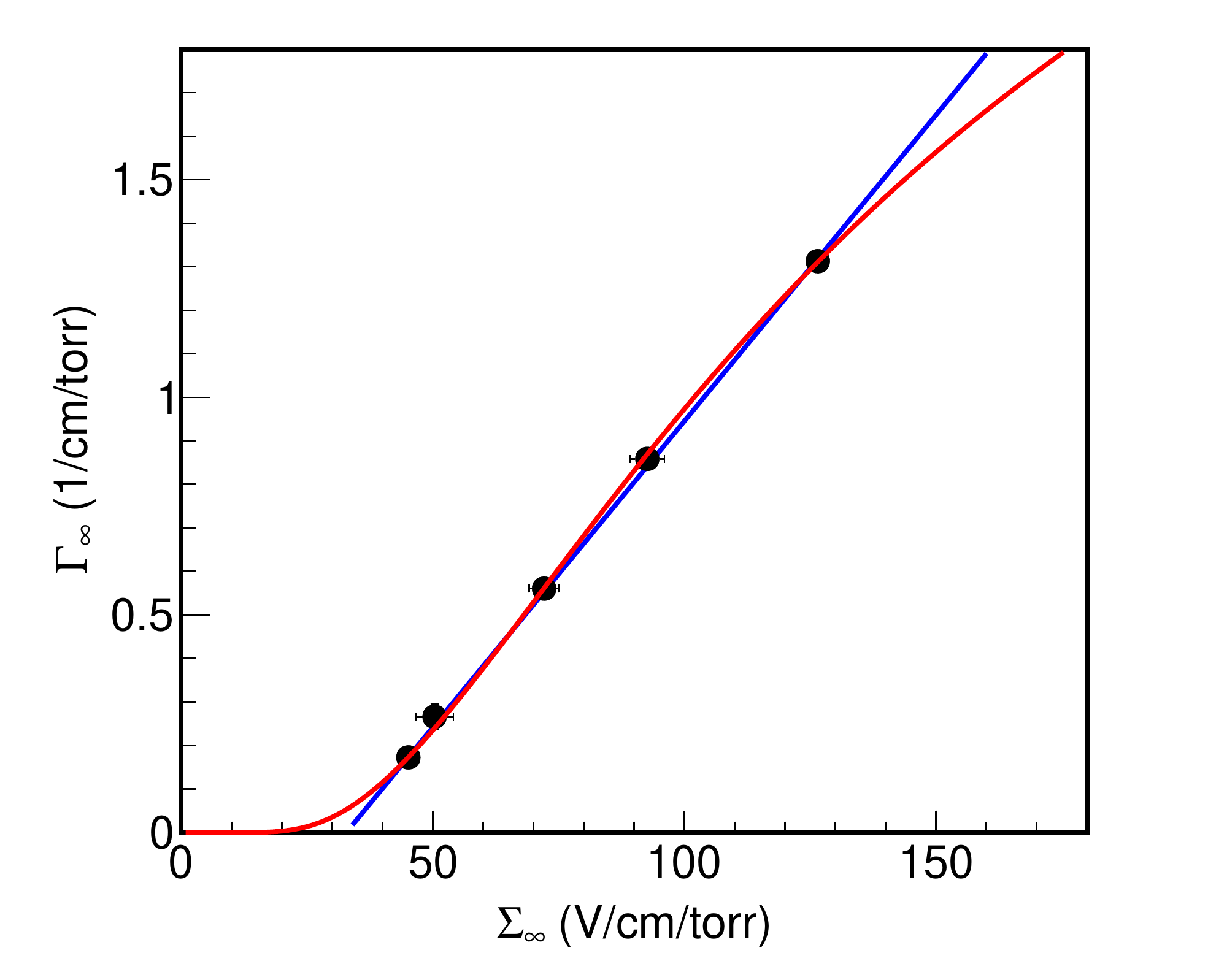}
  \includegraphics[width=\linewidth]{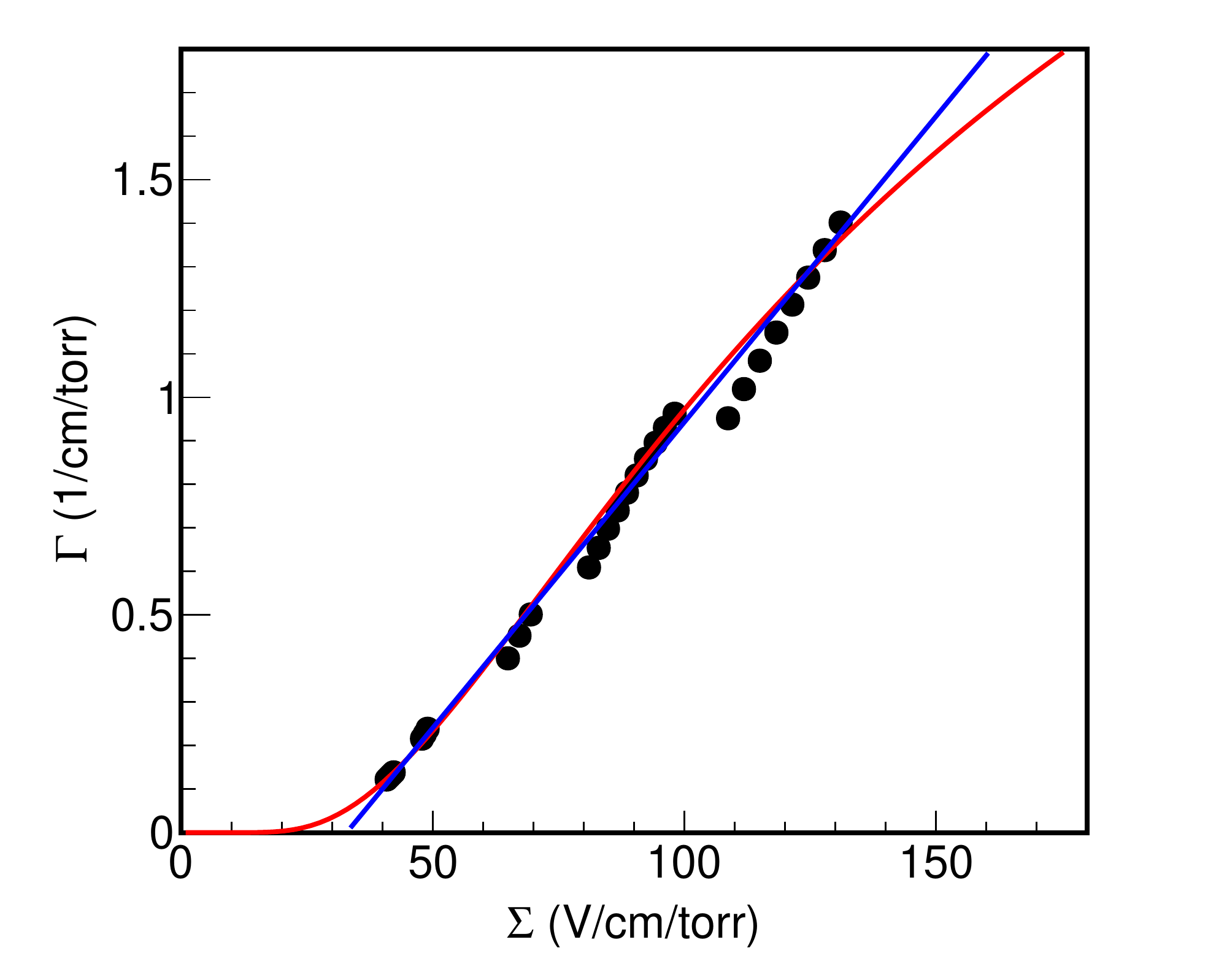}
  \setlength{\abovecaptionskip}{0pt}
  \caption{Top: Asymptotic RFTC $\Gamma_{\infty}$ versus asymptotic reduced field $\Sigma_{\infty}$ for five $^4$He:CO$_2$ (70:30) data sets.  The solid lines are fits of Eq.~\ref{eqn:red_gain_field} (blue online) and Eq.~\ref{eqn:reduced_gain} with $m=0$ (red online) to the data.  Bottom: RFTC $\Gamma$ versus reduced field $\Sigma$ for the same data sets.  The solid lines are overlays of the lines from the top plot.}
  \setlength{\belowcaptionskip}{0pt}
\label{fig:heco2_reduced_gain}
\end{figure}

As discussed in Section~\ref{sec:1THGEM}, we consider five of the six data sets shown in Fig.~\ref{fig:heco2_gain_res}.  The top plot in Fig.~\ref{fig:heco2_reduced_gain} shows the asymptotic RFTC versus asymptotic reduced field.  The solid lines are fits to the data of Eq.~\ref{eqn:red_gain_field} (blue online) and Eq.~\ref{eqn:reduced_gain} with $m=0$ (red online).  While it is assumed that $m=1$ when fitting Eq.~\ref{eqn:gain_voltage} to the individual data sets to obtain the asymptotic values, here we allow $m=0$. The results are summarized in Table \ref{tab:analysis_par}.  When $m=0$ in Eq.~\ref{eqn:reduced_gain} a value of $35 \pm 2$ is returned for $B/A$.  For reference, a W-value of \SI{34.4}{eV} \cite{Sharma:1998xw} was used to determine the gain values shown in Fig.~\ref{fig:heco2_gain_res}, and this illustrates the internal consistency of the method.  

The bottom plot in Fig.~\ref{fig:heco2_reduced_gain} shows the RFTC versus reduced field values for the five He:CO$_2$ data sets.  The solid lines are overlays of the fitted lines from the top plot to show where the asymptotic values fall within the full data sets.  We see that the many of the lower gain data points fall below either model prediction.  This implies that an effect that becomes more efficient at higher reduced fields in each setup is not accounted for in either model, and this could be related to the charge transfer efficiencies previously discussed.  However, a detector noise term that is capacitive in nature could affect the gain as well, with the lower gain values suffering more.  If the true low gain values were closer to the model predictions, then the corresponding PH resolution values would be lower.  This would result in the PH resolution versus gain data to `flatten' towards lower gains, and would be consistent with what has been observed in a similar GEM-based detector with a highly-segmented, low-noise, pixel ASIC readout \cite{Jaegle:2019jpx}.

 %to assess the range of functional forms that the asymptotic data could acquire in this parameter space.  The argument is that the ideal functional form for the gas mixture is more closely represented by the asymptotic values than by the individual data sets.  This is evident from the bottom plot in Fig.~\ref{fig:heco2_reduced_gain}, which shows the RFTC versus reduced field values for the five He:CO$_2$ data sets.  

Whatever the case, both Eqs. \ref{eqn:reduced_gain} with $m=0$ and \ref{eqn:red_gain_field} describe the asymptotic values well.  While constraining $m$ is not possible with this data, we note that future measurements could allow for this and discuss a few implications.  Recall that if $m=1$, the collision cross section between electrons in the avalanche is proportional to $E/p$, and $\alpha$ is proportional to the avalanching field strength in the GEM.  From this follows the exponential dependence of the gain on $V_G$ for a given setup at a given gas pressure.  If $m=0$, then $\alpha$ is independent of the field strength in the GEM.  This implies that the collision cross section between electrons is independent of the electron energy.  At least for electron energies below \SI{1}{keV}, this is not true \cite{Brook_1978}.  However, it is not clear that $m=1$ throughout the parameter space, and if $m \neq 1$ there will be deviations from the strictly exponential dependence of the gain on $V_G$.  These deviations are not drastic within the typically reduced field operating range of GEM-based detectors, but do increase with the reduced field.  Another outcome of $m=1$ is that the gain is predicted to increase without bound, whereas $m \neq 1$ indicates that the gain will indeed approach a limit.  Even though a limit is not explicitly considered in either model, it is expected as recombination effects increase and space charge issues arise.  Avalanche gain will increase until the Raether limit \cite{Raether} is reached.  

Using high gain data, we compared two limiting cases of Eq.~\ref{eqn:reduced_gain}.  One is setting $m=0$, which provides a measurement for the effective ionization potential of the gas mixture in the form of $B/A$ \cite{AOYAMA1985125, Engel}.  The other is the threshold model Eq.~\ref{eqn:red_gain_field}, which is equivalent to setting $m=1$ in Eq.~\ref{eqn:reduced_gain}  and adding a voltage offset.  Finally, we emphasize the robust nature of the relationship between $\Gamma_{\infty}$ and $\Sigma_{\infty}$, i.e. $\Sigma_{\infty}$ is a robust predictor of  $\Gamma_{\infty}$.  With a given GEM setup and gas pressure, this leads to the well known principle: the GEM voltage is a robust predictor of the effective gain.

\subsection{Avalanche variance and asymptotic reduced field}\label{sec:chi_heco2}

%\subsubsection{The avalanche distribution and reduced quantities}\label{sec:chi}

While Eq.~\ref{eqn:reduced_gain} forgoes many secondary effects, it still provides useful insight, and we can approach the avalanche variance discussion in a similar manner.  We consider only the effect of ionization, and combine the gain and avalanching field strength into a single quantity.  This quantity is related to the avalanche variance \cite{Alkhazov:1970fx}, which is the largest remaining contributor to the PH (energy) resolution. 

%This distance will be inversely proportional to the avalanching field strength within a given GEM, $V_G / n_gt$, and will tend make the avalanche distributions more peaked \cite{1958ZPhy..151..563S}.  

Consider the minimum distance a free electron must travel in order to initiate an avalanche.  We can then multiply this minimum distance by $\alpha$, which is a measurement of the number of electrons per unit length in the avalanche, to obtain a new quantity 

\begin{equation}
 \chi \sim \alpha \: \Bigg(\frac{n_g t}{V_G} \Bigg) = \frac{\textrm{ln}(G)}{n_gt} \Bigg(\frac{n_g t}{V_G} \Bigg) = \frac{\Gamma}{\Sigma}. 
%= A~ \Sigma ~^{m-1} e^{-B~ \Sigma~^{m-1}} % \frac{\textrm{ln}(G)}{V_G} %= A~ \Bigg(\frac{V_G}{n_gpt}\Bigg)^{m-1} \textrm{exp}~ \Bigg(-B~ \Bigg( \frac{n_gpt}{V_G} \Bigg)^{1-m} \Bigg)
\label{eqn:chi_def}
\end{equation}     
Using Eq.~\ref{eqn:gain_gem} for $\alpha$, we see that  $\chi$ is proportional to the RFTC divided by the reduced field.  Then, using Eq.~\ref{eqn:reduced_gain} we can determine its dependence on the reduced field  

\begin{equation}
\chi \sim A~ \Sigma ~^{m-1} e^{-B~ \Sigma~^{m-1}}. 
% \frac{\textrm{ln}(G)}{V_G} %= A~ \Bigg(\frac{V_G}{n_gpt}\Bigg)^{m-1} \textrm{exp}~ \Bigg(-B~ \Bigg( \frac{n_gpt}{V_G} \Bigg)^{1-m} \Bigg)
\label{eqn:chi_reduced_field}
\end{equation}     
Eq.~\ref{eqn:chi_def} shows that $\chi$ is proportional to the RFTC divided by the reduced field.  The constant of proportionality is a model parameter that represents the ionization potential of the gas.  Here, we use the returned value of $B/A$ listed in Table \ref{tab:analysis_par}.  However, we are not as interested in the value of $\chi$ as we are in its functional dependence on the reduced field.  The reason is that the avalanche variance will decrease to a minimum value as $\chi$ increases to a maximum \cite{Alkhazov:1970fx}.  

\begin{figure}[h!]
  \centering
  \includegraphics[width=\linewidth]{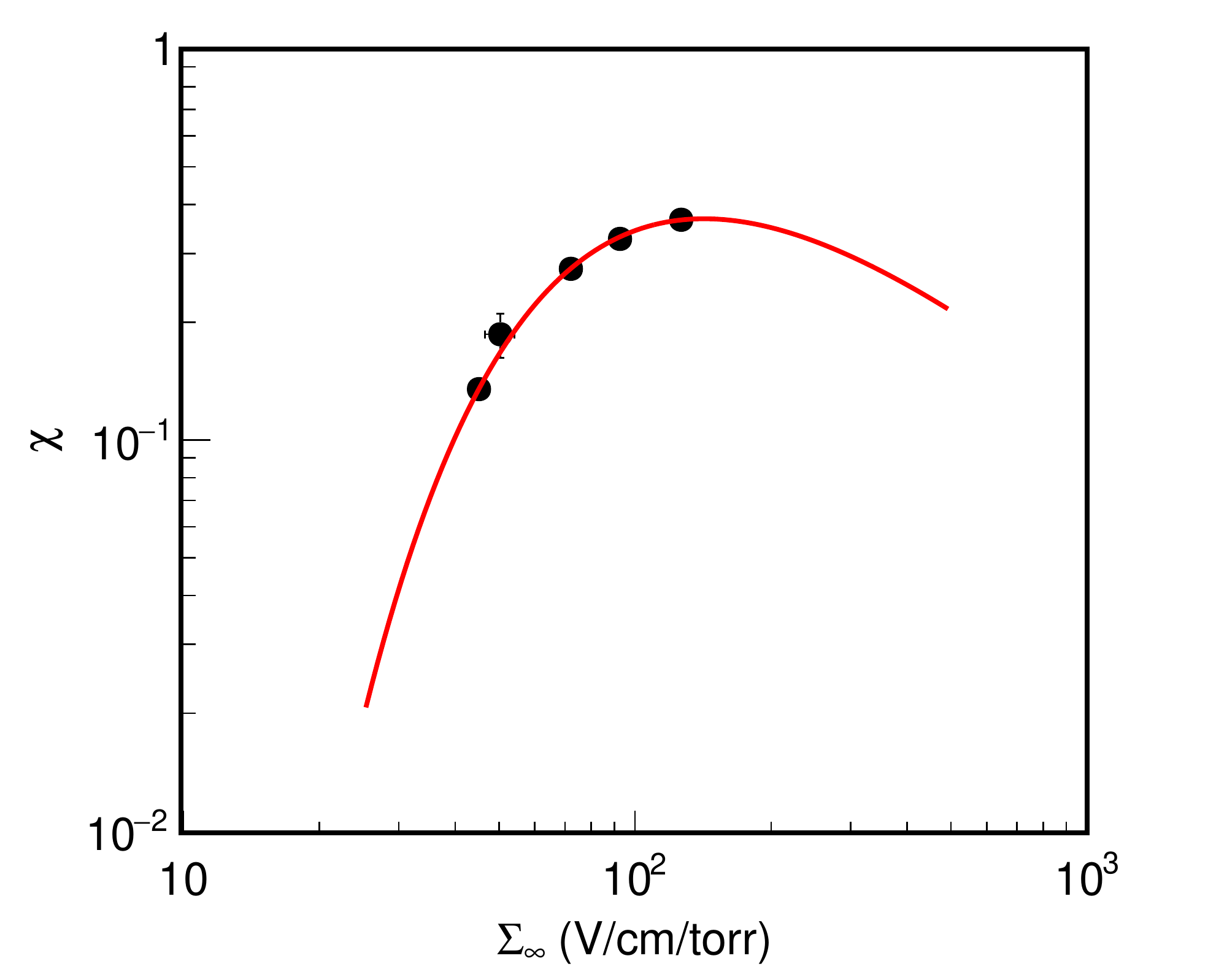}
  \caption{Parameter $\chi$ versus asymptotic reduced field $\Sigma_{\infty}$ for five $^4$He:CO$_2$ (70:30) data sets.  The solid line (red online) is a fit of Eq.~\ref{eqn:chi_reduced_field} to the data.  The minimum value of the avalanche distribution variance should occur at the maximum value of $\chi$.}
  \label{fig:he_chi_field}
\end{figure}

Fig.~\ref{fig:he_chi_field} shows $\chi$ versus asymptotic reduced field for the five He:CO$_2$ data sets.  The solid line is a fit of Eq.~\ref{eqn:chi_reduced_field} to the data points where $A$, $B$, and $m$ are fixed to their values returned from fitting Eq.~\ref{eqn:reduced_gain} with $m=0$ to the asymptotic reduced quantities in the top plot of Fig.~\ref{fig:heco2_reduced_gain}, and only a scale factor is left as a free parameter.  The maximum $\chi$ value occurs at approximately \SI{143}{V/cm/torr} for this gas mixture, indicating that this is the reduced field at which the avalanche variance will reach its minimum value.

More specifically, what is shown in Ref.~\cite{Alkhazov:1970fx} is that for Legler's model, and others, the avalanche variance has a complicated dependence on $\chi$.  As the field strength increases, the avalanche distribution changes from an exponential distribution to something closer to a Polya distribution \cite{1958ZPhy..151..563S}.  Since $\Gamma_{\infty}$ and $\Sigma_{\infty}$ are essentially proportional, we can infer that the avalanche variance will decrease towards a minimum value for some value of $\Gamma_{\infty}$ as well.  As a result, the energy resolution should approach a minimum value with increasing $\Gamma_{\infty}$.

%The avalanche variance decreases towards a minimum as $\chi$ increases towards a maximum and, 
%
%since $\Gamma_{\infty}$ and $\Sigma_{\infty}$ are essentially prportion  so we can use Fig.~\ref{fig:he_chi_field} 
%
%%A Polya distribution is peaked near the mean multiplication factor and has a variance less than unity.  
%
%In addition to decreasing toward a minimum as $\chi$ increases towards a maximum, the avalanche variance changes more at lower values of $\chi$ than it does near its minimum.  
%
%This translates into the following: If the only remaining gain, or reduced field, dependent term that is contributing to the PH resolution is the avalanche variance, then at lower reduced field values the PH resolution will change more than at higher values near the maximum $\chi$ value (minimum avalanche variance value).  
%
%In addition, the PH resolution should reach a minimum value near the same reduced field value where $\chi$ is maximized, and the avalanche variance is minimized.

%A feature of many charge avalanches models within uniform electric fields is that the avalanche variance will decrease as the reduced field increases.  Given the relationship between , we would expect the same behavior from it.

%%
\subsection{Asymptotic resolution versus asymptotic RFTC}\label{sec:heco2_fund_res}

%%
%\subsubsection{The minimum PH resolution as the energy resolution}\label{sec:min_res}

Our discussion about $\chi$ allows us to determine at what reduced field value the minimum PH resolution should occur, but it doesn't tell us anything about its actual value.  We use the asymptotic PH resolution $b$ and the asymptotic RFTC $\Gamma_{\infty}$ to propose a relationship analogous to Eq.~\ref{eqn:gain_res}, the relationship between the PH resolution and the gain:  

%With the detector noise term suppressed at high gain, we are left to conclude that any further decrease in the PH resolution is a result of the avalanche variance decreasing with higher reduced field values.

%To proceed, after applying corrections to the PH resolution from the stability data, we will return to the end of Section~\ref{sec:gain_res_curve} and Eq.~\ref{eqn:red_gain_res} for a final word about the minimum obtainable energy resolution in GEM-based detectors.
%
%With the detector noise term suppressed at high gain, we are left to conclude that any further decrease in the resolution is a result of the avalanche variance decreasing at higher reduced field values.  In Section~\ref{sec:heco2_analysis}, we will show using the He:CO$_2$ data that the asymptotic RFTC is essentially linear with the asymptotic reduced field, analogous to the relationship between the gain and the total GEM voltage, across a large parameter space.  In Section~\ref{sec:heco2_fund_res}, after correcting for the stability, we will also show that the asymptotic resolution has a relationship with the asymptotic RFTC that is analogous to the relationship between the PH resolution and the gain, i.e. Eq.~\ref{eqn:gain_res}.  In anticipation we write

%
\begin{equation}
b \equiv R_{\infty} = \sqrt  { \; \Bigg( \frac {\mathcal{A}}{\Gamma_{\infty}} \Bigg)^2 + \mathcal{B} \; ^2}.
%b \equiv \frac{\sigma_E}{G_{\infty}} = \sqrt  { \; \Bigg( \frac {\mathcal{A}}{\Gamma_{\infty}} \Bigg)^2 + \mathcal{B} \; ^2},
\label{eqn:red_gain_res}
\end{equation}     
$\mathcal{A}$ is a term partially attributed to the avalanche variance being suppressed at higher asymptotic RFTC.  The detector noise term in Eq.~\ref{eqn:gain_res} was observed to be independent of the gain.  However, we know the avalanche variance remains a function of the gain and reduced field, so $\mathcal{A}$ is likely a function $\Gamma_{\infty}$ and the reduced field.  Although the analogy with Eq.~\ref{eqn:gain_res} is not perfect, Eq.~\ref{eqn:red_gain_res} still provides useful insight.  In particular, after applying stability corrections to the PH resolution, $\mathcal{B}$ is the minimum PH resolution.

%%
%\subsubsection{Variance of the avalanche distribution and the Fano factor}\label{sec:fano_factor}

%Our discussion about $\chi$ allows us to determine at what reduced field value the minimum avalanche variance value, and hence the minimum PH resolution, should occur but it doesn't yet tell us anything about its actual value.  To proceed, after applying corrections to the PH resolution from the stability data, we will return to the end of Section~\ref{sec:gain_res_curve} and Eq.~\ref{eqn:red_gain_res} for a final word about the minimum obtainable energy resolution in GEM-based detectors.

%to estimate the minimum energy resolution achievable in GEM-based detectors we first use the stability data to correct the measured PH resolution values.  

As discussed in Section~\ref{sec:stability}, we define a two hour period bounded by the two vertical red lines in Fig.~\ref{fig:stability} that follows the initial stabilization (charge-up) period to better estimate minimum PH resolution values for certain setups, when the gain is stable and the resolution is closest to its minimum.  Note that three of the data sets were recorded near the minimum of stability runs, so the correction procedure is only performed on the 2-tGEM and 1-THGEM (\SI{380}{torr}) data sets.  The correction procedure follows:  

\begin{itemize}

\item{For a given setup, average the PH resolution values obtained from the five minute spectra within the two hour period containing the minimum PH resolution.}

\item{Determine the correction.  For a given setup, the stability run is performed at a single $V_G$, so we take the difference in quadrature between the average resolution obtained in the above step and the resolution measured in Fig. \ref{fig:heco2_gain_res} at the same $V_G$.}

\item{Each setup now has a correction factor.  The correction factor is subtracted in quadrature from all the PH resolution values measured with that setup in Fig. \ref{fig:heco2_gain_res}.  This results in corrected PH resolution values that better represent the minimum value achievable with that setup.}  

\item{Using the corrected PH resolutions values we obtain corrected asymptotic resolution values, denoted $R_{\infty}^*$.}

\end{itemize}

\begin{figure}[h!]
  \centering
  \includegraphics[width=\linewidth]{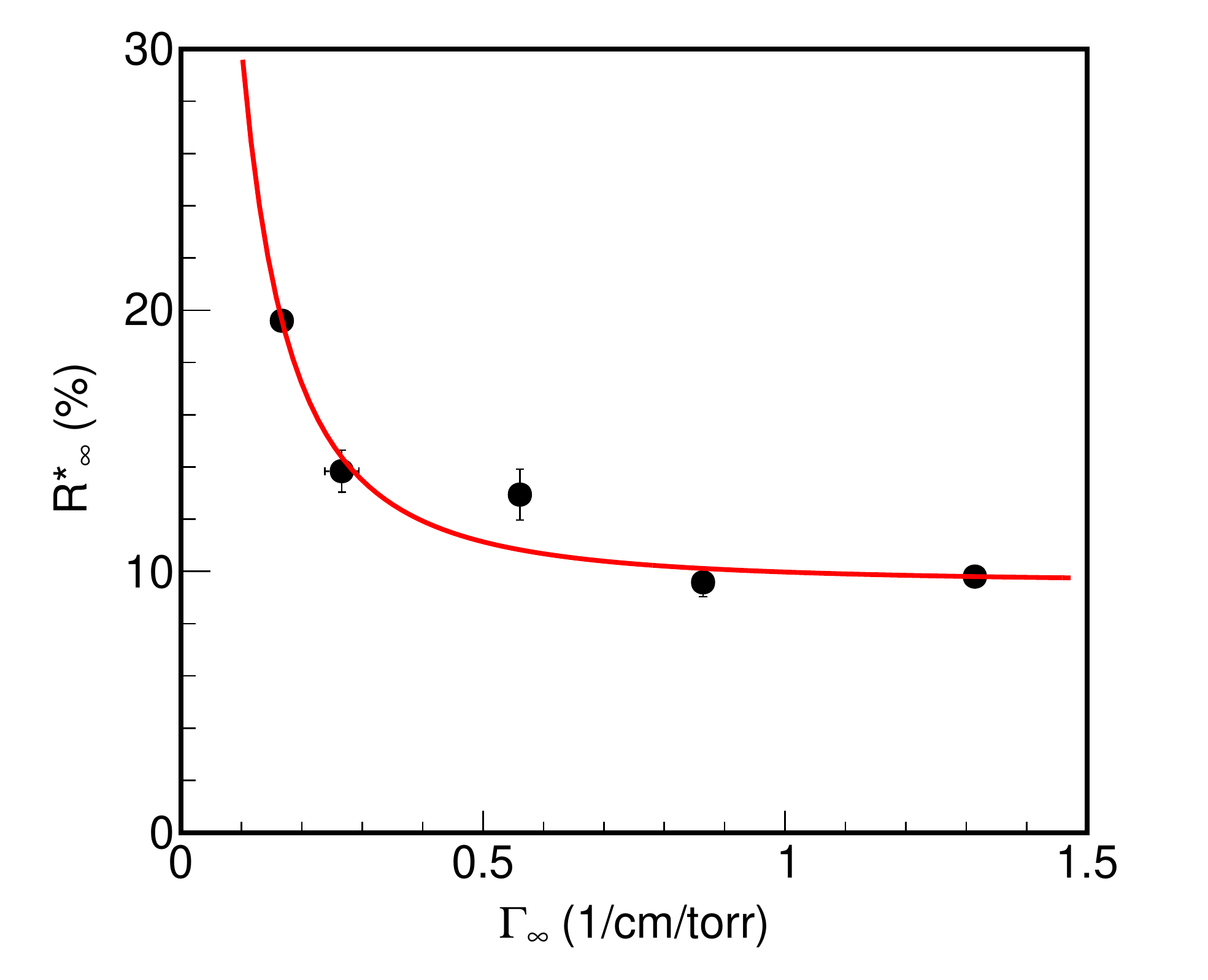}
  \caption{Asymptotic PH resolution versus asymptotic RFTC $\Gamma_{\infty}$ for five $^4$He:CO$_2$ (70:30) data sets.  The solid line (red online) is a fit of Eq.~\ref{eqn:red_gain_res} to the data.  After optimizing for the drift field and transfer efficiencies, the asymptotic resolution approached in this figure can be regarded as the minimum achievable energy resolution at a given ionization energy for a given gas mixture in GEM-based detectors.}
  \label{fig:he_res_red_gain}
\end{figure}
%

%9.55843e+00   9.59537e-02
%2.85412e+00   8.66719e-02

Fig.~\ref{fig:he_res_red_gain} shows the stability corrected asymptotic resolution versus the asymptotic RFTC, which looks as predicted.  The solid line is a result of fitting Eq.~\ref{eqn:red_gain_res} to the data points.  The values obtained for $\mathcal{A}$ and $\mathcal{B}$ are $(2.85 \pm 0.09) \times 10^{-2}$ \SI{}{1/cm/torr} and (9.6 $\pm$ 0.1)\%, respectively.  We note that the resolution for the 2-tGEM setup is slightly greater than for the 3-tGEM setup, and this could be related to the difference in the transfer and collection fields mentioned previously.  These two data sets also used different gas bottles, albeit for the same company with the same specifications.  In Ref.~\cite{Vahsen:2014fba}, a similar double-thin-GEM setup achieves an asymptotic PH resolution of $\approx$ 8.8\% with the same proportion of He:CO$_2$ gas from a different vendor.  While it is not clear this difference is only due to the gas, it is clear the gas quality can have a substantial impact on the gain and energy resolution.  

%In the end, by minimizing the gain and field-dependent effects, the asymptotic resolution is observed to decrease to a minimum value as a function of the asymptotic RFTC.  Fig.~\ref{fig:he_res_red_gain} shows the stability corrected asymptotic PH resolution versus the asymptotic RFTC.  The solid line is a result of fitting Eq.~\ref{eqn:red_gain_res} to the data points.  The values obtained for $\mathcal{A}$ and $\mathcal{B}$ are $(2.5 \pm 0.1) \times 10^{-2}$ \SI{}{1/cm/torr} and (9.98 $\pm$ 0.09)\%, respectively.  

%%%
\subsection{Minimum energy resolution}\label{sec:min_res}

%With the detector, gain, and reduced field-dependent effects minimized, we can be more explicit about the Fano factor and avalanche variance.  We can also consider the dependence of the energy resolution on the initial ionization energy.

Since the detector noise term has been suppressed, only the Fano factor and the avalanche variance remain as contributors to the energy resolution.  If each electron avalanche is assumed to be statistically independent, which is true for most of the proportional region, we can express the charge multiplication within the avalanche in terms of a single-electron avalanche factor, $A$.  This allows us to write the fractional resolution as \cite{Alkhazov:1970fx} \cite{Knoll:2010xta}

\begin{equation}
R_{\infty}^2 = \Bigg( \frac{\sigma_{n_0}}{n_0} \Bigg)^2 + \frac{1}{n_0} \Bigg( \frac{\sigma_A}{\overline{A}} \Bigg)^2,
%\Bigg( \frac{\sigma_G}{G} \Bigg)^2 = \Bigg( \frac{\sigma_{n_0}}{n_0} \Bigg)^2 + \frac{1}{n_0} \Bigg( \frac{\sigma_A}{\overline{A}} \Bigg)^2,
\label{eqn:gain_res_def2}
\end{equation}
where ${\overline{A}}$ and $n_0$ are the average multiplication of a single electron and the number of electrons in the primary ionization cloud, respectively.  

The Fano factor, $F = \sigma_{n_0}^2 / n_0$, \cite{PhysRev.72.26} is usually introduced to account for the fluctuations in the primary ionization.  We denote the fractional avalanche variance as $f = ( \sigma_A / \overline{A} )^2$ and $W = E_i / n_0$, where $E_i$ is the incident particle energy, is known as the W-value, or W-factor.  Equation \ref{eqn:gain_res_def2} can then be written as

\begin{equation}
R_{\infty} ^2 = W\frac{(F + f)}{E_i}.
%\Bigg (\frac{\sigma_G}{G} \Bigg)^2 = W\frac{(F + f)}{E_i},
\label{eqn:gain_res_fano_2}
\end{equation}
For a given incident particle energy, $W$, $F$, and $f$ are the fundamental experimental parameters that govern the achievable lower limit of energy resolution in ionization avalanching detectors.  

$W$ is the average amount of energy required to create an electron-ion pair and is dependent on the energy of the incident particle, but constant above a few keV \cite{Bronic}.  There is also a dependance on the type of incident radiation as heavily ionizing particles experience straggling, and not all of their kinetic energy results in ionization.  For gases, typical values are $\approx$ \SI{30}{eV}.  $F$ quantifies how much of the incident particle's energy went into ionization and it approaches unity as this energy decreases toward the ionization potential of the gas \cite{Bronic}.  $f$ quantifies the variance in the size of the avalanches initiated by single electrons.  Equations \ref{eqn:gain_res_def2} and \ref{eqn:gain_res_fano_2} will be true regardless of the gain or reduced field, but lower gains generally produce avalanches with larger $f$ values.

As discussed previously, the data points in Fig.~\ref{fig:he_res_red_gain} are the energy resolutions of the detectors, and $\mathcal{B}$ is regarded as the \textit{asymptotic} energy resolution value for this gas mixture and ionization energy.  $\mathcal{B}$~$\approx$~9.6\% for \SI{5.9}{keV} and, by doing this with a single energy source, we are controlling for changes of the Fano factor with incident particle energy.  Using Eq.~\ref{eqn:gain_res_fano_2} we obtain $F + f \approx 1.6$. 

While it depends on the exact value of the Fano factor, we can estimate $f \approx$ 1.4 (as a minimum).  Most avalanche distribution models show a lower fractional variance than 1.4, by as much as a factor of two or more, at similar reduced field values.  Ref.~\cite{Schindler:2010los} simulates many gas mixtures and they all show the avalanche variance continuing to decrease beyond what we can interpret here.  Recall from Section~\ref{sec:drift}, the drift field is not optimized to minimize the resolution and slightly better resolutions should be possible at higher drift fields, but not enough to account for what we see here.  This suggests that there may still be some optimization to be done, and correcting for the transfer efficiencies could yield additional improvements.  Ref.~\cite{Guedes:2003eq} shows there is indeed an effect on the energy resolution from the collection field strengths.  A more complete study would include another measurement(s) up to, and beyond the maximum $\chi$ value to see the resolution response, possibly including a single tGEM setup.  Additional source energies would allow one to look for deviations in the expected behavior as a function of energy.  

Finally, we refer back to the corrected 3-tGEM and 2-tGEM asymptotic resolutions (two data points on the right of Fig.~\ref{fig:he_res_red_gain}).  These multi-GEM setups consist of different numbers of GEMs, yet produce nearly the same asymptotic resolutions.  This highlights that the most important quantity for determining the energy resolution is $\Gamma_{\infty}$.  From Fig.~\ref{fig:heco2_reduced_gain}, the most important quantity for determining $\Gamma_{\infty}$ is $\Sigma_{\infty}$.  These two statements can be regarded as the main outcomes of this work with He:CO$_2$.

%%%%
\subsection{Remark about distributing voltage in multi-GEM setups}

In principle, how a given voltage is distributed amongst multiple GEMs should not affect the gain, however observations disagree \cite{Bachmann:1999xc}.  This is presumably due to the transfer efficiencies changing, and could be compensated by appropriately adjusting the voltages.  Thus, $\Gamma_{\infty}$ would increase proportionally with $\Sigma_{\infty}$ regardless of how the voltage is distributed.  Fig.~\ref{fig:he_res_red_gain} explicitly shows that operating at higher RFTC produces a better resolution than lower RFTC.  Given the nearly linear relationship of $\Sigma_{\infty}$ and $\Gamma_{\infty}$, this means operating at a higher average reduced field is better for resolution, but only until the flat part of Fig.~\ref{fig:he_res_red_gain} is reached.  For multiple GEM setups, it is better to have each GEM operating in the flat regime of Fig.~\ref{fig:he_res_red_gain} than not.  So for a given gain, equally distributing the voltage amongst the GEMs will give the highest $\Sigma_{\infty}$, hence highest $\Gamma_{\infty}$ for each GEM.  This will ensure that each GEM is operating nearest its lowest (best) resolution.

%%%
\section{Measurements and results with SF$_6$}\label{sec:sf6}

This section presents the SF$_6$ gain and resolution results.  All measurements are performed with a collimated $^{55}$Fe source (\SI{5.9}{keV}) in room temperature gas with 1-THGEM.

\subsection{Using pulse-heights to measure gain in SF$_6$}\label{sec:sf6_PH}

Compared to electrons, NIs are more massive and therefore have lower drift velocities.  The slow arrival of the ionization cloud can cause some of the charge to go undetected, resulting in a larger variance in the measured gain.  In the setup used here, we estimate less than $\approx$~10\% of the charge in any single event is lost due to this effect.

%A Cremat CR-110 charge sensitive preamplifier, with a decay time of $\SI{140}{\mu s}$, was used for the SF$_6$ gain measurements.  This means that after $\SI{140}{\mu s}$, $\approx$ 37\% of the output voltage signal has decayed.  Using the existing simulation framework used for the work in Ref.~\cite{Vahsen:2020pzb}, we simulated one-thousand \SI{5.9}{keV} electrons in SF$_6$ at a pressure of \SI{40}{torr} drifting over a \SI{5}{cm} drift length.  In SF$_6$ at a pressure of \SI{40}{torr}, with a drift field of \SI{500}{V/cm}, the drift velocity is $\approx$ $\SI{0.05}{mm/ \mu s}$ \cite{Phan:2016veo}.  If we assume a uniform distribution of the ionization in time then, for a \SI{4}{mm} event, the signal will arrive steadily over $\approx$ \SI{80}{\text{$\mu$}s}.  This means that by the time the end of the charge signal is collected by the preamplifier, the beginning of the output voltage signal has decayed by $\approx$ 57\% of 37\%, or $\approx$ 21\%.  Every part of the signal arriving later will be affected less until the last part of the signal is not affected at all.  This also only happens for the longest events in $z$, of which there are few.  We then estimate that the signals for the longest events in $z$ are $\approx$ 90\% collected (half of 21\% is not collected).  Thus the SF$_6$ results presented here will represent a lower bound on the gain (by a maximum of $\approx$ 10\%, i.e. for the longest events).}

A Cremat CR-110 charge-sensitive preamplifier, with a decay time constant of $\SI{140}{\mu s}$, was used for the SF$_6$ gain measurements.  Using the simulation framework of Ref.~\cite{Vahsen:2020pzb}, we simulated one-thousand \SI{5.9}{keV}~electrons in SF$_6$ at a gas pressure of \SI{40}{torr} drifting over a \SI{5}{cm}~drift length with a drift field of \SI{500}{V/cm}, which corresponds to a drift velocity of $\approx$ $\SI{0.05}{mm/ \mu s}$ \cite{Phan:2016veo}.  For an electron recoil track (charge cluster) measuring \SI{4}{mm} along the drift direction, and accounting for the output voltage signal decay, we estimate that $\approx$~90\% of the charge will be integrated by the preamplifier.  The simulated results show that $\approx$~95\% of the electron recoil tracks are shorter than \SI{4}{mm} along the drift direction.  Therefore, we claim the SF$_6$ results presented here represent a lower bound on the gain by a maximum of 10\% (for the longest events).

%%%%%%%%%%%%%%%%%%%%%%%. SF6 table
\begin{table*}[ht]
\centering
 %\begin{tabular}{l | l | c | >{\centering\arraybackslash}p{1.8cm} | >{\centering\arraybackslash}p{2.0cm} | >{\centering\arraybackslash}p{2.6cm} r}
  %\begin{tabular}{ >{\centering\arraybackslash}p{1.2cm} |  >{\centering\arraybackslash}p{1.8cm}  |  >{\centering\arraybackslash}p{1.8cm}  |  >{\centering\arraybackslash}p{2.2cm}   | c | c | c | c  r}
  \begin{tabular}{ >{\centering\arraybackslash}p{1.2cm}  >{\centering\arraybackslash}p{1.8cm}   >{\centering\arraybackslash}p{1.8cm}   >{\centering\arraybackslash}p{2.2cm}  cccc}
\toprule 
  Pressure (torr) & Drift field (\SI{}{V/cm}) & GEM field (\SI{}{kV/cm}) & Collection field (\SI{}{V/cm}) & $V_1$ (\SI{}{V}) & $V_2$ (\SI{}{V}) & $G_{\infty} = 100 \times a$ & $b$ (\SI{}{\%}) \\
  \hline
\noalign{\vskip 2mm}
   %20 & 283 & 19.1 & 1558 & $(4.0 \pm 0.2) \times 10^2$ & $108 \pm 6$ & $n/a$ & $34.5 \pm 0.6$ \rule{0pt}{3ex} \\ %\hline
   40 & 499 & 22.6 & 1842 & $(6.2 \pm 0.2) \times 10^2$ & $84 \pm 6$ & $(3.4 \pm 0.7) \times 10^4$ & $44 \pm 5$  \\ %hline
\bottomrule
   \end{tabular}
   
   \setlength{\abovecaptionskip}{6pt}
  \caption{Detector settings and fit results for $^{55}$Fe measurements with pure SF$_6$ at \SI{40}{torr} using 1-THGEM.  The GEM and collection field are proportional to the total GEM voltage, $V_G$.  The values shown are for the highest $V_G$.  The drift field is held constant.}
  \setlength{\belowcaptionskip}{0pt}
\label{tab:sf6}
\end{table*}
%%%%%%%%%%%%%%%%%%%%%%%%%%.  SF6 table
%%%%%%%%%%%%%%%%%%%%%%%%%%%%%%%%%%%

\subsection{SF$_6$ measurements with 1-THGEM} \label{sec:sf6_meas}

To initiate an avalanche, the electrons must first be stripped from the NIs.  This requires a large reduced field, which is easier to produce at low gas pressures.  THGEMs have demonstrated robust performance at low gas pressures \cite{Shalem:2005ix}, and the 1-THGEM setup was constructed to pursue NI gas gain measurements at low pressure.  These are our first results with SF$_6$.  

\begin{figure}[h!]
  \centering
  \includegraphics[width=\linewidth]{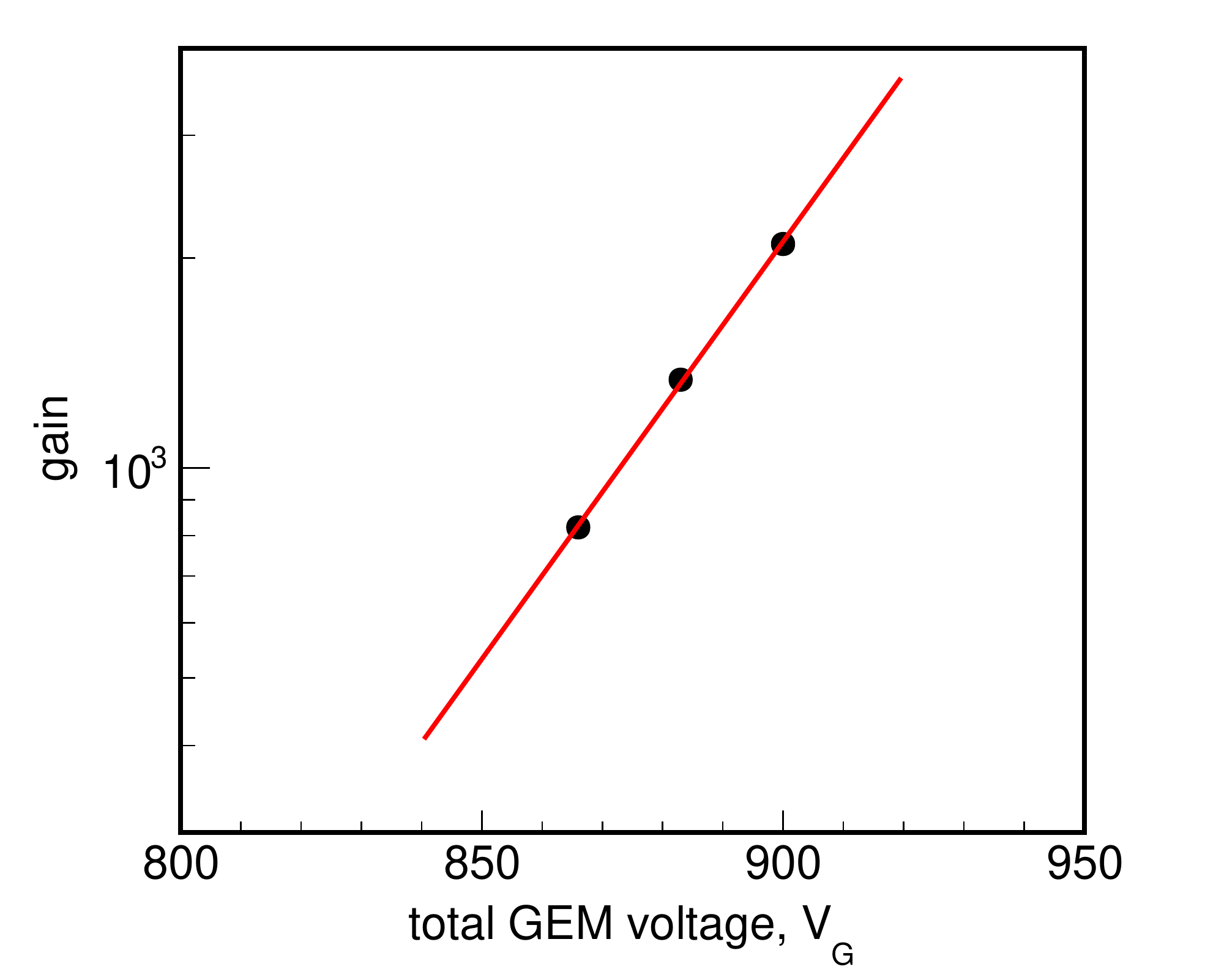}
  \includegraphics[width=\linewidth]{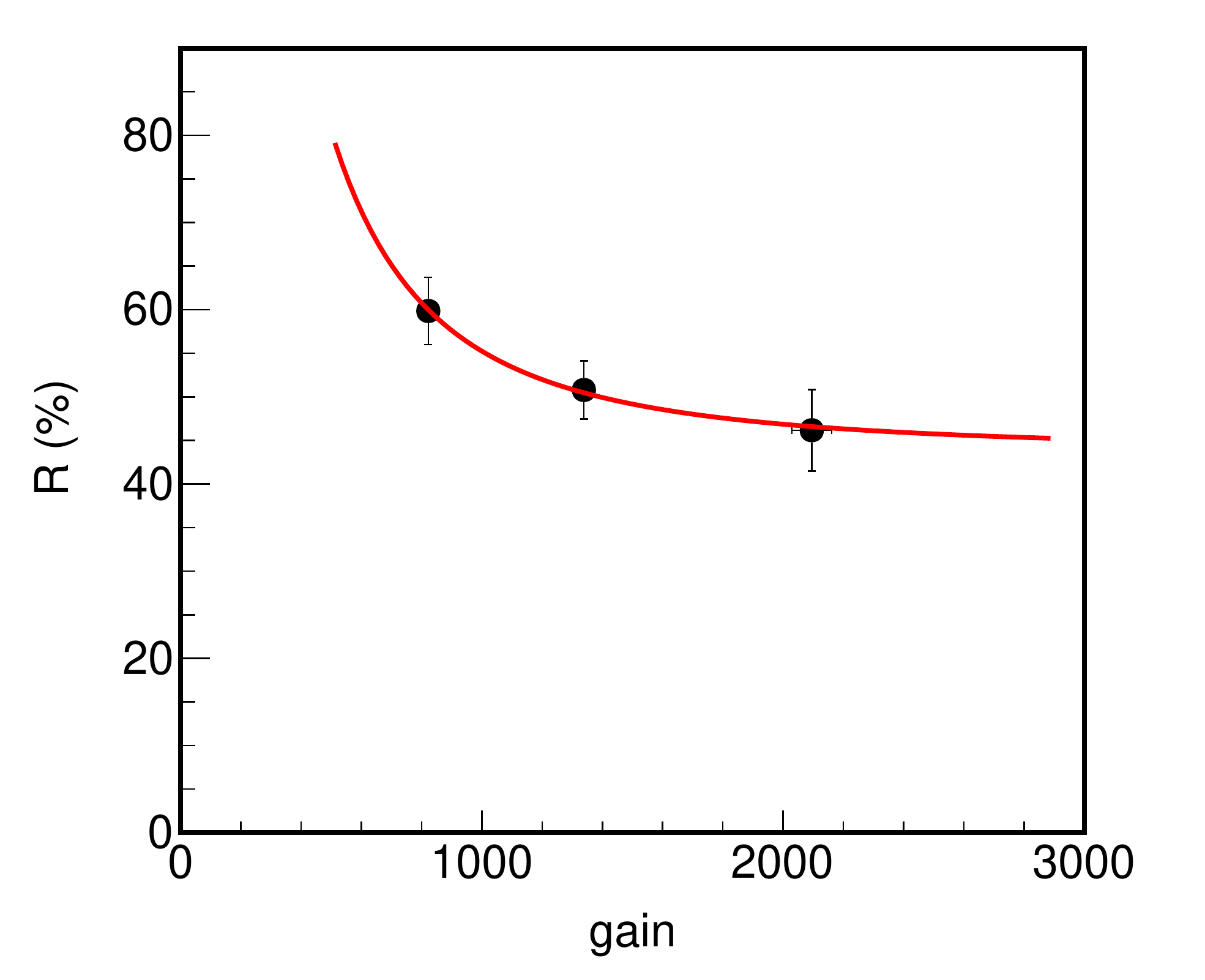}
  \caption{Results of $^{55}$Fe measurements with pure SF$_6$ at \SI{40}{torr} using 1-THGEM.  Top:  Gain versus total GEM voltage. Bottom:  PH resolution versus gain.  The solid lines (red online) are fits of Eqs. \ref{eqn:gain_voltage} (top) and \ref{eqn:gain_res} (bottom) to the data.}
  \label{fig:sf6_gain_res}
\end{figure}
%

%The resulting track lengths along $z$ forms a Gaussian distribution with $\mu \approx$ \SI{1.9}{mm} and $\sigma \approx$ \SI{0.7}{mm}.  

%This means that the extent in $z$, of nearly all events, is $<$ \SI{4}{mm}, and 95\% of them extend $<$ \SI{3.3}{mm}.  

%This respectively corresponds to $\approx$ 21\% and $\approx$ 17\% of the event charge not being collected.  

%Thus the following results represent a lower bound on the gain (by a maximum of $\approx$ 21\%).

%This means that after $\SI{140}{\mu s}$, $\approx$ 37\% of the collected charge has decayed away.  

%This means that $\approx$ 26\% of the charge from that event will have decayed away by the time the full event is collected.  

%In SF$_6$ at a pressure of \SI{40}{torr}, with a drift field of \SI{500}{V/cm}, the drift velocity is $\approx$ $\SI{0.05}{mm/ \mu s}$ \cite{Phan:2016veo}.  

%A track extending $\approx$ $\SI{5}{mm}$ in $z$ will arrive over $\approx$ \SI{100}{\mu s}.

%%
%\subsubsection{SF$_6$ gain and resolution measurements}

The lower right panel in Fig.~\ref{fig:spectra} shows a PH spectrum obtained with SF$_6$ at a pressure of \SI{40}{torr} with a measured gain and resolution of $\approx$~2000 and 47\%, respectively.  The effect of material outgassing is drastic at low pressures, resulting in substantial degradation of the gas gain over time.  Gain degradation up to one order of magnitude within a five minute spectrum was observed.  To counteract the degradation, constant gas flow was maintained at the maximum rate the system would allow, approximately \SI{200}{ml/min} at a gas pressure of \SI{40}{torr}.  This reduced the gain degradation to a few percent per hour.  However, measurements at 20 and \SI{30}{torr} were more affected and, although we do not present them here, similar gain values were obtained.

Fig.~\ref{fig:sf6_gain_res} shows the gain versus GEM voltage (top) and PH resolution versus gain (bottom).  The gain is similar to that of an electron drift gas such as He:CO$_2$, but the PH resolution is substantially worse for SF$_6$ despite an $\mathcal{O}(10)$ larger reduced field.  The poor resolution is at least partially due to recombination occurring in the collection region.  For the same reason we claim the SF$_6$ gain measurements are lower bounds, the PH resolution measurements are taken as upper bounds.  Analysis of the full SF$_6$ waveforms would likely yield improvements to the PH resolution measurements and provide further insight.

Gain measurements with SF$_6$ were attempted with the other setups listed in Table \ref{tab:dcube_dimensions}; however, only THGEMs produced measurable gain.  As discussed in Ref.~\cite{Phan:2016veo}, if the mean-free-path for detachment is large enough, then THGEMs may be required to produce measurable gains.  A larger detachment mean-free-path will also cause lower gain for the same reduced field compared with other gases (electron-drift), which is observed.

\section{Summary and final remarks}

%%%%%%%%%%%%%%%%%%%%%%%%%%%%%%%%%%%

GEMs are becoming increasingly popular in experimental particle physics.  A comprehensive understanding of their gain and how it affects energy resolution is needed when designing GEM-based detectors.  We presented the combined results of extensive gas detector optimization studies for directional nuclear recoil detection, including our first results with SF$_6$, a Negative Ion (NI) gas.  %SF$_6$ is particularly interesting for the directional dark matter community which seeks to lower diffusion in the drift region, thus enhancing a detector's ability to record individual nuclear recoil ionization tracks and extract a definitive directional dark matter signature.  

%However regardless of the exact application, the results shown here provide a good overview of the gain, its effect on the energy resolution, and the contributing experimental parameters when considering any GEM-based detector.

Using a large data set obtained with a $^4$He:CO$_2$ (70:30) gas mixture and multiple experimental setups, we showed that the gas (avalanche) gain and energy resolution from all setups can be described together with a simple formalism.  We showed the (nearly) linear relationship between the asymptotic, reduced, first Townsend coefficient ($\Gamma_{\infty}$) and the asymptotic, reduced, electric field averaged over GEMs ($\Sigma_{\infty}$) is robust over a large parameter space.  We also showed that $\Gamma_{\infty}$ is the most important quantity for determining the minimum energy resolution in GEM-based detectors.  One can expect to measure an energy resolution ($\sigma{_E}/E$) of roughly 9.5\% with an optimized GEM-based detector using a $^4$He:CO$_2$ (70:30) gas mixture at an ionization energy of \SI{5.9}{keV}.  For many electron drift gas mixtures, multi-GEM (two or three) assemblies offer the best performance and flexibility.  A three-GEM assembly requires operation at a higher total voltage than a two-GEM assembly, but achieves higher gains.  Both assemblies produce gains of $\mathcal{O}$(10$^4$) and energy resolutions near the minimum achievable with this technology.

There are many factors that affect the gain and energy resolution, including GEM mechanical parameters, electric field ratios, transfer efficiencies, and gas purity.  Some of these effects can be corrected for, and many have secondary effects on the gain and energy resolution.  The methods presented in this work offer a simple tool for interpreting, predicting, and optimizing the gain and energy resolution of GEM-based detectors.

\section*{Acknowledgements}

We thank the machine shop at the University of Hawaii, Department of Physics \& Astronomy for their help with manufacturing detector components.  We thank Majd Ghrear for his useful feedback on the manuscript.  We acknowledge support from the U.S. Department of Homeland Security under Award Number 2011-DN-077-ARI050-03 and the U.S. Department of Energy under Award Numbers DE-SC0007852, DE-SC0010504, and DE-SC0018034.

\section{References}

\bibliography{/Users/tthorpe/Documents/papers/papers_in_progress/submitted/gain_paper/sources.bib}{}

\providecommand{\href}[2]{#2}\begingroup\raggedright\begin{thebibliography}{10}

\bibitem{Sauli:1997qp}
F.~Sauli, \emph{{GEM: A new concept for electron amplification in gas
  detectors}}, \href{https://doi.org/10.1016/S0168-9002(96)01172-2}{\emph{Nucl.
  Instrum. Meth. A} {\bfseries 386} (1997) 531}.

\bibitem{Nygren:1978rx}
J.~N. Marx and D.~R. Nygren, \emph{{The Time Projection Chamber}},
  \href{https://doi.org/10.1063/1.2994775}{\emph{Phys. Today} {\bfseries 31N10}
  (1978) 46}.

\bibitem{ALICETPC:2020ann}
{\scshape ALICE TPC} collaboration, \emph{{The upgrade of the ALICE TPC with
  GEMs and continuous readout}},
  \href{https://doi.org/10.1088/1748-0221/16/03/P03022}{\emph{JINST} {\bfseries
  16} (2021) P03022} [\href{https://arxiv.org/abs/2012.09518}{{\ttfamily
  2012.09518}}].

\bibitem{Colaleo:2015vsq}
M.~A. Akl et~al., \emph{{CMS Technical Design Report for the Muon Endcap GEM
  Upgrade}}, .

\bibitem{Vahsen:2014fba}
S.~Vahsen, M.~Hedges, I.~Jaegle, S.~Ross, I.~Seong, T.~Thorpe et~al.,
  \emph{{3-D tracking in a miniature time projection chamber}},
  \href{https://doi.org/10.1016/j.nima.2015.03.009}{\emph{Nucl. Instrum. Meth.
  A} {\bfseries 788} (2015) 95}
  [\href{https://arxiv.org/abs/1407.7013}{{\ttfamily 1407.7013}}].

\bibitem{Jaegle:2019jpx}
I.~Jaegle et~al., \emph{{Compact, directional neutron detectors capable of
  high-resolution nuclear recoil imaging}},
  \href{https://doi.org/10.1016/j.nima.2019.06.037}{\emph{Nucl. Instrum. Meth.
  A} {\bfseries 945} (2019) 162296}
  [\href{https://arxiv.org/abs/1901.06657}{{\ttfamily 1901.06657}}].

\bibitem{Lewis:2018ayu}
P.~Lewis et~al., \emph{{First Measurements of Beam Backgrounds at SuperKEKB}},
  \href{https://doi.org/10.1016/j.nima.2018.05.071}{\emph{Nucl. Instrum. Meth.
  A} {\bfseries 914} (2019) 69}
  [\href{https://arxiv.org/abs/1802.01366}{{\ttfamily 1802.01366}}].

\bibitem{Vahsen:2011qx}
S.~Vahsen, H.~Feng, M.~Garcia-Sciveres, I.~Jaegle, J.~Kadyk et~al., \emph{{The
  Directional Dark Matter Detector ($D^{3}$)}},
  \href{https://doi.org/10.1051/eas/1253006}{\emph{EASPUB} {\bfseries 53}
  (2012) 43} [\href{https://arxiv.org/abs/1110.3401}{{\ttfamily 1110.3401}}].

\bibitem{Kim:2008zzi}
T.~Kim, M.~Freytsis, J.~Button-Shafer, J.~Kadyk, S.~E. Vahsen and W.~A. Wenzel,
  \emph{{Readout of TPC tracking chambers with GEMs and pixel chip}},
  \href{https://doi.org/10.1016/j.nima.2008.02.049}{\emph{Nucl. Instrum. Meth.
  A} {\bfseries 589} (2008) 173}.

\bibitem{Vahsen:2014mca}
S.~Vahsen, K.~Oliver-Mallory, M.~Lopez-Thibodeaux, J.~Kadyk and
  M.~Garcia-Sciveres, \emph{{Tests of gases in a mini-TPC with pixel chip
  readout}}, \href{https://doi.org/10.1016/j.nima.2013.10.029}{\emph{Nucl.
  Instrum. Meth. A} {\bfseries 738} (2014) 111}.

\bibitem{Lewis:2014poa}
P.~Lewis, S.~Vahsen, I.~Seong, M.~Hedges, I.~Jaegle and T.~Thorpe,
  \emph{{Absolute Position Measurement in a Gas Time Projection Chamber via
  Transverse Diffusion of Drift Charge}},
  \href{https://doi.org/10.1016/j.nima.2015.03.024}{\emph{Nucl. Instrum. Meth.
  A} {\bfseries 789} (2015) 81}
  [\href{https://arxiv.org/abs/1410.1131}{{\ttfamily 1410.1131}}].

\bibitem{Vahsen:2020pzb}
S.~E. Vahsen et~al., \emph{{CYGNUS: Feasibility of a nuclear recoil observatory
  with directional sensitivity to dark matter and neutrinos}},
  \href{https://arxiv.org/abs/2008.12587}{{\ttfamily 2008.12587}}.

\bibitem{Vahsen:2021gnb}
S.~E. Vahsen, C.~A.~J. O'Hare and D.~Loomba, \emph{{Directional recoil
  detection}},
  \href{https://doi.org/10.1146/annurev-nucl-020821-035016}{\emph{Ann. Rev.
  Nucl. Part. Sci.} {\bfseries 71} (2021) 189}
  [\href{https://arxiv.org/abs/2102.04596}{{\ttfamily 2102.04596}}].

\bibitem{Mayet:2016zxu}
F.~Mayet et~al., \emph{{A review of the discovery reach of directional Dark
  Matter detection}},
  \href{https://doi.org/10.1016/j.physrep.2016.02.007}{\emph{Phys. Rept.}
  {\bfseries 627} (2016) 1} [\href{https://arxiv.org/abs/1602.03781}{{\ttfamily
  1602.03781}}].

\bibitem{Battat:2016pap}
J.~B.~R. Battat et~al., \emph{{Readout technologies for directional WIMP Dark
  Matter detection}},
  \href{https://doi.org/10.1016/j.physrep.2016.10.001}{\emph{Phys. Rept.}
  {\bfseries 662} (2016) 1} [\href{https://arxiv.org/abs/1610.02396}{{\ttfamily
  1610.02396}}].

\bibitem{Morgan:2004ys}
B.~Morgan, A.~M. Green and N.~J. Spooner, \emph{{Directional statistics for
  WIMP direct detection}},
  \href{https://doi.org/10.1103/PhysRevD.71.103507}{\emph{Phys. Rev. D}
  {\bfseries 71} (2005) 103507}
  [\href{https://arxiv.org/abs/astro-ph/0408047}{{\ttfamily
  astro-ph/0408047}}].

\bibitem{Spergel:1987kx}
D.~N. Spergel, \emph{{The Motion of the Earth and the Detection of Wimps}},
  \href{https://doi.org/10.1103/PhysRevD.37.1353}{\emph{Phys. Rev. D}
  {\bfseries 37} (1988) 1353}.

\bibitem{Phan:2016veo}
N.~Phan, R.~Lafler, R.~Lauer, E.~Lee, D.~Loomba, J.~Matthews et~al., \emph{{The
  novel properties of SF$_6$ for directional dark matter experiments}},
  \href{https://doi.org/10.1088/1748-0221/12/02/P02012}{\emph{JINST} {\bfseries
  12} (2017) P02012} [\href{https://arxiv.org/abs/1609.05249}{{\ttfamily
  1609.05249}}].

\bibitem{garfield}
R.~Veenhof, ``Garfield++ webpage at \uppercase{CERN}.''
  \url{https://garfieldpp.web.cern.ch/garfieldpp/}.

\bibitem{Martoff:2000wi}
C.~Martoff, D.~Snowden-Ifft, T.~Ohnuki, N.~Spooner and M.~Lehner,
  \emph{{Suppressing drift chamber diffusion without magnetic field}},
  \href{https://doi.org/10.1016/S0168-9002(99)00955-9}{\emph{Nucl. Instrum.
  Meth. A} {\bfseries 440} (2000) 355}.

\bibitem{SnowdenIfft:1999hz}
D.~Snowden-Ifft, C.~Martoff and J.~Burwell, \emph{{Low pressure negative ion
  drift chamber for dark matter search}},
  \href{https://doi.org/10.1103/PhysRevD.61.101301}{\emph{Phys. Rev. D}
  {\bfseries 61} (2000) 101301}
  [\href{https://arxiv.org/abs/astro-ph/9904064}{{\ttfamily
  astro-ph/9904064}}].

\bibitem{Snowden-Ifft:2014taa}
D.~P. Snowden-Ifft, \emph{{Discovery of multiple, ionization-created CS2 anions
  and a new mode of operation for drift chambers}},
  \href{https://doi.org/10.1063/1.4861908}{\emph{Rev. Sci. Instrum.} {\bfseries
  85} (2014) 013303}.

\bibitem{Battat:2014van}
{\scshape DRIFT} collaboration, \emph{{First background-free limit from a
  directional dark matter experiment: results from a fully fiducialised DRIFT
  detector}}, \href{https://doi.org/10.1016/j.dark.2015.06.001}{\emph{Phys.
  Dark Univ.} {\bfseries 9-10} (2015) 1}
  [\href{https://arxiv.org/abs/1410.7821}{{\ttfamily 1410.7821}}].

\bibitem{Phan:2016zvy}
N.~S. Phan, \emph{{Extending the Reach of Directional Dark Matter Experiments
  Through Novel Detector Technologies}}, Ph.D. thesis, New Mexico U., 2016.

\bibitem{Baracchini:2017ysg}
E.~Baracchini, G.~Cavoto, G.~Mazzitelli, F.~Murtas, F.~Renga and S.~Tomassini,
  \emph{{Negative Ion Time Projection Chamber operation with SF$_6$ at nearly
  atmospheric pressure}},
  \href{https://doi.org/10.1088/1748-0221/13/04/P04022}{\emph{JINST} {\bfseries
  13} (2018) P04022} [\href{https://arxiv.org/abs/1710.01994}{{\ttfamily
  1710.01994}}].

\bibitem{Ishiura:2019ebd}
H.~Ishiura, R.~Veenhof, K.~Miuchi and I.~Tomonori, \emph{{MPGD simulation in
  negative-ion gas for direction-sensitive dark matter searches}},
  \href{https://doi.org/10.1088/1742-6596/1498/1/012018}{\emph{J. Phys. Conf.
  Ser.} {\bfseries 1498} (2020) 012018}
  [\href{https://arxiv.org/abs/1907.12729}{{\ttfamily 1907.12729}}].

\bibitem{Ikeda:2020pex}
T.~Ikeda, T.~Shimada, H.~Ishiura, K.~D. Nakamura, T.~Nakamura and K.~Miuchi,
  \emph{{Development of a negative ion micro TPC detector with SF$_6$ gas for
  the directional dark matter search}},
  \href{https://doi.org/10.1088/1748-0221/15/07/P07015}{\emph{JINST} {\bfseries
  15} (2020) P07015} [\href{https://arxiv.org/abs/2004.09706}{{\ttfamily
  2004.09706}}].

\bibitem{Thorpe:2018irh}
T.~N. Thorpe, \emph{{ \href{http://hdl.handle.net/10125/62685}{Gain resolution
  studies and first dark matter search with novel 3D nuclear recoil
  detectors}}}, Ph.D. thesis, Hawaii U., 2018.

\bibitem{Chechik:2004wq}
R.~Chechik, A.~Breskin, C.~Shalem and D.~Mormann, \emph{{Thick GEM-like hole
  multipliers: Properties and possible applications}},
  \href{https://doi.org/10.1016/j.nima.2004.07.138}{\emph{Nucl. Instrum. Meth.
  A} {\bfseries 535} (2004) 303}
  [\href{https://arxiv.org/abs/physics/0404119}{{\ttfamily physics/0404119}}].

\bibitem{AOYAMA1985125}
T.~Aoyama, \emph{Generalized gas gain formula for proportional counters},
  \href{https://doi.org/https://doi.org/10.1016/0168-9002(85)90817-4}{\emph{Nuclear
  Instruments and Methods in Physics Research Section A: Accelerators,
  Spectrometers, Detectors and Associated Equipment} {\bfseries 234} (1985) 125
  }.

\bibitem{osti_4345702}
W.~Diethorn, \emph{A methane proportional counter system for natural
  radiocarbon measurements (thesis)}, .

\bibitem{WILLIAMS1962229}
A.~Williams and R.~Sara, \emph{Parameters affecting the resolution of a
  proportional counter},
  \href{https://doi.org/https://doi.org/10.1016/0020-708X(62)90115-1}{\emph{The
  International Journal of Applied Radiation and Isotopes} {\bfseries 13}
  (1962) 229 }.

\bibitem{Zastawny_1966}
A.~Zastawny, \emph{Gas amplification in a proportional counter with carbon
  dioxide}, \href{https://doi.org/10.1088/0950-7671/43/3/314}{\emph{Journal of
  Scientific Instruments} {\bfseries 43} (1966) 179}.

\bibitem{Bachmann:1999xc}
S.~Bachmann, A.~Bressan, L.~Ropelewski, F.~Sauli, A.~Sharma and D.~Mormann,
  \emph{{Charge amplification and transfer processes in the gas electron
  multiplier}},
  \href{https://doi.org/10.1016/S0168-9002(99)00820-7}{\emph{Nucl. Instrum.
  Meth. A} {\bfseries 438} (1999) 376}.

\bibitem{Shalem:2006iw}
C.~Shalem, R.~Chechik, A.~Breskin and K.~Michaeli, \emph{{Advances in thick
  GEM-like gaseous electron multipliers. Part I: Atmospheric pressure
  operation}}, \href{https://doi.org/10.1016/j.nima.2005.12.241}{\emph{Nucl.
  Instrum. Meth. A} {\bfseries 558} (2006) 475}
  [\href{https://arxiv.org/abs/physics/0601115}{{\ttfamily physics/0601115}}].

\bibitem{Hallermann:2010zz}
L.~Hallermann, \emph{{Analysis of GEM properties and development of a GEM
  support structure for the ILD Time Projection Chamber}}, Ph.D. thesis,
  Hamburg U., Hamburg, 2010.
\newblock 10.3204/DESY-THESIS-2010-015.

\bibitem{Lewis:2021mgp}
P.~M. Lewis, M.~T. Hedges, I.~Jaegle, J.~Schueler, T.~N. Thorpe and S.~E.
  Vahsen, \emph{{Primary track recovery in high-definition gas time projection
  chambers}}, \href{https://doi.org/10.1140/epjc/s10052-022-10283-3}{\emph{Eur.
  Phys. J. C} {\bfseries 82} (2022) 324}
  [\href{https://arxiv.org/abs/2106.15829}{{\ttfamily 2106.15829}}].

\bibitem{Correia:2014vla}
{\scshape CERN RD-51} collaboration, \emph{{A dynamic method for charging-up
  calculations: the case of GEM}},
  \href{https://doi.org/10.1088/1748-0221/9/07/P07025}{\emph{JINST} {\bfseries
  9} (2014) P07025} [\href{https://arxiv.org/abs/1401.4009}{{\ttfamily
  1401.4009}}].

\bibitem{Alexeev:2015kda}
M.~Alexeev et~al., \emph{{The gain in Thick GEM multipliers and its
  time-evolution}},
  \href{https://doi.org/10.1088/1748-0221/10/03/P03026}{\emph{JINST} {\bfseries
  10} (2015) P03026}.

\bibitem{Brucken:2021tze}
E.~Br\"ucken, J.~Heino, T.~Hild\'en, M.~Kalliokoski, V.~Litichevskyi,
  R.~Turpeinen et~al., \emph{{Hole Misalignment and Gain Performance of Gaseous
  Electron Multipliers}},
  \href{https://doi.org/10.1016/j.nima.2021.165271}{\emph{Nucl. Instrum. Meth.
  A} {\bfseries 1002} (2021) 165271}
  [\href{https://arxiv.org/abs/2103.07944}{{\ttfamily 2103.07944}}].

\bibitem{Hilden:2018isz}
T.~E. Hilden, J.~E. Brucken, D.~Varga and M.~Vargyas, \emph{{GEM foil gain
  prediction}}, \href{https://doi.org/10.22323/1.322.0010}{\emph{PoS}
  {\bfseries MPGD2017} (2019) 010}.

\bibitem{Sharma:1998xw}
A.~Sharma, \emph{{Properties of some gas mixtures used in tracking detectors}},
  .

\bibitem{Brook_1978}
E.~Brook, M.~F.~A. Harrison and A.~C.~H. Smith, \emph{Measurements of the
  electron impact ionisation cross sections of he, c, o and n atoms},
  \href{https://doi.org/10.1088/0022-3700/11/17/021}{\emph{Journal of Physics
  B: Atomic and Molecular Physics} {\bfseries 11} (1978) 3115}.

\bibitem{Raether}
H.~Raether, \emph{Electron avalanches and breakdown in gases}. Butterworths,
  01, 1964.

\bibitem{Engel}
A.~Engel, \emph{Ionized Gases}. AIP-Press, 01, 1994.

\bibitem{Alkhazov:1970fx}
G.~Alkhazov, \emph{{Statistics of electron avalanches and ultimate resolution
  of proportional counters}},
  \href{https://doi.org/10.1016/0029-554X(70)90818-9}{\emph{Nucl. Instrum.
  Meth.} {\bfseries 89} (1970) 155}.

\bibitem{1958ZPhy..151..563S}
H.~{Schlumbohm}, \emph{{Zur Statistik der Elektronenlawinen im ebenen Feld.
  III}}, \href{https://doi.org/10.1007/BF01338427}{\emph{Zeitschrift fur
  Physik} {\bfseries 151} (1958) 563}.

\bibitem{Knoll:2010xta}
G.~Knoll, \emph{{Radiation Detection and Measurement (4th ed.)}}. John Wiley,
  Hoboken, NJ, 2010.

\bibitem{PhysRev.72.26}
U.~Fano, \emph{Ionization yield of radiations. ii. the fluctuations of the
  number of ions}, \href{https://doi.org/10.1103/PhysRev.72.26}{\emph{Phys.
  Rev.} {\bfseries 72} (1947) 26}.

\bibitem{Bronic}
K.~{Bronić}, \emph{{W values and Fano factors for electrons in rare gases and
  rare gas mixtures.}}, {\emph{Ionizing Radiation (Hoshasen)} (1998) }.

\bibitem{Schindler:2010los}
H.~Schindler, S.~F. Biagi and R.~Veenhof, \emph{{Calculation of gas gain
  fluctuations in uniform fields}},
  \href{https://doi.org/10.1016/j.nima.2010.09.072}{\emph{Nucl. Instrum. Meth.
  A} {\bfseries 624} (2010) 78}.

\bibitem{Guedes:2003eq}
G.~P. Guedes, A.~Breskin, R.~Chechik and D.~Mormann, \emph{{Effects of the
  induction-gap parameters on the signal in a double-GEM detector}},
  \href{https://doi.org/10.1016/S0168-9002(02)01924-1}{\emph{Nucl. Instrum.
  Meth. A} {\bfseries 497} (2003) 305}.

\bibitem{Shalem:2005ix}
C.~K. Shalem, R.~Chechik, A.~Breskin, K.~Michaeli and N.~Ben-Haim,
  \emph{{Advances in thick GEM-like gaseous electron multipliers. Part II:
  Low-pressure operation}},
  \href{https://doi.org/10.1016/j.nima.2005.12.219}{\emph{Nucl. Instrum. Meth.
  A} {\bfseries 558} (2006) 468}
  [\href{https://arxiv.org/abs/physics/0601119}{{\ttfamily physics/0601119}}].

\end{thebibliography}\endgroup

\bibliographystyle{JHEP}

%%%
\appendix

\section{Reduced quantities with different ionization energies in Ar:CO$_2$ (70:30)}\label{sec:arco2}

In Ref.~\cite{Vahsen:2014fba}, a setup similar to 2-tGEM was used to measure gain with an Ar:CO$_2$ (70:30) gas mixture.  Events from $^{55}$Fe (\SI{5.9}{keV} X-rays) and $^{210}$Po (\SI{5.3}{MeV} $\alpha$-particles) were measured.  Due to the lack of precise information about the $^{210}$Po source location with respect to the sensitive volume, the measured ionization energy from the $\alpha$-particles was estimated to be approximately \SI{4}{MeV} with a large uncertainty.  The spectra associated with the $\alpha$-particles were integrated for \SI{30}{minutes} as the source rate was much lower than that of the $^{55}$Fe source. 

\begin{figure}[h!]
  \centering
  \includegraphics[width=\linewidth]{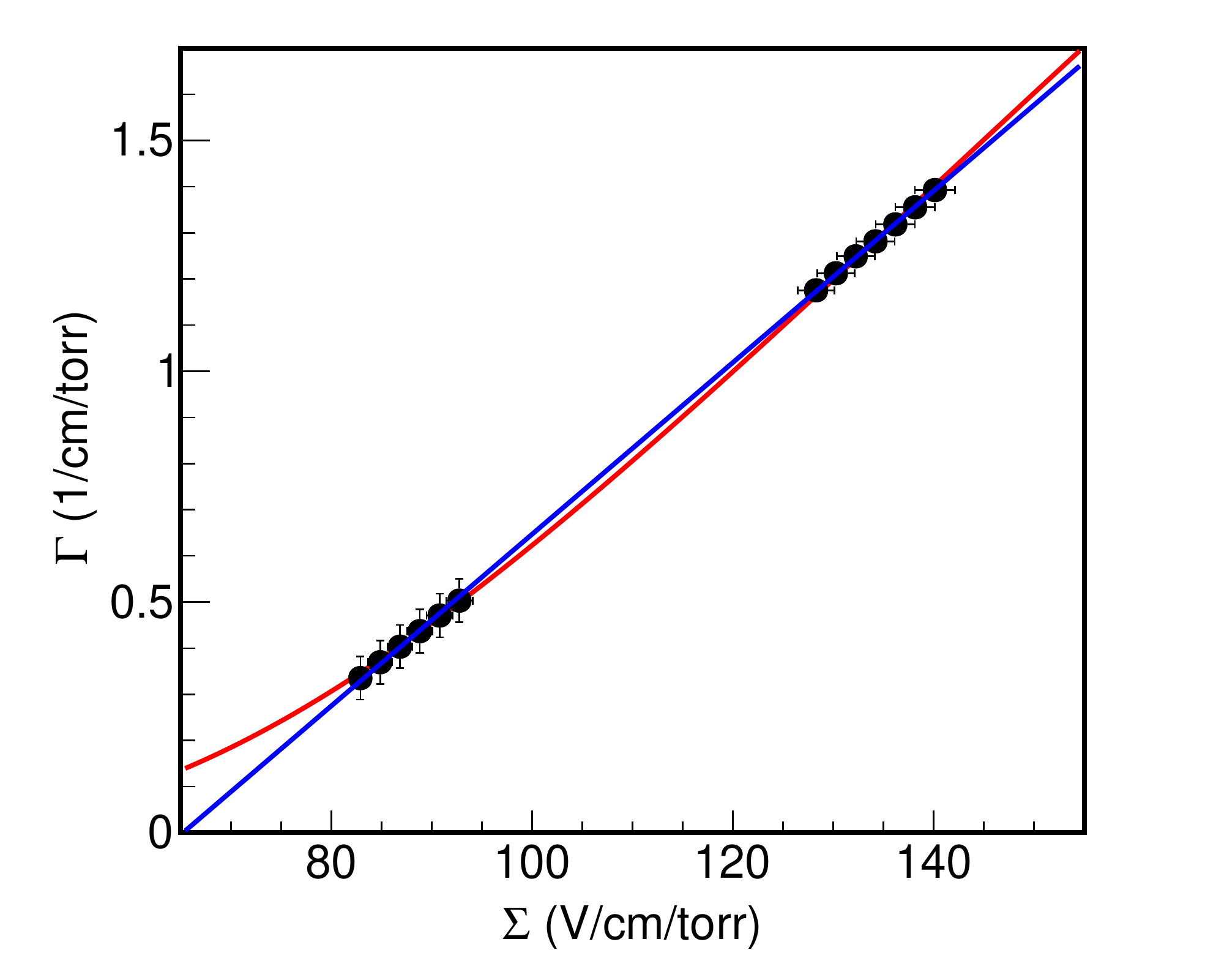}
  \caption{Reduced First Townsend Coefficient (RFTC) versus reduced field for gas gain measured with an Ar:CO$_2$ (70:30) gas mixture at atmospheric pressure.  Data recorded with two different energy sources, approximately \SI{4}{MeV} collected with alpha particles from $^{210}$Po (left grouping) and \SI{5.9}{keV} X-rays from $^{55}$Fe (right grouping), using a double thin $(\SI{50}{\text{$\mu$}m})$ GEM detector setup described in Ref.~\cite{Vahsen:2014fba} .  The solid lines are fits to the data of Eq.~\ref{eqn:red_gain_field} (blue online) and Eq.~\ref{eqn:reduced_gain} with $m=0$ (red online).}
  \label{fig:arco2_reduced_gain}
\end{figure}

Fig.~\ref{fig:arco2_reduced_gain} shows the RFTC versus the reduced field for the two different ionization energies measured in Ar:CO$_2$ at atmospheric pressure.  The group of data points on the left are the signals measured from the $\alpha$-particles, which produce roughly three orders of magnitude more ionization than the \SI{5.9}{keV} X-rays (group of data points on the right).  The solid lines are fits of Eq.~\ref{eqn:red_gain_field} (blue online) and Eq.~\ref{eqn:reduced_gain} with $m=0$ (red online) to the data.  For reference, a W-value of \SI{28.05}{eV} \cite{Sharma:1998xw} is used to determine the gain values discussed in Ref.~\cite{Vahsen:2014fba}.  It should be mentioned that the electric fields in the transfer and collection regions are approximately 40\% lower for the $\alpha$-particle measurements than for the \SI{5.9}{keV} X-rays.  This is because these fields are proportional to the GEM fields, which are lower for the $\alpha$-particle measurements.  This could be introducing some transfer inefficiencies resulting in lower measured gain than if these fields were held to the same values as with the $^{55}$Fe measurements.  This may explain why the lower gain values from each data set are not falling below either model, as observed in the He:CO$_2$ data. 

Regarding the PH resolution, as evident in Eq.~\ref{eqn:gain_res_fano_2} the asymptotic PH resolution will decrease with the square root of the ionization energy.  This is discussed more in Refs. \cite{Vahsen:2014fba} and \cite{Thorpe:2018irh}.  Over the range of reduced field considered here, the avalanche variance does change and is likely larger for the low gain values associated with the $\alpha$-particle measurements compared with the high gains with those from $^{55}$Fe.  In further analyses with PH resolution values obtained with different ionization energies, changes in the avalanche variance should be accounted for, along with any changes in the ionization energy and resulting changes of the Fano factor. 

\end{document}